\documentclass[onecolumn, linenumber]{aa} 

\usepackage{lineno}
\usepackage{epsfig}          
\usepackage{natbib}         
\usepackage{amssymb}        
\usepackage{pstricks}
\usepackage[varg]{txfonts}

\usepackage{graphicx}
\usepackage{graphics}
\graphicspath{{png/}}

 \bibpunct{(}{)}{;}{a}{}{,} 

\newcommand{\be}{\begin{equation}} \newcommand{\ee}{\end{equation}} \newcommand{\bea}{\begin{eqnarray}} \newcommand{\eea}{\end{eqnarray}} \newcommand{\ba}{\begin{array}} \newcommand{\ea}{\end{array}}     \newcommand{\bit}{\begin{itemize}} \newcommand{\eit}{\end{itemize}} \newcommand{\ben}{\begin{enumerate}} \newcommand{\een}{\end{enumerate}}  

\begin{document}

\title{Meridional Circulation Dynamics in a Cyclic Convective Dynamo}

\author{D. Passos
        \inst{1, 2, 3}
    \and M. Miesch \inst{4} \and G. Guerrero \inst{5} \and P. Charbonneau
        \inst{3}}

\institute{  CENTRA,  Instituto Superior T\'ecnico, Universidade de Lisboa,
              Av. Rovisco Pais, 1049-001 Lisboa, Portugal\\
 \email{dariopassos@ist.utl.pt}
\and
 Departamento de F\'\i sica, Universidade do Algarve,
       Campus de Gambelas, 8005-139 Faro, Portugal
\and
 D\'epartment de Physique, Universit\'e de Montr\'eal, C.P. 6128,
       Centre-ville, Montr\'eal, Qc, Canada  H3C-3J7
\and
    High Altitude Observatory,
        NCAR, Boulder CO 80301-2252, USA
\and
    Physics Department, Universidade Federal de Minas Gerais,
        Av. Antonio Carlos, 6627, Belo Horizonte, MG, 31270-901, Brazil
}

\titlerunning{Meridional Circulation Dynamics in a Cyclic Convective Dynamo}

\abstract {Surface observations indicate that the speed of the solar
meridional circulation
in the photosphere varies in anti-phase with the solar cycle.
The current explanation for the source of this variation is that inflows into active regions alter the global surface pattern of the meridional circulation.
When these localized inflows are integrated over a full hemisphere, they contribute to the slow down of the axisymmetric poleward horizontal component.
The behavior of this large scale flow deep inside the convection zone remains largely unknown.
Present helioseismic techniques are not sensitive enough to capture the dynamics of this weak large scale flow.
Moreover, the large time of integration needed to map the meridional circulation inside the convection zone, also masks some of the possible dynamics on shorter timescales.
In this work we examine the dynamics of the meridional circulation that emerges from a 3D MHD global simulation of the solar convection zone.
Our aim is to assess and quantify the behavior of meridional circulation
deep inside the convection zone, where the cyclic large-scale magnetic
field can reach considerable strength. Our analyses indicate that the
meridional circulation morphology and amplitude are both highly influenced by
the magnetic field, via the impact of magnetic torques on the global
angular momentum distribution. A dynamic feature induced by these magnetic torques is the development of a prominent upward flow at mid latitudes in
the lower
convection zone that occurs near the equatorward edge of the toroidal bands
and that peaks during cycle maximum.
Globally, the dynamo-generated
large-scale magnetic field drives variations in the meridional flow, in stark contrast to the
conventional kinematic flux transport view of the magnetic field being advected passively
by the flow. }

\keywords{Dynamo -- Magnetohydrodynamics (MHD)-- Sun: magnetic fields -- Sun: activity -- Sun: interior -- Sun: evolution}

\maketitle

\section{Introduction}

The solar magnetic cycle and the magnetohydrodynamic (MHD) dynamo that powers
it are the drivers of space weather, the latter impacting technological assets
ranging from power lines networks to satellite communications.
The solar dynamo is a MHD system in which the flow of plasma across a
pre-existing magnetic field induces more magnetic field, offsetting Ohmic
dissipation and in a manner leading to cyclic polarity reversals of the
large-scale magnetic component. The bulk of this field induction takes place
in the sun's convection zone (CZ), where convective turbulence and
differential rotation act as energy reservoir and primary inductive
flows. A detailed understanding of the dynamics of these flows,
including the nonlinear backreaction of the Lorentz force associated with
the dynamo-generated magnetic field, is therefore crucial for understanding
the dynamo process at a fully dynamical level.

The total velocity field can be decomposed into turbulent motions (from
granulation to the larger convective scales) and large-scale, globally
coherent motions organized on the scale of the sun itself. The latter can
be subdivided into (differential) rotation, fluid motion in the azimuthal
direction, $\mathbf{\hat{e}}_\phi$, and meridional circulation (MC),
motions in the  \{$\mathbf{\hat{e}}_r$, $\mathbf{\hat{e}}_\theta$\} plane.

Since the initial proposals of mean-field solar dynamo theory
\citep{Parker1955, Steenbeck1966}, differential rotation has been identified
as a critical ingredient for magnetic field amplification, through the so
called $\Omega$ effect. Thanks to the advances in helioseismology over the
last decades, this flow is now mapped for the interior of the Sun with a high
degree of confidence for a wide range of latitudes and depths
\citep{Howe2009}. Analytical profiles that closely match the observed
differential rotation profile are nowadays used in several kinematic
mean-field dynamo models \cite[see review by][]{Charbonneau2010}.

Increasing attention has been given to the meridional circulation following
the development of surface magnetic flux transport models
\citep{Wang1989, Wang1990} and flux transport (FT) dynamos
\cite[e.g.][]{vanBallegooijen1988, Choudhuri1995, Dikpati1999}. It also gained
particular visibility as a key ingredient in the Babcock-Leighton dynamo
framework (BL) \citep{Babcock1961, Leighton1969, Wang1991},
since it carries the product of decayed active regions from low latitudes
to the poles in these models. It is believed that this magnetic flux transport
is what eventually reverses the polar field and switches the polarity of the
solar magnetic dipole.

Observational data of the deep meridional flow is challenging
to obtain because of its relatively low amplitude when compared to the
omnipresent background of convective turbulence. This implies that the
inclusion of the MC profile in
dynamo models is prone to more uncertainties (when compared to the
differential rotation) given the unavailability of a
complete mapping throughout the convection zone. Until recently the MC was
only known with confidence in the near surface layers where a mean poleward
flow of around 15$\sim$20 m s$^{-1}$ is observed \citep{Duval1979, Giles1997,
Gizon2003, Ulrich2010}.
In the past, this led mean-field modelers to adopt simple MC profiles
(one cell per hemisphere) using extrapolations based on mass conservation in
the solar interior \cite[e.g.][]{vanBallegooijen1988}.
Nowadays, improved helioseismic
techniques allow the measurement of the MC between $\pm$ 60º in
latitude and down to 0.75 R$_\odot$.
Nevertheless, these extended measurements
are not yet conclusive as different groups obtain different radial
profiles for the MC \cite[cf.][]{Zhao2013,Jackiewicz15,Rajaguru2015}. One
thing is almost certain: the meridional circulation profile is more
complex than the commonly used one-cell-per-hemisphere configuration,
with the possibility of having multiple cells stacked in radius and
in latitude.

The first evidence of variations in the amplitude of the
MC came from the correlation tracking measurements of surface features
presented by  \cite{Komm1993}. Inspired by these
early results, it was later proposed that coherent variations
(on the cycle time scale) in the MC strength could
directly account for the
variations of the amplitude of the solar cycle
\cite[e.g.][]{Nandy2004, Passos2008, Lopes2009, Karak2010}.
Evidence that the MC actually exhibits this type of moderate
variations came from the magnetic feature tracking measurements
of \cite{Hathaway2010} and \cite{Upton2014}.  These authors showed that the surface poleward
meridional flow varies in anti-phase with the solar cycle, decreasing its
amplitude around solar maximum and speeding up again towards
solar minimum.
This type of solar-cycle related large scale flow perturbation is also visible
in the rotation through the torsional oscillations \citep{Howard1980}.
This suggests that the source mechanism for torsional oscillations and MC
cyclic variations has a magnetic origin.
Recent observations of the equatorward drift velocity
of the sunspot butterfly wings, impose strong constraints on the variation of
the meridional circulation at the storage depth of the toroidal flux
tubes which, upon destabilization and buoyant rise across CZ, will emerge
as active regions. According to \cite{Hathaway2011} the drift velocity is
independent of cycle strength, which means that either the MC is constant
at the depth where the toroidal field accumulates or it might not be directly
responsible for the equatorward migration of the activity belt.

The MC role in dynamo theory is that of a transport mechanism of magnetic
flux between different regions. In BL flux-transport dynamo models
it transports magnetic flux towards the equator at the base of the
CZ, and toward the poles at the surface. In models operating in the advection
dominated regime it also couples the bottom of the CZ, where the
strong radial shear is located, with the photospheric layers where
the Babcock-Leighton mechanism operates. However, the relevance of
this coupling mechanism can be deemed less relevant if faster
transport processes are considered (e.g. buoyancy, turbulent
pumping or high magnetic diffusivity). For this reason,
and because of different parameterizations,
the impact that MC amplitude variations
have in the cycle strength varies between models operating in the
advection or diffusion dominated regimes
\citep{Yeates2008, Lopes2009, Nandy2011, Karak2014}.
Most of the times the authors loosely attribute these changes to the Lorentz
force feedback.
These models typically run in the so called kinematic regime, i.e. the
background velocity field (differential rotation
and MC) induces and organizes the magnetic field, but the latter does
not feedback into the velocity field.
A few exceptions in this respect are
the non-kinematic mean-field simulations of \cite{Rempel06} and the
low order dynamo simulations of \cite{Passos2012} which both include
magnetic feedback onto the MC.

An alternative to the Lorentz force feedback to explain why the
MC varies, had its origins in the
magnetic feature tracking measurements of \cite{Meunier1999}. The author
identified variations in the surface meridional flow that can be
associated with local inflows into active regions.
Based on earlier ideas by H.C. Spruit and on these measurements,
\cite{Cameron2010, Cameron2012} showed that, in the context of the surface
flux transport (SFT) model framework, inflows into active regions could
account for the global variations measured in the surface meridional flow.
The cyclic surface MC variations are characterized by a weakening
of the meridional flow on the poleward sides of the active region belt.
This can be interpreted as the joint action of inflows toward
the sunspot areas superimposed on a mean poleward meridional flow \citep{Hathaway2014}.
SFT simulations also have shown that variations in the amplitude
of the MC can have an important impact in the amplitude of the next solar
cycle because it determines the amount of polar field that is available for
the next-cycle production \citep{Jiang2013, Upton2014, Martin-Belda2016}.
The bottom line is that for both types of models, axisymmetric FT and SFT,
the dynamics of the MC is a key element that needs to be
better understood.

From a fluid dynamics point of view, the MC arises from angular momentum
redistribution in the presence of rotation, convection and
thermal gradients \citep[e.g.][]{Tassoul1982, Rudiger1989, Tassoul2000}.
Therefore, one of the most appropriate ways to study the MC is by means of
global 3D MHD simulations.  Such simulations allow to disentangle the
complex chain of interactions that the various physical processes create.
Several authors have investigated the physical origin of differential rotation
and MC profiles obtained in global simulations
\cite[for a detailed review see][]{Miesch2005}.
From analysis based on the balances of the angular momentum and the
thermal wind balance, there seems to be an agreement to the conjecture that the
meridional motions appear as the consequence of the turbulent transport
of angular momentum \citep[e.g.][hereafter FM15]{Guerrero2013,Gastine14,Featherstone2015}.
However, these studies were conducted in purely hydrodynamic simulations
without considering the effects of magnetism.

Here we present an analysis that follows a similar
methodology as in \cite{Brun2011}, \cite{MieschHindman2011} and
FM15 but for a 3D MHD global simulation
that develops a large-scale magnetic field cycle \citep[][henceforth PC14]{PC14}.
The MC profile obtained in this model was presented in \cite[][henceforth PCM15]{PCM15}
 and exhibits interesting parallels to the
helioseismic profile measured by \cite{Zhao2013}.
Another motivation for this work comes from the early realization that this
meridional flow varies in intensity along
the magnetic cycle as noticed in \cite{Passos2012}.

In this work we describe how the large-scale magnetic field interacts
and modifies the MC profile in this simulation and
extrapolate our findings to the solar case.
The paper is organized as follows. In sections 2 and 3 we present the
3D model used and describe the MC profile obtained. In section 4 we conduct
an angular momentum balance analysis and study the role of the large scale
magnetic field in inducing variations in the MC through gyroscopic pumping (GP).
In section 5 we develop an equation for the meridional force balance in the
presence of magnetic fields and study the contribution from its various
terms to the variations in the MC cell structure. We conclude with a
discussion about the relevance of our findings to the actual case of the Sun.

\section{The model}

The results presented here are obtained through a dynamics analysis to
the \textit{EULAG millennium simulation} described in PC14, an Implicit Large-Eddy
Simulation (ILES) of global solar convection produced with the EULAG-MHD
code \citep{SmolarCharb2013}.
The model solves the MHD extension of the anelastic equations of
\citet{LippsHemler1982} in a spherical shell defined between
$0.61 < r/R_\odot < 0.96$. The convection is driven via a volumetric
heating/cooling term in the energy equation. The domain is
gravitationally-stratified, rotates at the solar rate, and includes a convectively stable fluid layer underlying the convection zone. The governing
equations read:
\bea
    \frac{D \mathbf{u}}{D t} &=& -\nabla \pi'
            - \mathbf{g}\frac{\Theta'}{\Theta_o}
            + 2\mathbf{u} \times \mathbf{\Omega}
            +\frac{1}{\mu \rho_o}
            (\mathbf{B} \cdot \nabla) \mathbf{B}\, , \label{eq:momentum}\\
    \frac{D \Theta'}{D t} &=& - \mathbf{u}\cdot \nabla \Theta_e
            -\alpha \Theta' + \mathcal{H} \, , \\
    \frac{D \mathbf{B}}{Dt} &=& (\mathbf{B}\cdot \nabla) \mathbf{u}
            -\mathbf{B} (\nabla \cdot \mathbf{u}) \, , \\
    \nabla \cdot (\rho_o \mathbf{u}) &=& 0 \, , \\
    \nabla \cdot \mathbf{B} &=& 0 \, ,
\eea
where $D/Dt\equiv \partial/\partial t+\mathbf{u}\cdot\nabla$
is the convective (Lagrangian) derivative,
$\mathbf{u}$ is the flow velocity, $\mathbf{B}$ is the magnetic field,
$\Theta$ is the potential temperature,
$\mathbf{\Omega}=\Omega_0 (\sin\theta,\cos\theta,0)$
is the angular velocity with $\theta$ being the latitude,
$\mu$ the magnetic permeability, $\mathcal{H}$ symbolizes radiative diffusion,
$\mathbf{g}$ is the gravitational acceleration, $\rho_o$ the density
stratification, $\alpha=1/\tau$ defines the time-scale of the Newtonian
cooling term that drives convection and $\pi'$ is a
density-normalized pressure
perturbation that subsumes centrifugal forces and magnetic pressure.
Prime quantities represent perturbations with respect to an
arbitrarily selected
ambient state (denoted by the subscript \textit{'e'}). Quantities related
with the basic state of the anelastic asymptotic expansion are denoted by
the sub-script \textit{'o'}. A more detailed description of the model and its
parameters can be found in \cite{Ghizaru2010, Racine2011, SmolarCharb2013, Cossette2014, Guerrero2016}.

The numerical scheme used in the EULAG-MHD simulations does not consider
any explicit dissipative terms. The MPDATA algorithm introduces a sub-grid
scale numerical viscosity at the minimal level required to maintain
stability \citep{Margolin2006}. This effectively maximizes the Reynolds numbers and turbulence levels of the model for a given grid size. For the
grid resolution used in this simulation, \citet{Strugarek2016} estimates an
effective viscosity of the order of $1\times10^{12}$ cm$^2$ s$^{-1}$
for the convection zone.
These authors show that even in an average sense this dissipation
is strongly scale-dependent, varying by more than a factor of 10 between
global and convective scales (see in particular their Fig. 5 and accompanying discussion). Using this information and the rms velocity in the bulk of
the CZ we can say, as a conservative estimate, that this simulation is operating in a Reynolds number between 40 and 50.
\citet{Strugarek2016} also show that in this simulation the Prandtl number is slightly higher than 1 which is consistent with the values assumed for ILES simulations. Taking an average value for the vorticity in the middle of the convection zone and the solar rotation rate we can estimate a Rossby number of approximately 0.02. Comparison to equivalent dimensionless numbers computed
from simulations using explicit dissipation is meaningful only up to a point.
The MPDATA algorithm at the core of EULAG introduces dissipation in a
spatiotemporally intermittent manner, in response to the development of
strong local gradient in advected variables.

The low viscosity regime allows us to reproduce a tachocline by
placing a convectively stable layer at the bottom of the domain. The
shear layer develops naturally and persists for a time scale much longer
than the dynamo cycle period.  Including the tachocline in the model allows
to reproduce some of the features of the solar magnetism, the most
important being the generation of a deep seated strong toroidal magnetic
field which seems to govern the field dynamics in the entire domain.
For instance, it drives the development of magneto-shear instabilities and
the generation of non-axisymmetric turbulent modes in the stable layer
that might influence the dynamo period \citep{Lawson2015,Guerrero2016}.
Furthermore, \cite{Guerrero2016b} argue that the magnetic tension
due to the large scale field at the tachocline
induces the speed-up and slow-down of the axial motions, ultimately
establishing a torsional oscillations pattern \citep[see also][]{Beaudoin2013}.

The \textit{millennium simulation} was performed on a mesh of
$N_\phi \times N_\theta \times N_r = 128 \times 64 \times 47$. This
relatively low resolution allows for a long integration time, generating
a solution extending over more than 1600 years which spans about 41 polarity
reversals (half magnetic cycle with a period of $\sim$40 yr).  The numerical experiments carried out by \citet{Strugarek2016} show that for this
resolution EULAG-MHD solutions agree with those obtained with the pseudo-spectral code ASH, which in turn has been accurately benchmarked against four other pseudospectral codes \citep{Jones2011}.
While this simulation generates rotational torsional oscillations
of solar-like amplitude \citep[see][]{Beaudoin2013}, it does
not reproduce the near-surface shear layer
\citep{MieschHindman2011, Hotta2015, Guerrero2016b}, nor the inflows
associated with active regions \citep[e.g.][]{Cameron2012,Martin-Belda2016}.
Therefore the relevance of these effects for the global
meridional circulation is not addressed here.

\section{Flows in the meridional plane}

To highlight the influence that the magnetic field has on
the dynamics of the large-scale meridional flows, we compare first the
morphology of the MC of the \textit{millennium simulation} (MHD) with a purely
hydrodynamic (HD) analog simulation spanning nearly 330 yr  for the same
resolution and physical parameters.
To ensure that the time intervals in which we will
study the flow dynamics is not influenced by the initial conditions,
we excluded the first 82 years of the HD and 873 years of the MHD simulated data from this analysis.

\begin{figure*}[htb]
        \centering
        \includegraphics[height=7 cm]{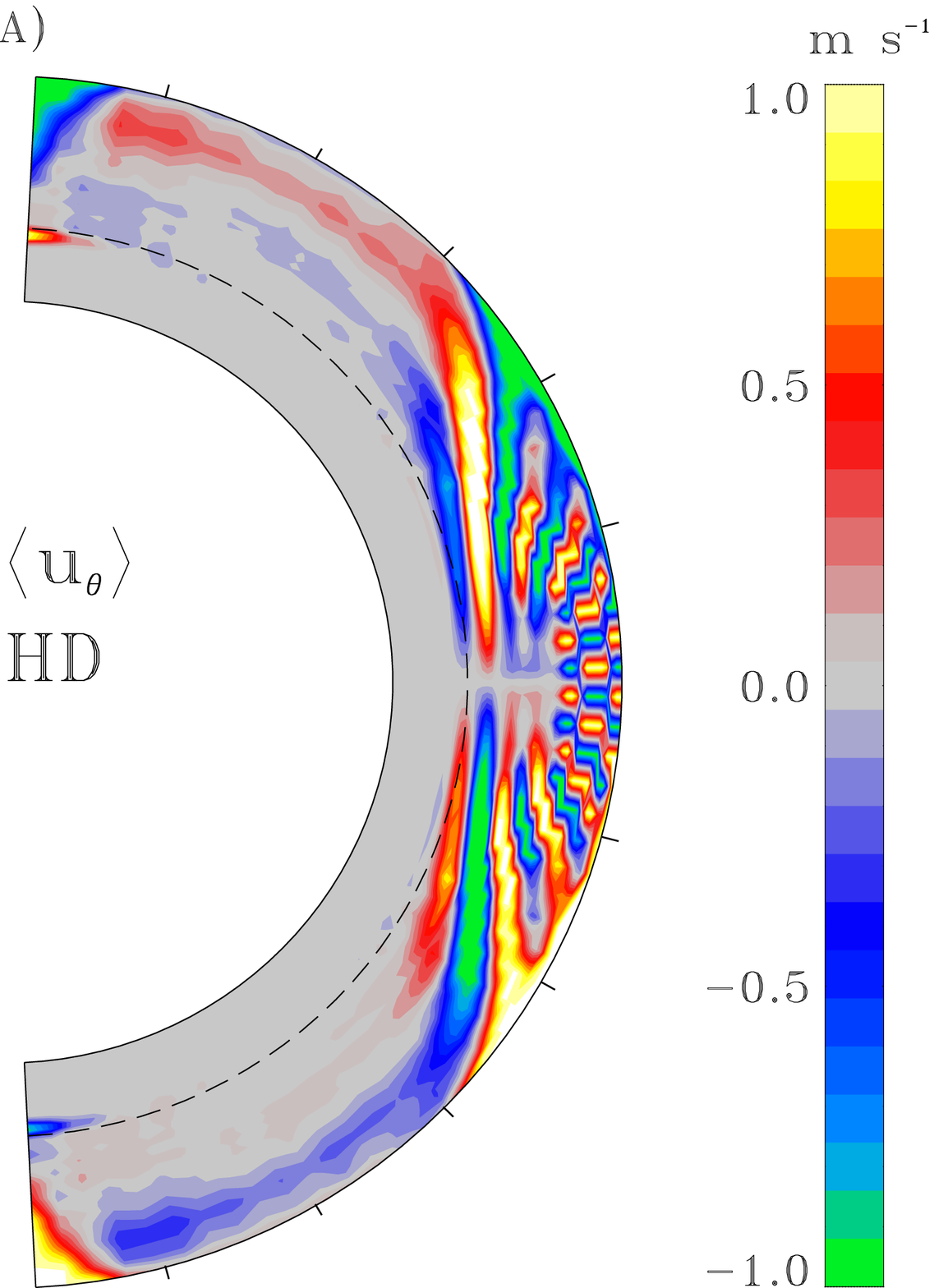}
        \hspace{1cm}
        \includegraphics[height=7 cm]{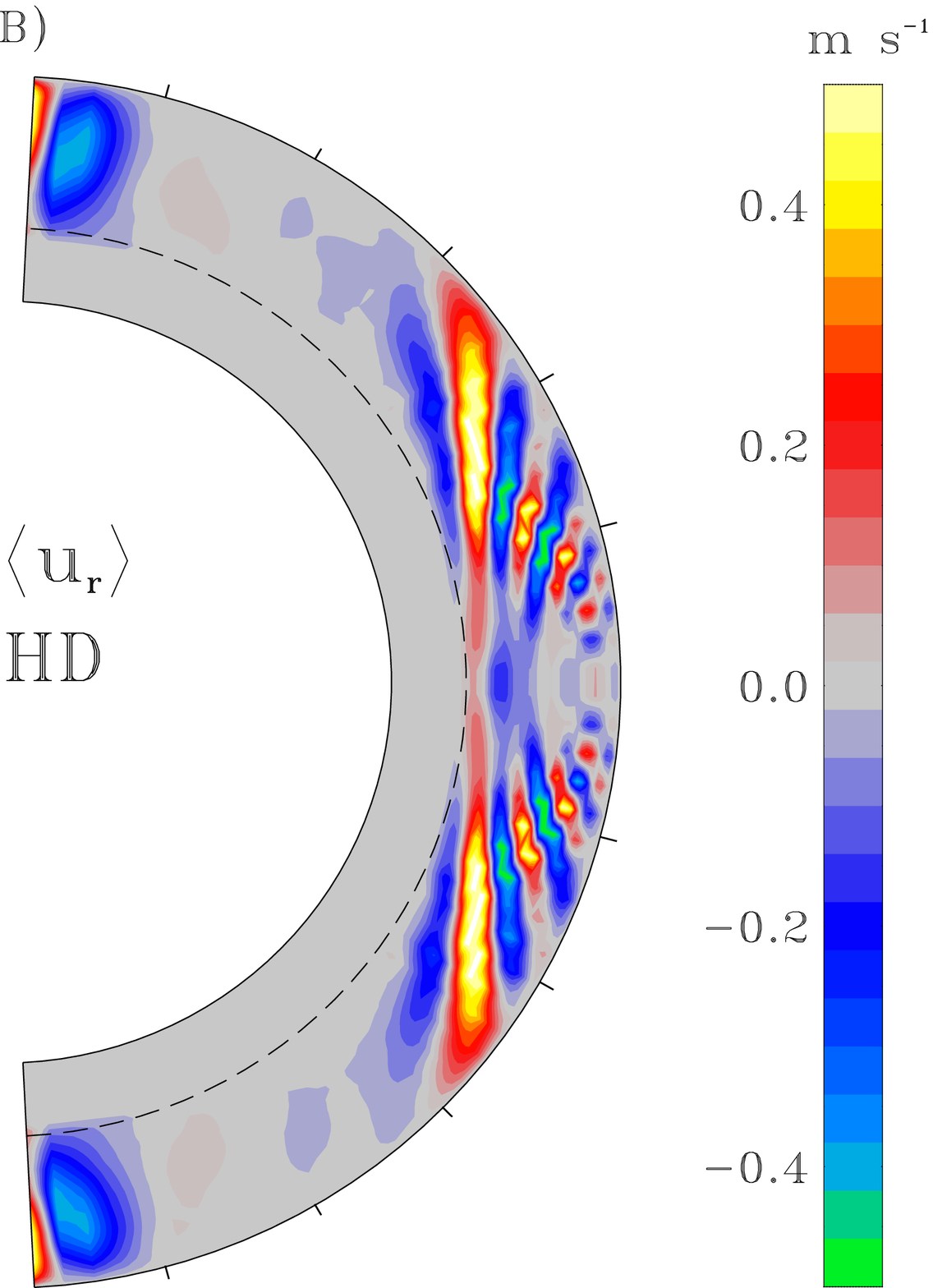}
        \hspace{1cm}
        \includegraphics[height=7 cm]{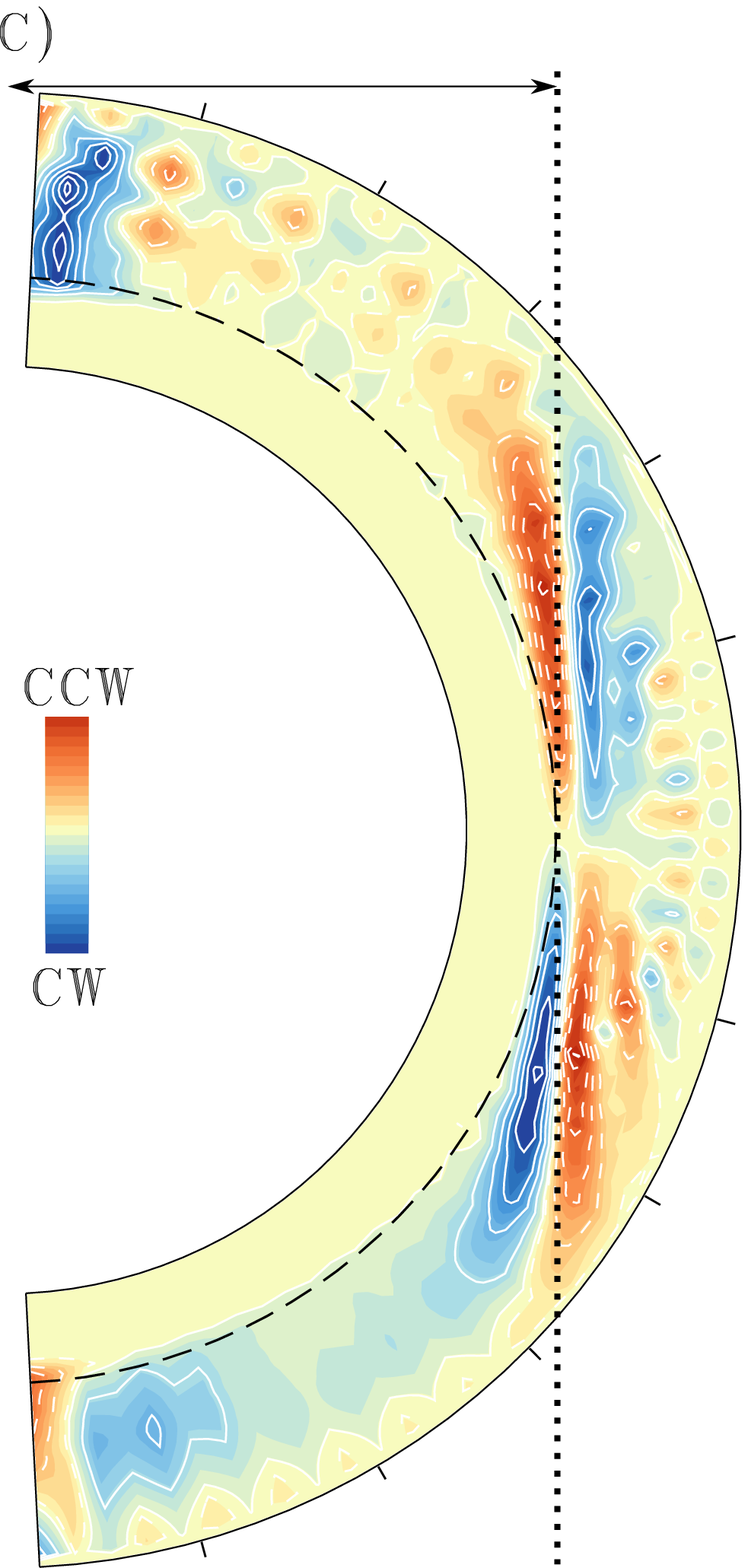}\\
        \vspace{0.8cm}
        \includegraphics[height=7 cm]{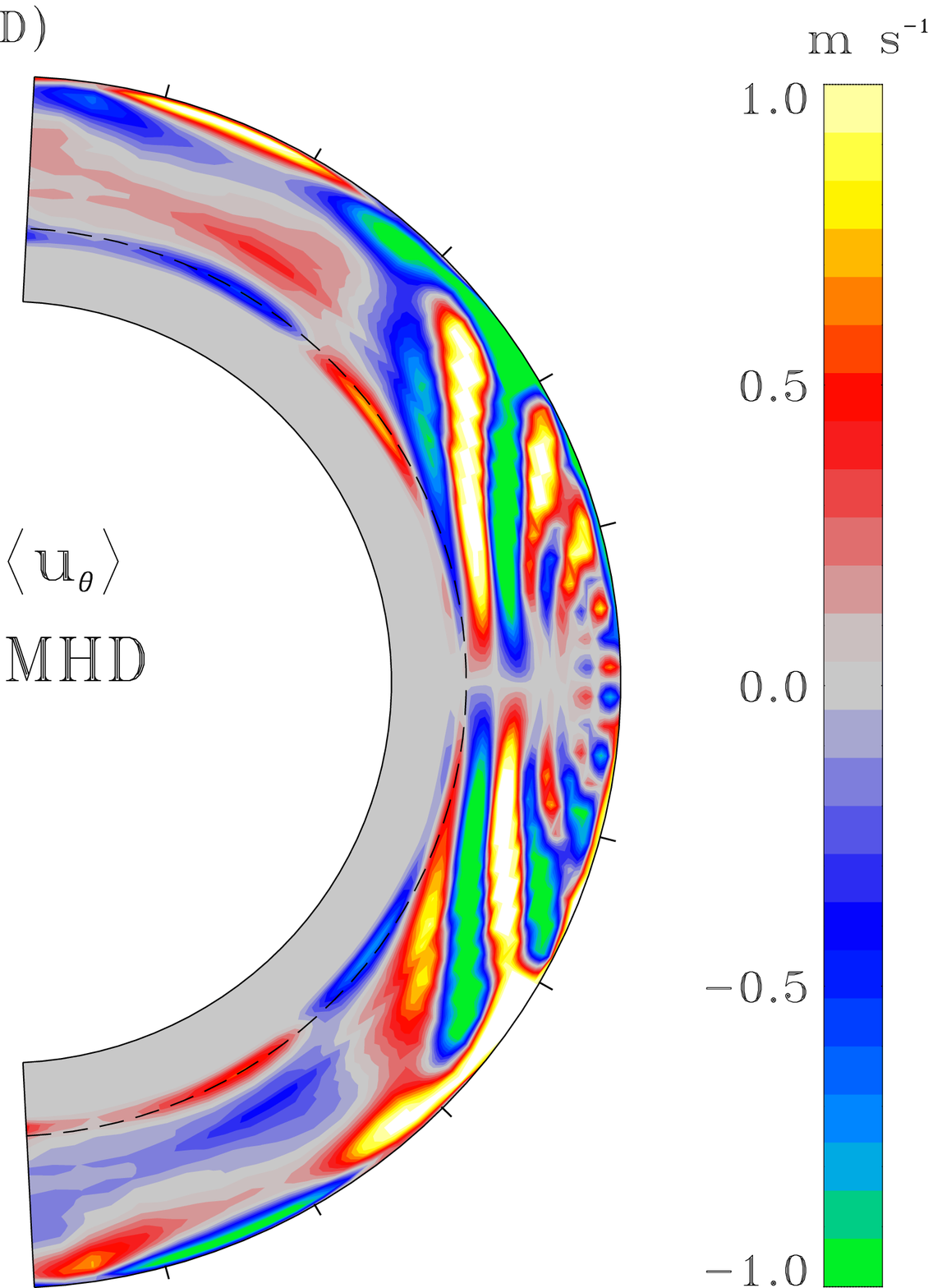}
        \hspace{1cm}
        \includegraphics[height=7 cm]{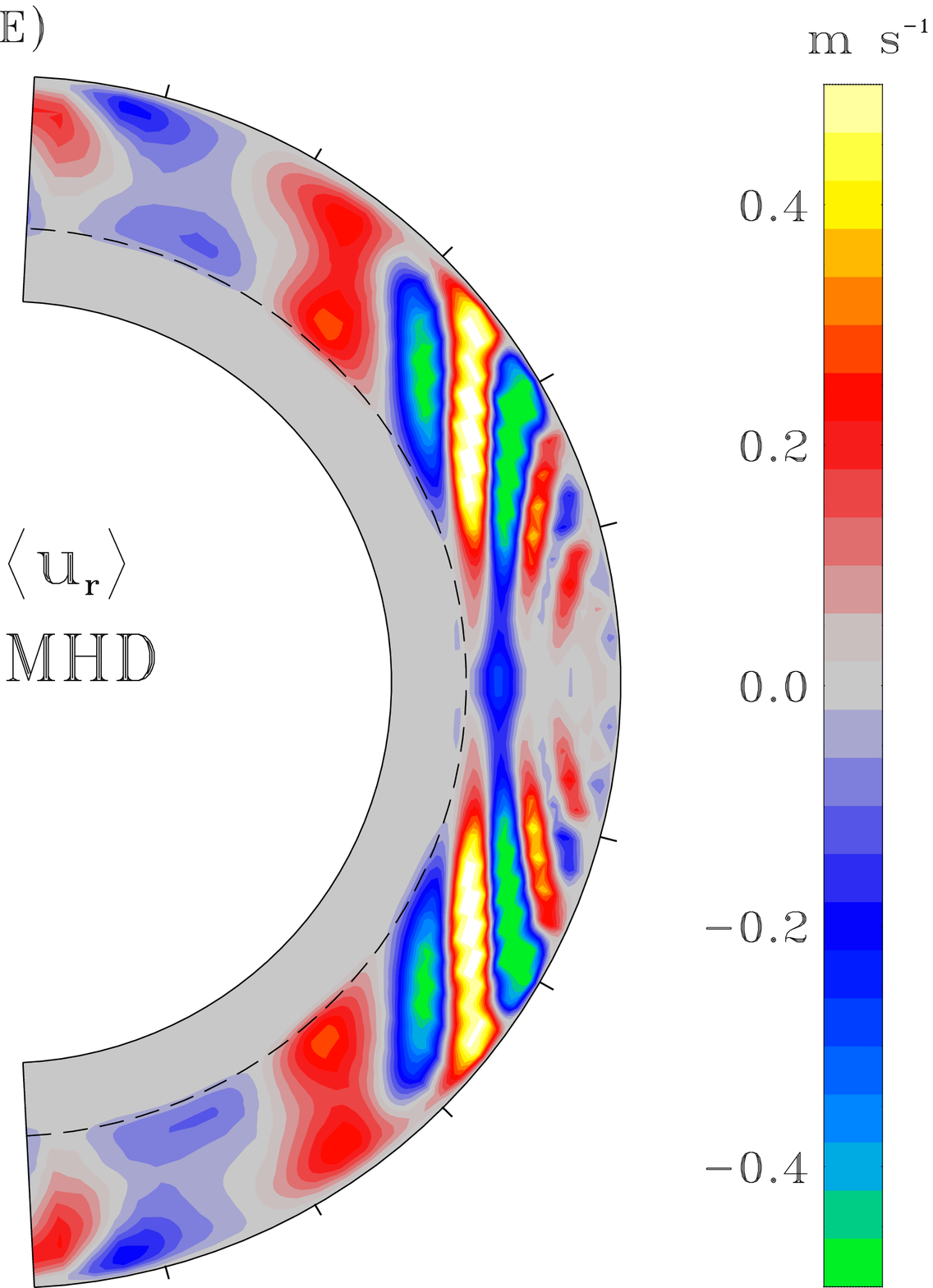}
        \hspace{1cm}
        \includegraphics[height=7 cm]{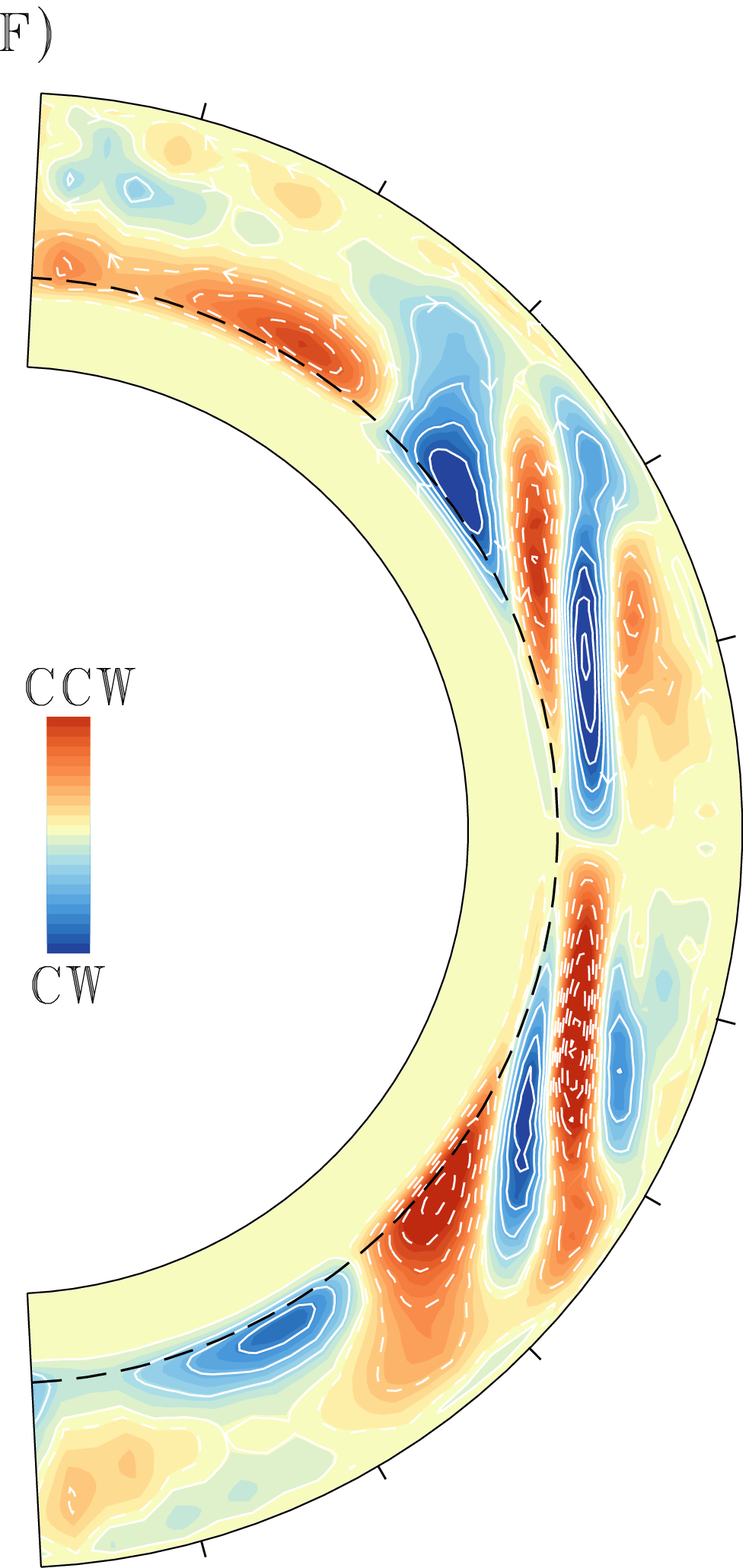}
        \caption{Meridional plane displays of time averaged $\langle u_r \rangle$, $\langle
        u_\theta \rangle$ and streamfunction. The top row is for a HD
        simulation, while the bottom row shows the same
        quantities for a MHD counterpart.
        For the HD (MHD) case, the temporal average corresponds to
        246 (406) yr.  In panels A)
        and D), red/yellow (blue/green) denotes flows towards the north (south).
        In panels B) and E) red (blue) denotes rising (sinking) flows.
        In panels C) and F) we present the streamlines of the MC.
        The red (blue) tones denotes circulation in the counterclockwise
        (clockwise) direction. The black dashed line indicates the depth
         of the base of the convection zone. Below it we have the stable layers. The vertical dotted line
         in C) indicates the position of the tangent cylinder for reference.
         At the top layer, thin ticks mark 15$^\circ$ intervals.
         Panel C) can be directly compared with Fig.\ 1g)
         of FM15.
        }
        \label{fig:vmoy}
\end{figure*}

As usual, the MC radial and latitudinal components, $\langle u_r \rangle$ and
$\langle u_\theta\rangle$, respectively, are obtained by zonally averaging
these components of the velocity field. In Fig.\ \ref{fig:vmoy} we present
these quantities and the corresponding streamfunction averaged over a time interval of $\sim$246 yrs (for the HD simulation) and $\sim$406 yrs (for
the MHD).

In both cases (HD and MHD), $\langle u_r \rangle$  and
$\langle u_\theta\rangle$ have a columnar-like morphology at low latitudes,
with several thin cells parallel to the rotation axis (in agreement with the
Taylor-Proudman constraint imposed by rotation). This behavior is
typical of other similar global MHD simulations \cite[see][for examples]{Miesch2005}.  The small scale pattern that appears near the
top layers at low latitudes is an artifact from both, the low numerical viscosity of the code, and the time average of multiple asymmetric
larger-scale flows over time.

Things become much more distinct between the two cases in the area contained
inside a cylinder tangent to the base of the convection zone (in the equator)
and parallel to the rotation axis (highlighted in Fig.\ \ref{fig:vmoy}C)
by the vertical dotted line and the horizontal arrow).

For the HD case, inside the tangent cylinder (TC) we observe two circulation
cells. One of them has counterclockwise rotation (in the north) and spreads
from the cylinder border to higher latitudes. This big cell is easier to
identify in the southern hemisphere of panel \ref{fig:vmoy}C. There is
another small cell, rotating clockwise (in the north) nearby the pole
(see panels \ref{fig:vmoy}A, \ref{fig:vmoy}B and \ref{fig:vmoy}C).

Since outside the TC the flows (axis-oriented cells) have
much higher velocities we purposefully used a low color saturation threshold
in these figures in order to reveal the MC components everywhere.
Otherwise, only the cells outside the tangent cylinder would be visible.

For the MHD case, on the other hand, Figs.\ \ref{fig:vmoy}D and
\ref{fig:vmoy}E show a different scenario. We notice that inside the
TC the MC pattern, is much more complex than in the HD case.
In this region, the flow morphology shows two main cells of opposite
circulation, with a prominent upflow between them at around $\pm 48^\circ$
latitude in the bottom half of the CZ. Near the poles we have
several smaller cellular structures (see Fig.\ \ref{fig:vmoy}F).

\subsection{Cyclic variations of the MC}

\begin{figure*}[htb]
  \centering
     \includegraphics[width=17 cm]{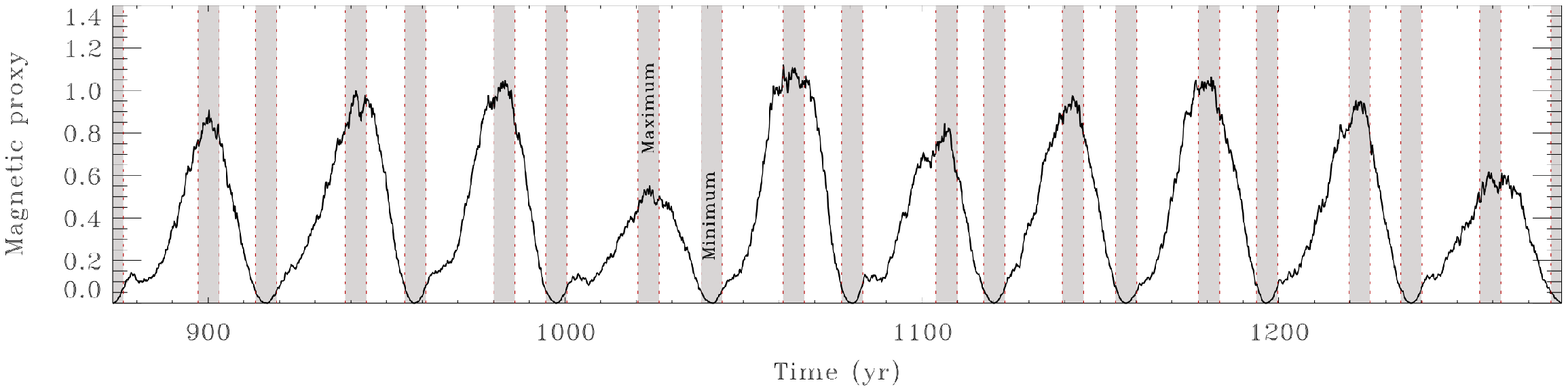}
     \caption{Magnetic cycle proxy (for the northern hemisphere).
     The vertical gray areas represent the maxima and minima phases considered.}  For reference we labeled these phases for simulated cycle
    number 24.
  \label{fig:bxmoy_proxy}
\end{figure*}

The morphology of the MC evolves as the magnetic cycle progresses and these
changes are significant especially inside the TC. This region
coincides with the location where the large scale magnetic cycle develops and
where other solar-like dynamic phenomena such as torsional oscillations
are observed \citep{Beaudoin2013}.
In order to investigate how the MC evolves with the magnetic cycle, we
consider a time interval of 10 sunspot-like cycles (half magnetic cycles),
spanning from cycle 21 to cycle 30 of the \textit{millennium simulation}
and covering a period of 406 yr (see Fig.\ \ref{fig:bxmoy_proxy}).

\begin{figure*}[htb]
  \centering
     \includegraphics[height=6.5 cm]{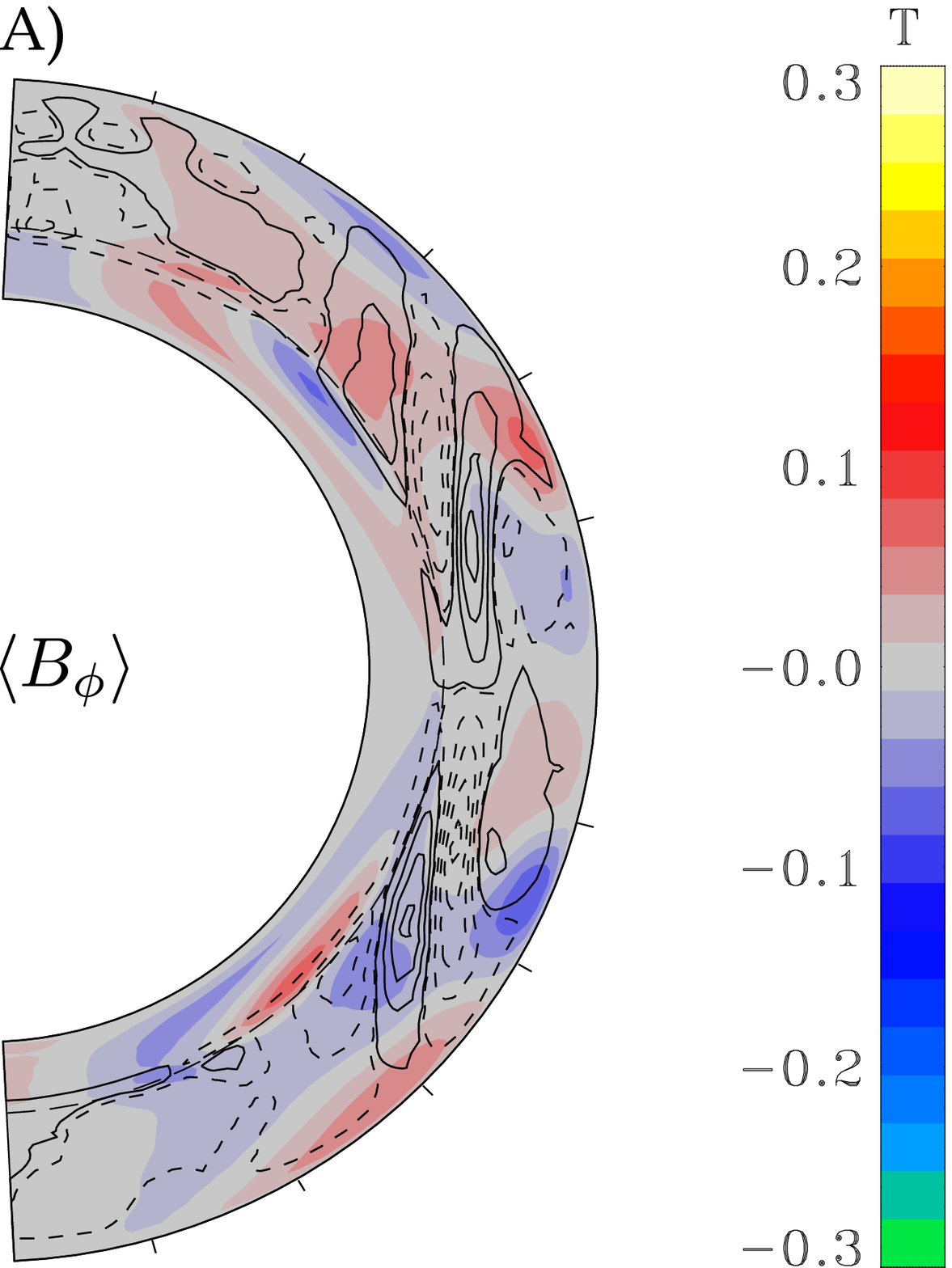}
     \hspace{1.3cm}
     \includegraphics[height=6.5 cm]{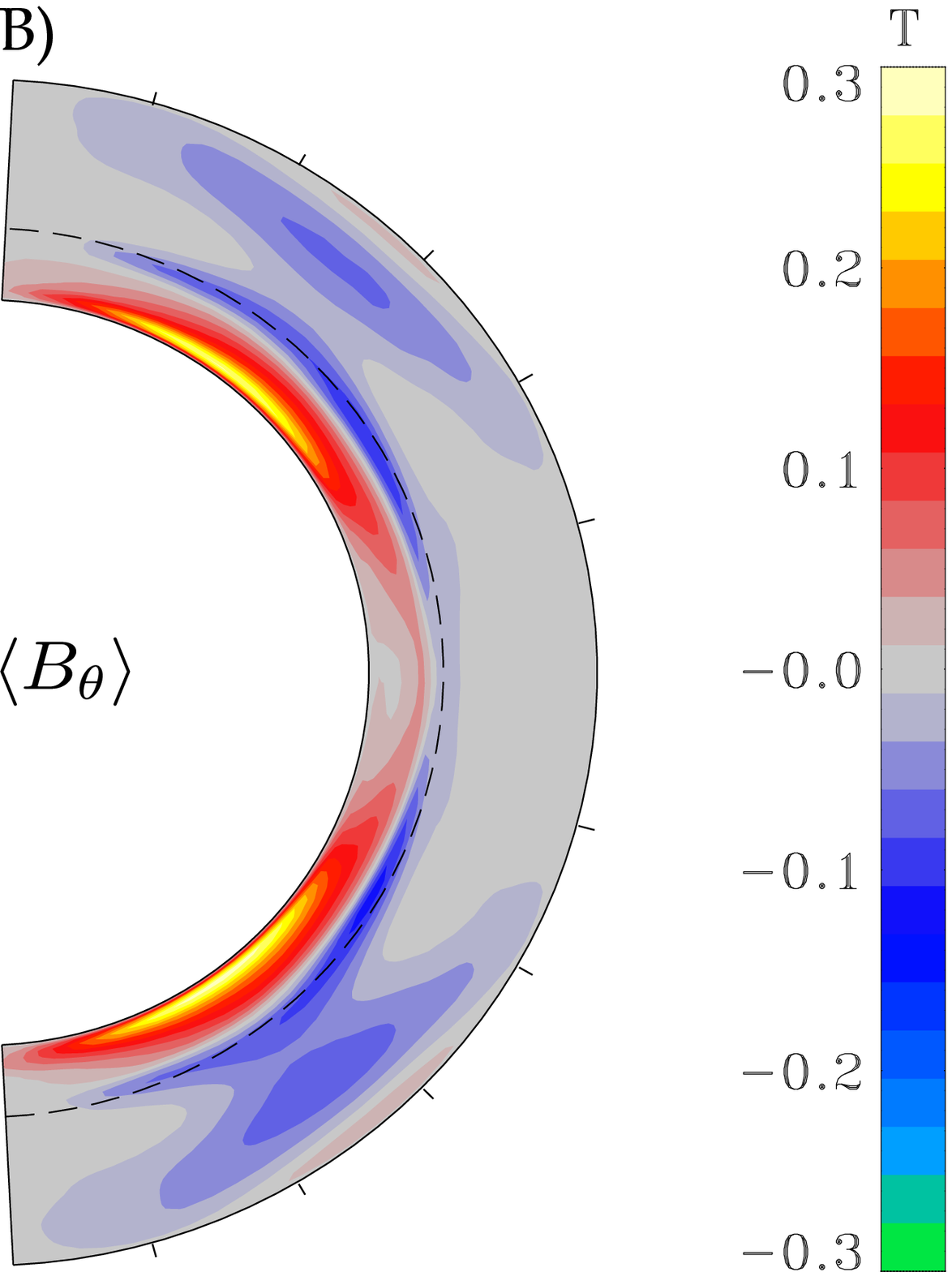}
     \hspace{1.3cm}
     \includegraphics[height=6.5 cm]{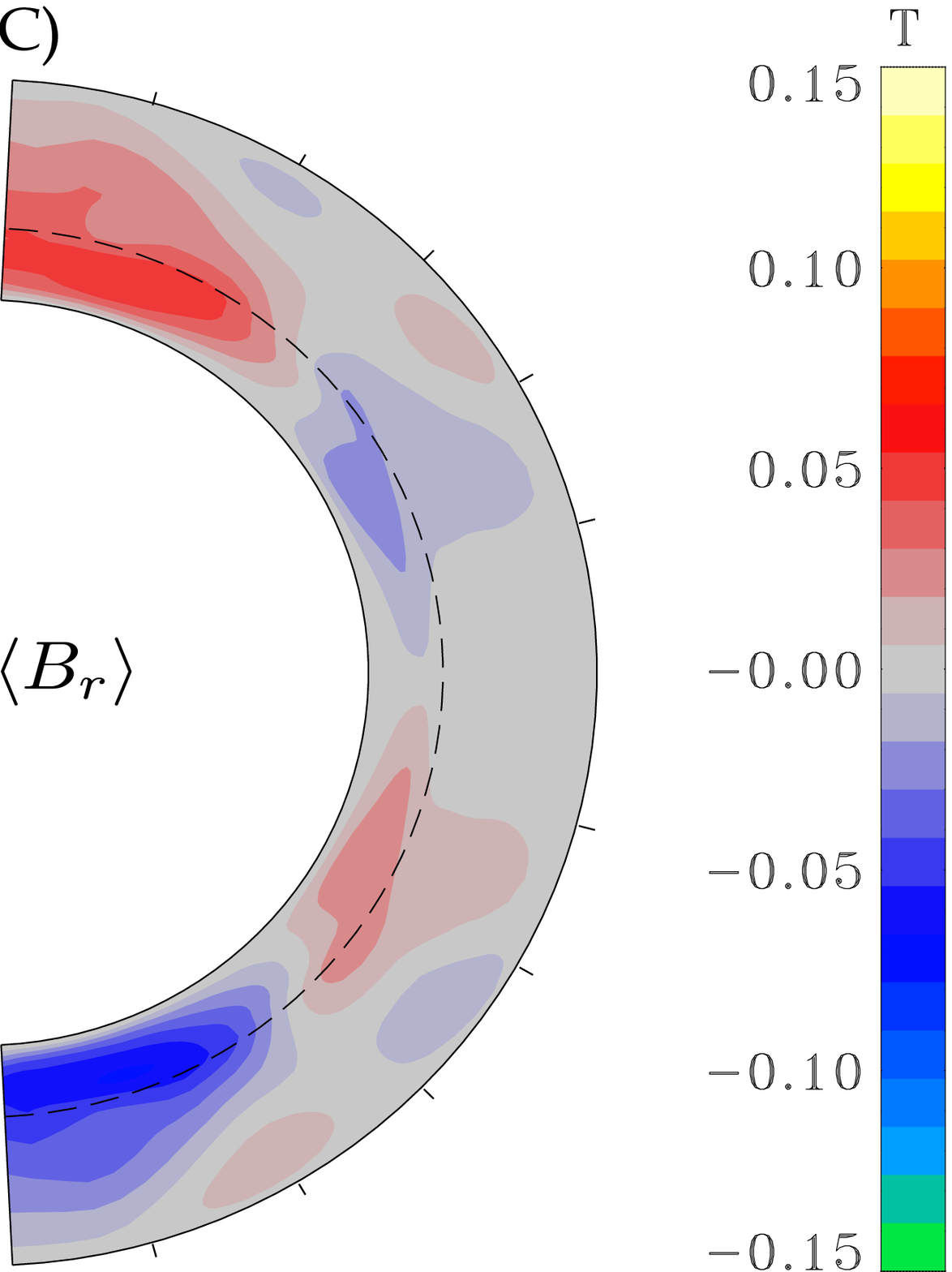}\\
     \vspace{0.5cm}
     \includegraphics[height=6.5 cm]{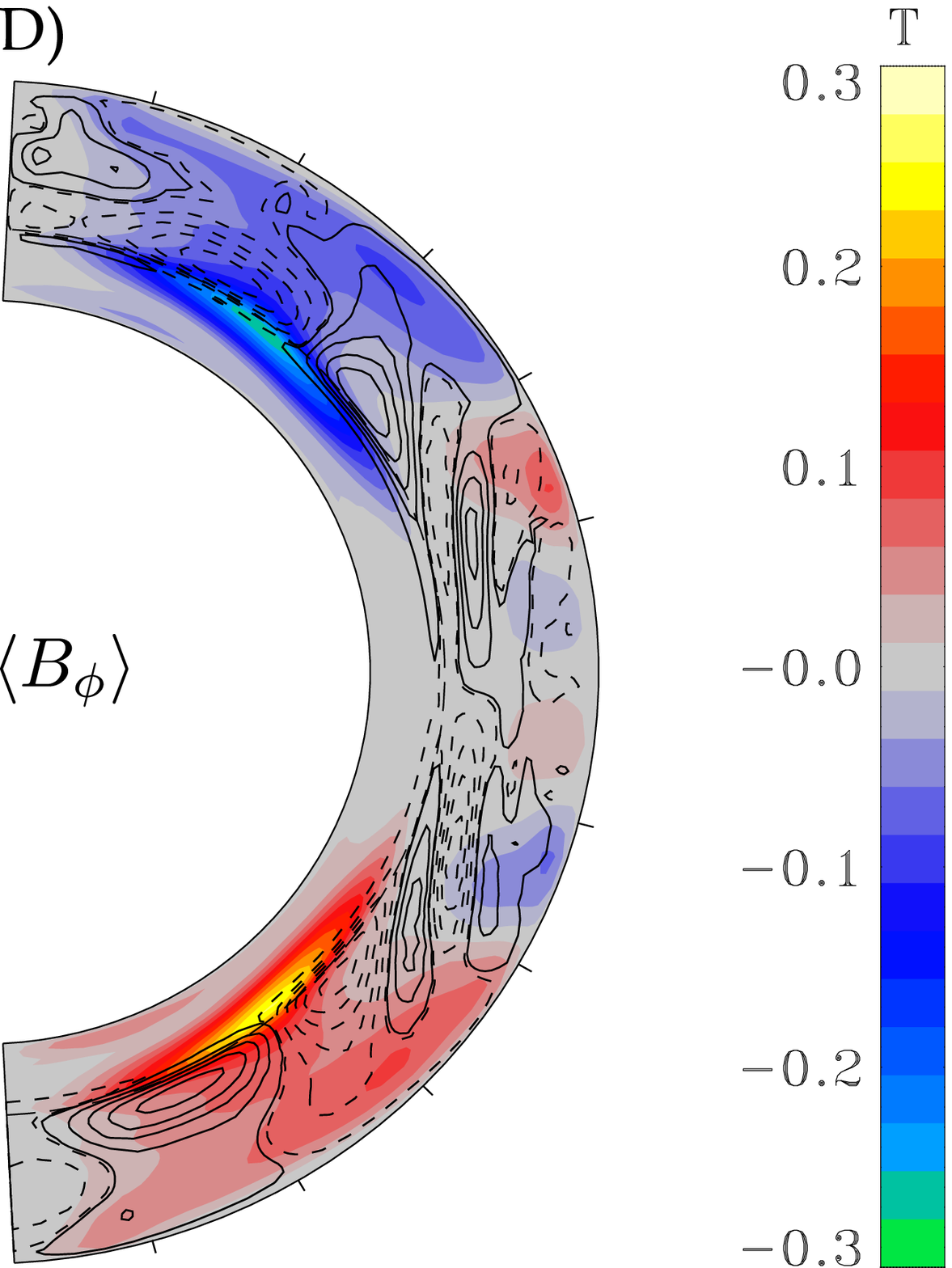}
     \hspace{1.3cm}
     \includegraphics[height=6.5 cm]{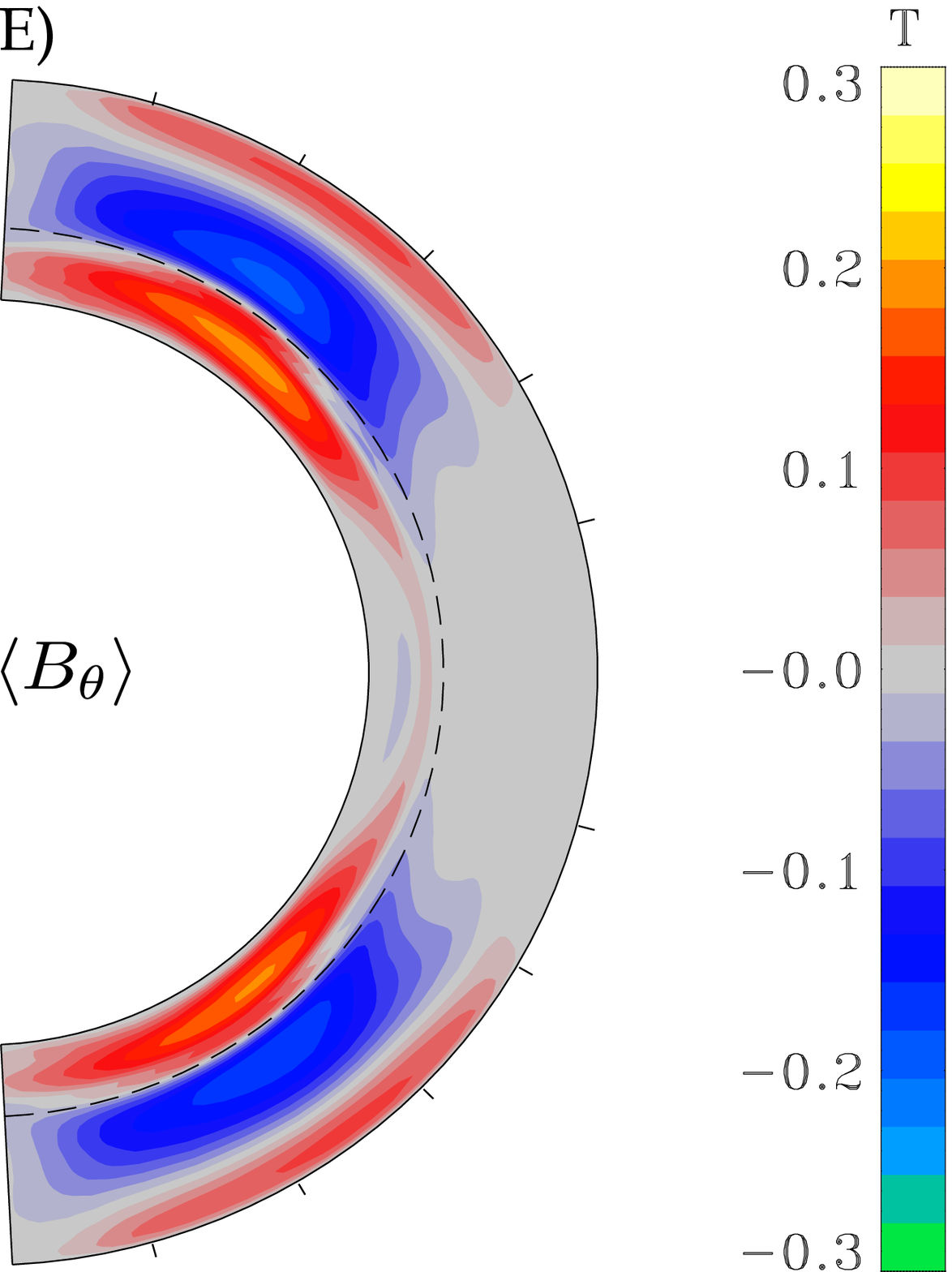}
     \hspace{1.3cm}
     \includegraphics[height=6.5 cm]{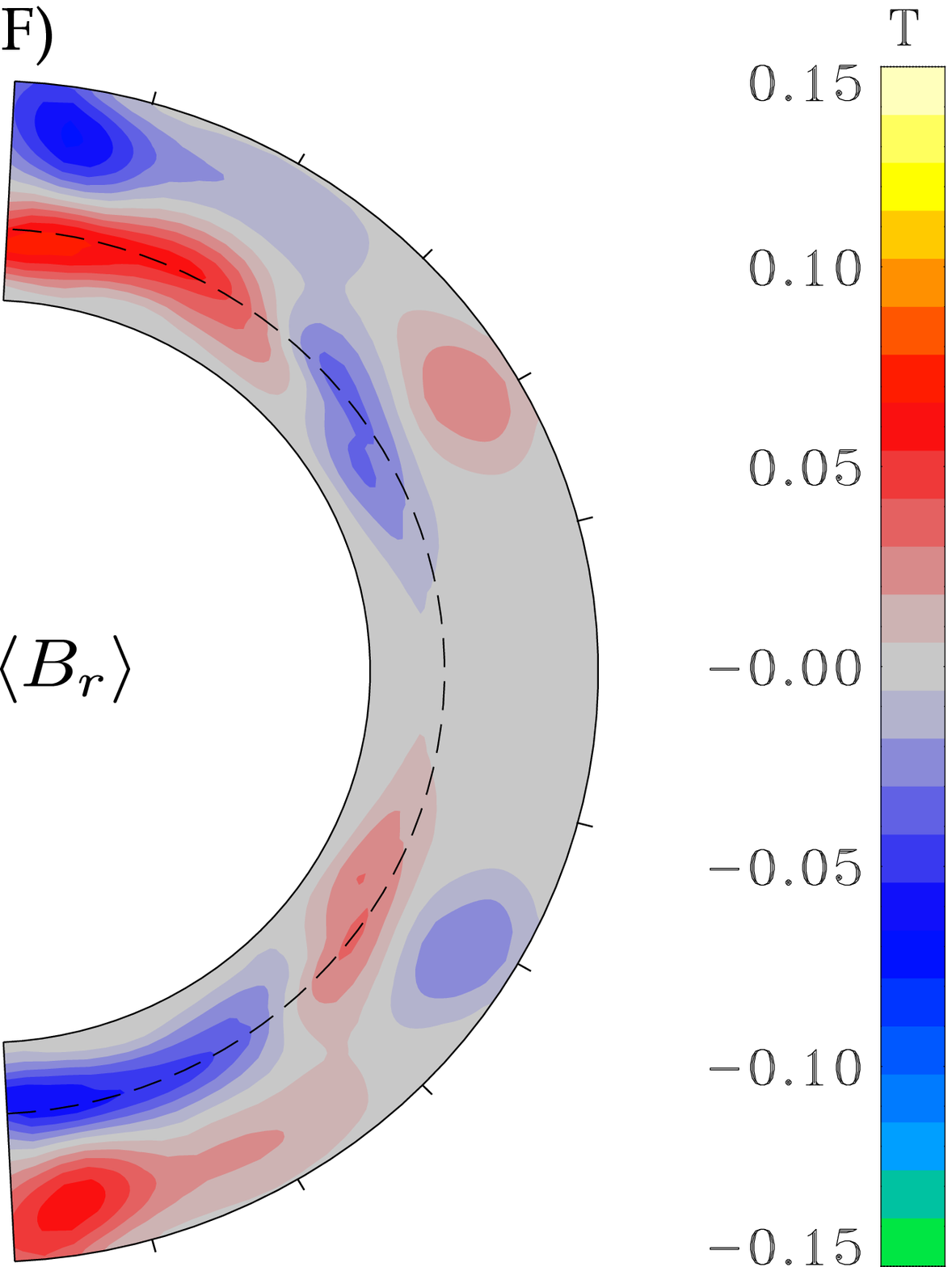}
     \caption{Zonally averaged magnetic field components
      sampled at the minimum (top row) and maximum (bottom row) of cycle 25.  Panels A) and D) show $\langle B_\phi \rangle$ with the MC
     streamfunction overploted, where dashed (solid) contour lines represent
     cells circulating in the counterclockwise (clockwise)
     direction. Panels  B) and E) show $\langle B_\theta \rangle$, and panels
      C) and F) show $\langle B_r \rangle$.}
  \label{fig:bmoy_25}
\end{figure*}

The proxy for the cycles presented here is the same one defined in PC14,
i.e., the normalized amplitude of $\langle B_\phi \rangle$ integrated
over an extended region (in r and $\theta$), centered around 48$^\circ$
near the base of the CZ, where this field component accumulates (here just
for the northern hemisphere).
In order to emphasize differences along the cycle, we compare
quantities averaged out around times of minimum and maximum. These
two epochs are labeled for cycle 24 in Fig.\ \ref{fig:bxmoy_proxy}.
We generically define the minima and maxima phases as intervals of $\pm 3$ yr
around the time of cycle minimum and maximum respectively. Fig.\
\ref{fig:bmoy_25} presents snapshots of the magnetic field components and
MC morphology taken at these two phases of the simulated cycle 25 (starting around t=1045 yr).

Outside the TC, the behavior of the MC seems to be driven
mainly by the same mechanism as in the HD case, since it still
maintains the well defined columnar patterns. In what follows we focus
our attention in the region inside the TC.

At maximum (Fig.\ \ref{fig:bmoy_25}D we have a counterclockwise rotating
cell between 48$^{\circ}$ and the pole in the bottom half of the CZ and
another smaller cell, clockwise rotating, in the upper half of the CZ
between 70$^{\circ}$ and the pole.
Between 25$^{\circ}$ and 46$^{\circ}$ (still inside the TC area),
there is a clockwise rotating cell that spreads through most depths
(and with a tendency to get oriented almost parallel to the rotation axis).
These mid latitude cells share an upflow section at around 48$^{\circ}$,
the latitude at which the zonally averaged toroidal field achieves
stronger values. During the minimum phase the two mentioned cells
above 48$^{\circ}$ get mixed and almost vanish while the lower latitude
one also decreases in strength.
Studies by \citet{Passos2012, Passos2016} indicate that the magnetic field
has an important role in modulating the horizontal component of the MC in
most of the CZ (specially inside the TC region). In the next section, we
explore the nature of the interaction between magnetic field and flows that
lead to cyclic variations in the MC.

\section{Mechanisms of MC spatio-temporal variations}

The acceleration of meridional flows is described by the zonal
component of the vorticity equation, which we will consider in Section 5.
A variety of global convection simulations, mean-field models, and theoretical
arguments consistently suggest that the dominant terms in this equation
are the Coriolis and baroclinic terms, which give rise to thermal wind balance
(TWB); see \citet{kitch95}, \citet{durne99}, \citet{brun02},
\citet{Miesch2006}, \citet{balbu09}, \citet{miesc09}, \citet{gasti14},
\citet{warne16}, \citet{kitch16}, and references therein.

However, the TWB equation itself is degenerate in the meridional flow.  It cannot be solved to yield the steady-state MC profile.  Global
circulations can be  driven by departures from TWB, which are thought to occur
in the upper and  lower boundary layers of the convection zone due to turbulent
stresses  \citep{balbu10, MieschHindman2011, warne16}.
This is commonly found in  mean-field models in which the meridional momentum
transport by the convective  Reynolds stress is modeled as a turbulent
diffusion \citep{dikpa14,kitch16}.  Here the poleward circulation in the solar surface layers is attributed to the axial gradient of the angular velocity, $\partial \Omega / \partial z$, which appears in the zonal vorticity equation through the Coriolis force (see Sec.\ 5).

A related possibility is the phenomenon of gyroscopic pumping (GP).  This occurs when a steady source of zonal momentum (an axial torque) induces a meridional flow by means of inertia.  This operates by means of the Coriolis term in the zonal vorticity equation ($\propto \partial \Omega / \partial z$) as noted in the previous paragraph but it does not require a sustained departure from TWB.  In a steady state, the amplitude and structure of the MC is largely determined by the nature of the axial torque, as described in detail by \citet{MieschHindman2011}.  GP has been definitively demonstrated by \citet{hayne91} and is thought to play an important role in the exchange of air masses between the stratosphere and troposphere \citep{holto95}.  It has been reproduced in the laboratory in the classic Plumb-McEwan experiment \citep[see discussion by][]{McIntyre98}. GP has also been invoked to account for the circulations in planetary atmospheres \citep{read86} and stellar radiative zones \citep{spieg92,fritt98,garau10,wood11}.

The maintenance of meridional flows by GP has been found in previous mean-field solar convection models by \citet{rempe05} and in previous global convection simulations by \citet{Brun2011}, FM15, \citet{gasti14}, \citet{Hotta2015}, and \citet{Guerrero2016b}.  In non-magnetic models, the source of the zonal torque is the angular momentum transport by the Reynolds stress.  However, magnetic torques can also induce a meridional flow in an analogous way.  In this section we investigate this mechanism in detail for our cyclic convective dynamo.

The angular velocity $\Omega$ in our model is defined as
\be
    \Omega = \frac{\langle u_\phi\rangle}{\lambda} + \Omega_0
\ee
where $\Omega_0=2.42405\times10^{-6}$ s$^{-1}$ is the angular velocity of the
coordinate system, and $\lambda=r\cos(\theta)$ is the so called momentum arm
with $\theta$ being the latitude.
The differences between the HD and MHD differential rotation profiles  are shown in detail by \citet{Beaudoin2013} (see their Fig. 2).

The redistribution of specific angular momentum,
$\mathcal{L} = \lambda^2 \Omega$,
plays a central role in the establishment and maintenance of mean flows.
Angular momentum transport by the convective Reynolds stress not only governs
the magnitude of the differential rotation, $\Delta \Omega$, but it also
regulates the structure and amplitude of the meridional circulation by means
of GP \citep{MieschHindman2011}.  In a steady state, the
meridional acceleration induced by the inertia of the differential rotation
is offset mainly by horizontal pressure gradients.  This can be expressed as
a balance between the Coriolis and baroclinic terms in the zonal vorticity
equation; the so-called thermal wind balance (TWB).
Any torque that disrupts this balance through a local acceleration or
deceleration of $\Omega$ will induce a meridional flow that will act to
restore the equilibrium profile of $\Omega$ that is consistent with TWB.

Thus, there are several ways in which magnetism can influence the MC.
The first is through the direct acceleration of the meridional flow due
to the mean\footnote{Azimuthally-averaged.} meridional Lorentz Force.
The second is by exerting a torque through the zonal component of the
mean Lorentz force that alters the rotation profile, $\Omega$.  This is
the mechanism of GP.
A third way for magnetism to influence the MC is by altering the convective
momentum and energy transport by means of the non-axisymmetric components
of the Lorentz force, namely the Maxwell stress.  In this section we will
demonstrate that, as in the non-magnetic convection simulations of FM15,
it is the second mechanism, GP, that largely accounts for the
structure and variability of the meridional flow that we see (Sec.\ 3).

The equation that describes the conservation of angular momentum in an
anelastic system (\ref{eq:dLdt}), \citep[see appendix of ][]{MieschHindman2011}
gives us some information about the physical mechanisms involved. This
equation is obtained by multiplying the zonal component of the momentum
equation by $\lambda$, and then averaging over longitude (indicated by angular
brackets,\ $\left< \,\right>$).
For an inviscid simulation like ours (neglecting numerical diffusion), this
yields
\be
    \rho_0 \frac{\partial \mathcal{L}}{\partial t}
       + \langle \rho_0 \mathbf{u}_m \rangle \cdot \nabla \mathcal{L}
       =  -\nabla \cdot \left( \mathbf{F}^{\mathrm{RS}}
       + \mathbf{F}^{\mathrm{MS}} + \mathbf{F}^{\mathrm{MT}} \right)
       \equiv \mathcal{F} \, ,
    \label{eq:dLdt}
\ee
where $\mathbf{u}_m = u_r\, \hat{\mathbf{e}}_r + u_\theta\,
\hat{\mathbf{e}}_\theta$ , and the terms on the r.h.s. include angular
momentum fluxes due to the Reynolds stresses, Maxwell stresses and
the large scale magnetic fields (magnetic tension). These fluxes are defined as
\bea
     \mathbf{F}^{\mathrm{RS\,\,}} &\equiv& \lambda
     \left( \langle \rho_0 u'_r u'_\phi\rangle\,
       \mathbf{\hat{e}_r} + \langle \rho_0 u'_\theta u'_\phi \rangle\,
       \mathbf{\hat{e}_\theta}
     \right) \,,\label{eq:Frs}\\
     \mathbf{F}^{\mathrm{MS}} &\equiv& - \frac{\lambda}{\mu_0}
     \left(\langle b'_r b'_\phi\rangle\, \mathbf{\hat{e}_r}
       + \langle b'_\theta b'_\phi \rangle\, \mathbf{\hat{e}_\theta}
     \right)\,, \label{eq:Fms}\\
     \mathbf{F}^{\mathrm{MT}} &\equiv& - \frac{\lambda}{\mu_0}
     \left(\langle b_\phi b_r \rangle\, \mathbf{\hat{e}_r}
       + \langle b_\phi b_\theta \rangle\, \mathbf{\hat{e}_\theta}
     \right) \,, \label{eq:Fmt}
 \eea
where we use the classical Reynolds
decomposition of the field and flow components
in zonal means (associated with large scale flows) and fluctuations
(associated with small scale turbulence), e.g., $u_\theta=\left<u_\theta\right> + u_\theta'$.
Equation (\ref{eq:dLdt}) tells us how angular momentum variations due to
local zonal forcings (net axial torque $\mathcal{F}$) can induce variations
in the meridional flow. The same approach was used by \cite{MieschHindman2011}
to study how the MC is established in the solar near surface shear layer.
It was also used by \cite{Brun2011} and FM15 to study how the MC is
established in hydrodynamic (non-magnetic) simulations of global convection
(with and without a tachocline).
These authors found that the Reynolds stress term has a preponderant
role in establishing the overall amplitude and morphology of the MC.
More recently, \cite{Guerrero2016b} used Eq. (\ref{eq:dLdt}) to identify
the main agent driving torsional oscillations in a global dynamo simulation.
Their results suggest that the magnetic tension at the bottom of the
convection zone induces axial torques that periodically speed-up and slow-down the angular velocity.

\subsection{Angular momentum balance in the HD case}

For comparison purposes we begin by performing this analysis to our HD
simulation. Since there are no magnetic fields involved, equation (\ref{eq:dLdt}) reduces to
\be
    \langle\rho_0\rangle \frac{\partial \mathcal{L}}{\partial t}
    + \langle \rho_0 \mathbf{u}_m \rangle \cdot \nabla \mathcal{L}
    =  -\nabla \cdot \left( \mathbf{F}^{\mathrm{RS}} \right)\, .
    \label{eq:dLdt_HD}
\ee

\begin{figure*}[htb]
  \centering
     \includegraphics[height=6.5 cm]{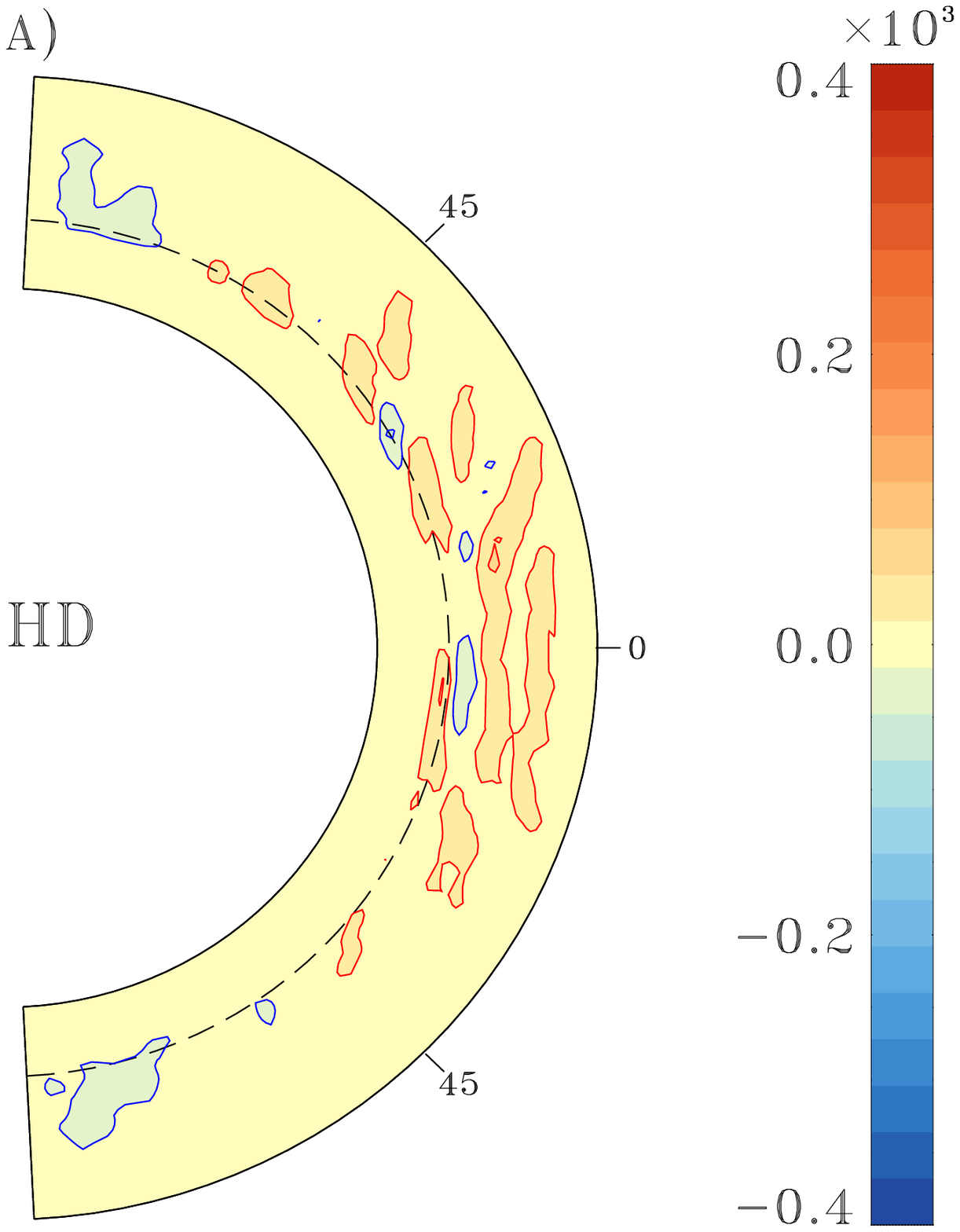}
     \includegraphics[height=6.5 cm]{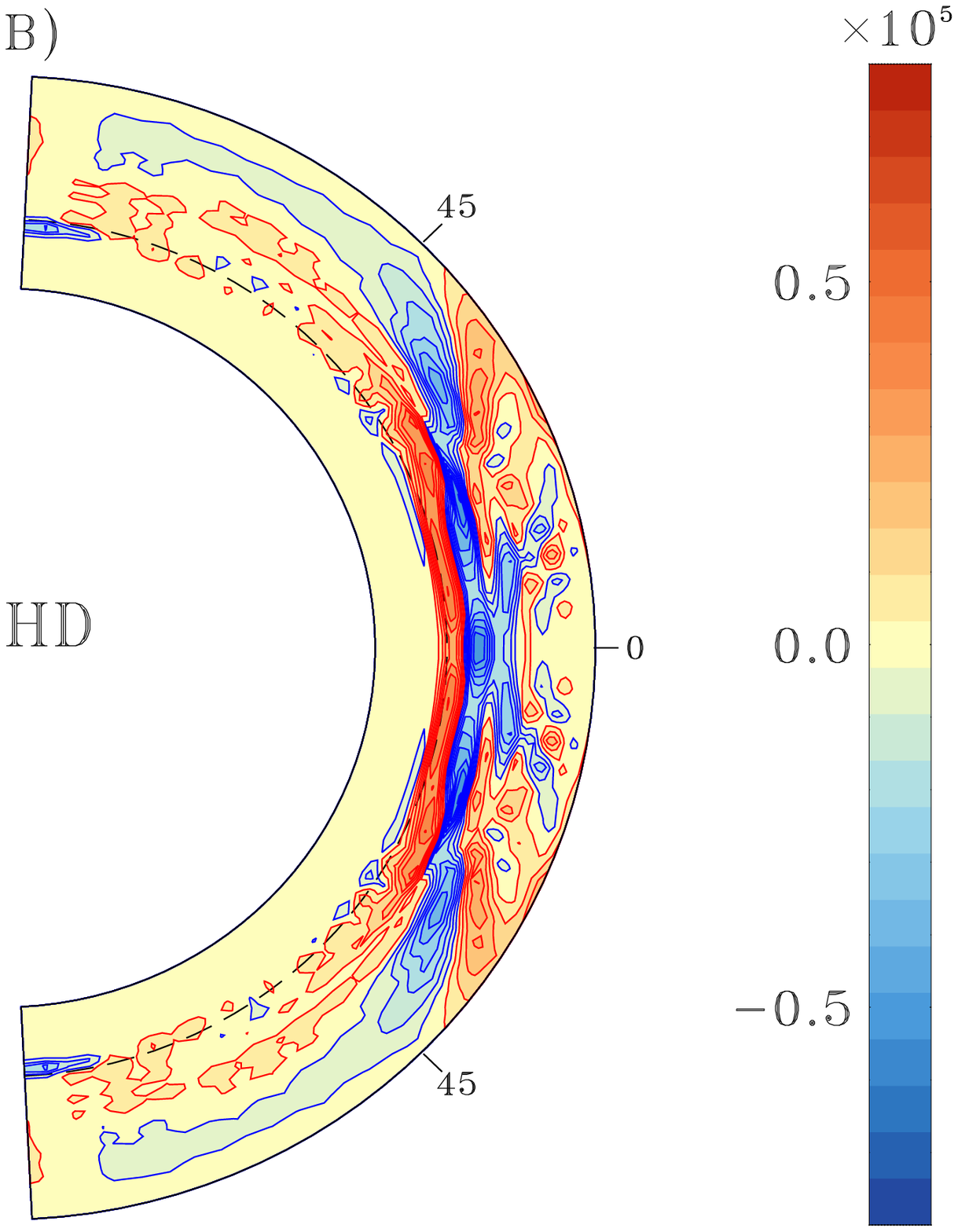}
     \includegraphics[height=6.5 cm]{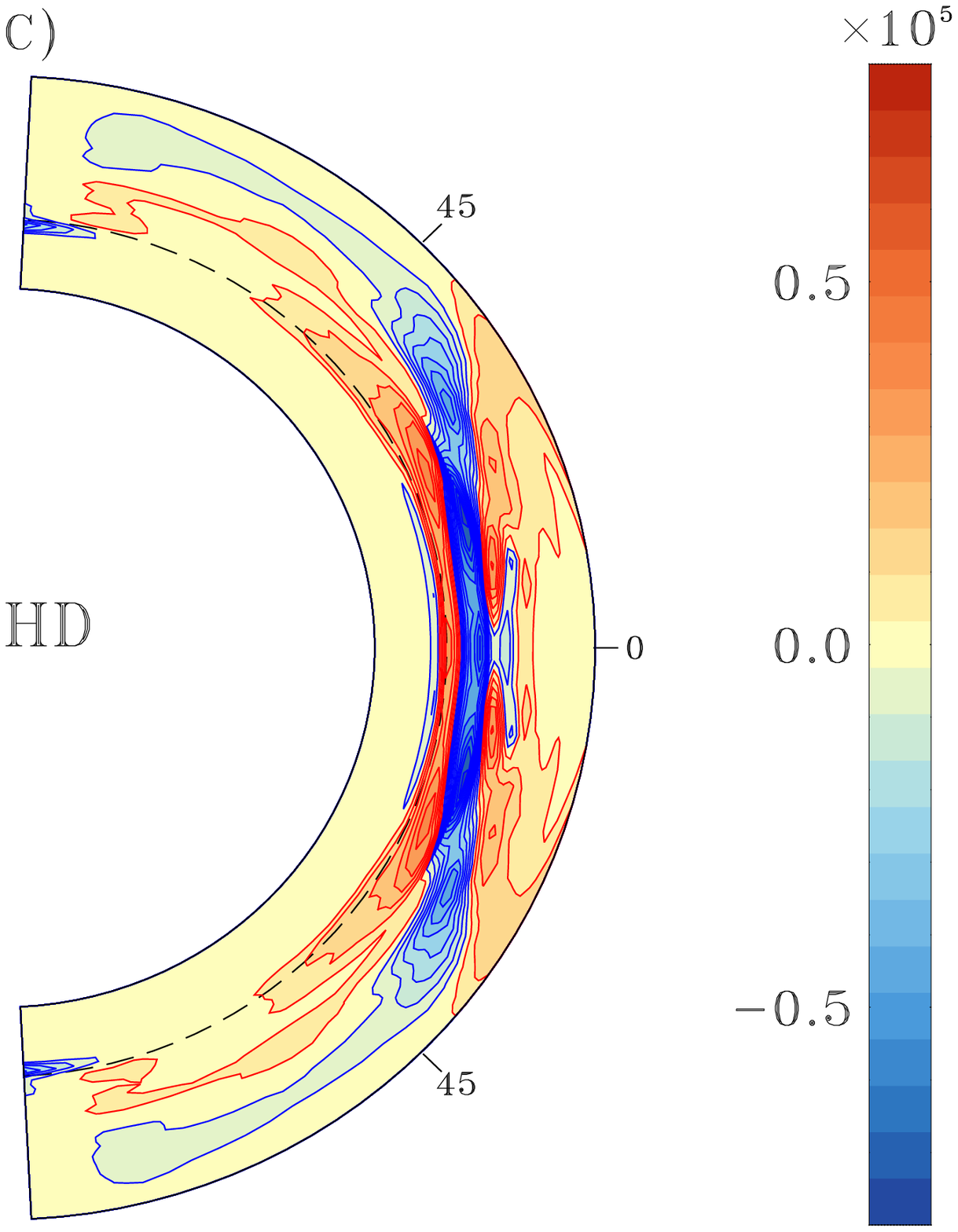}
     \caption{A) $\rho_0 \partial \mathcal{L}/\partial t$, B) $\langle \rho_0
     \mathbf{u}_m \rangle \cdot \nabla \mathcal{L}$ and C) $ -\nabla \cdot
     \left( \mathbf{F}^{\mathrm{RS}} \right)$ in kg m$^{-1}$ s$^{-2}$,
     averaged over 246 yr.
     Notice that B) and C) have the
     same scale ($10^5$) but A) is two orders of magnitude smaller
     ($10^3$), indicating that a steady state has been achieved. }
     \label{fig:allHD}
\end{figure*}

Fig.\ \ref{fig:allHD} shows the individual terms of Eq.
(\ref{eq:dLdt_HD}) averaged over a time interval of 246 yr.
When $\langle\rho_0\rangle \partial \mathcal{L} / \partial t$
is small compared to the other terms, we can assume that
$\mathcal{F}$ is due to the sole action of the Reynolds stresses. When
$\mathcal{F}>0$ (red lines and shades) the net torque is prograde inducing
a meridional flow away from the rotation axis.
While $\mathcal{F}<0$ (blue lines and shades), the net torque is retrograde
and induces a flow toward the rotation axis.

To interpret this figure we first focus on the Reynolds stress shown in
Fig.\ \ref{fig:allHD}C.  Outside the TC, below 45$^\circ$,
it exhibits a
divergence in the mid-lower CZ (blue) and a convergence in the upper CZ (red),
with contours approximately aligned with the rotation axis. This is consistent
with the work of FM15 and signifies the transport of angular momentum by sheared banana cells\footnote{However, we note a typo in
Fig.\ 8 of FM15. The caption is correct but the labels on frames $b$,
$c$, $g$, and $h$ are not; what is shown there is the divergence
$\nabla \cdot \mathbf{F}^{\mathrm{RS}}$, not the convergence
$-\nabla \cdot \mathbf{F}^{\mathrm{RS}}$.}.

As described by FM15, the divergence of $\mathbf{F}^{\mathrm{RS}}$ in the
lower CZ near the equator induces a prominent clockwise (CW) circulation
cell in the northern hemisphere (NH), immediately outside the TC.
This effect is also observed in our simulation (Fig.\ \ref{fig:vmoy}C).

In FM15 (and in the MHD simulation considered here (Fig.\ \ref{fig:vmoy}F),
this CW (blue) cell lies between two CCW (red) cells above and below.
However, in the HD simulation considered here, the upper cell at low latitudes
(outside the TC) is absent, with the circulation
in this region dominated by a series of smaller-scale cells. This difference
can be most likely attributed to different effective viscosities between both
models.

Inside the TC, the CCW cell that pervades most of the CZ in
the NH is somewhat less evident here but is also present. In Fig.\
\ref{fig:vmoy}C is easier to identify its antisymmetric counterpart in the
southern hemisphere (SH).
The different morphologies in this CCW cell found in FM15 and our HD
simulation can be partially attributed to the presence of the stable zone and
overshoot region, absent in  FM15. The inward angular momentum transport
by downflow plumes leads to a convergence of the angular momentum flux in
the overshoot region that persists to very low latitudes, establishing a
strong equatorward flow.
The high density in the overshoot region makes this a substantial contribution
to the mass flux and mass conservation largely accounts for the poleward
flow in the mid CZ.  This establishes the strong CCW circulation cells near
the base of the CZ. Similar results were seen in the penetrative convection
 simulations by \cite{Miesch2000}; see their Fig.\ 16$a$.

The close correspondence between panels B) and C) of Fig.\ \ref{fig:allHD}
and the small amplitude of A) indicate a statistically steady state.
Thus, the advection of angular momentum by the MC balances the transport
of angular momentum by the convective Reynolds stress.
Furthermore readjustments in $\mathcal{L}$ occur on the time scale of a
several days which means that any averaging over longer periods will result
only in a small residual.
The small differences between frames B and C can be attributed
to the contribution of numerical viscous fluxes
\citep[especially in the $\theta$ direction, see][]{Guerrero2016b}, to a
residual $\partial \mathcal{L}/\partial t$, to
the finite duration of the temporal averaging
and due to the different numerical methodologies used while running the
model and \textit{a posteriori} analysis. We will elaborate on this in the
next section.

\subsection{Angular momentum balance in the MHD case}

\begin{figure*}[htb]
  \centering
     \includegraphics[height=6.5 cm]{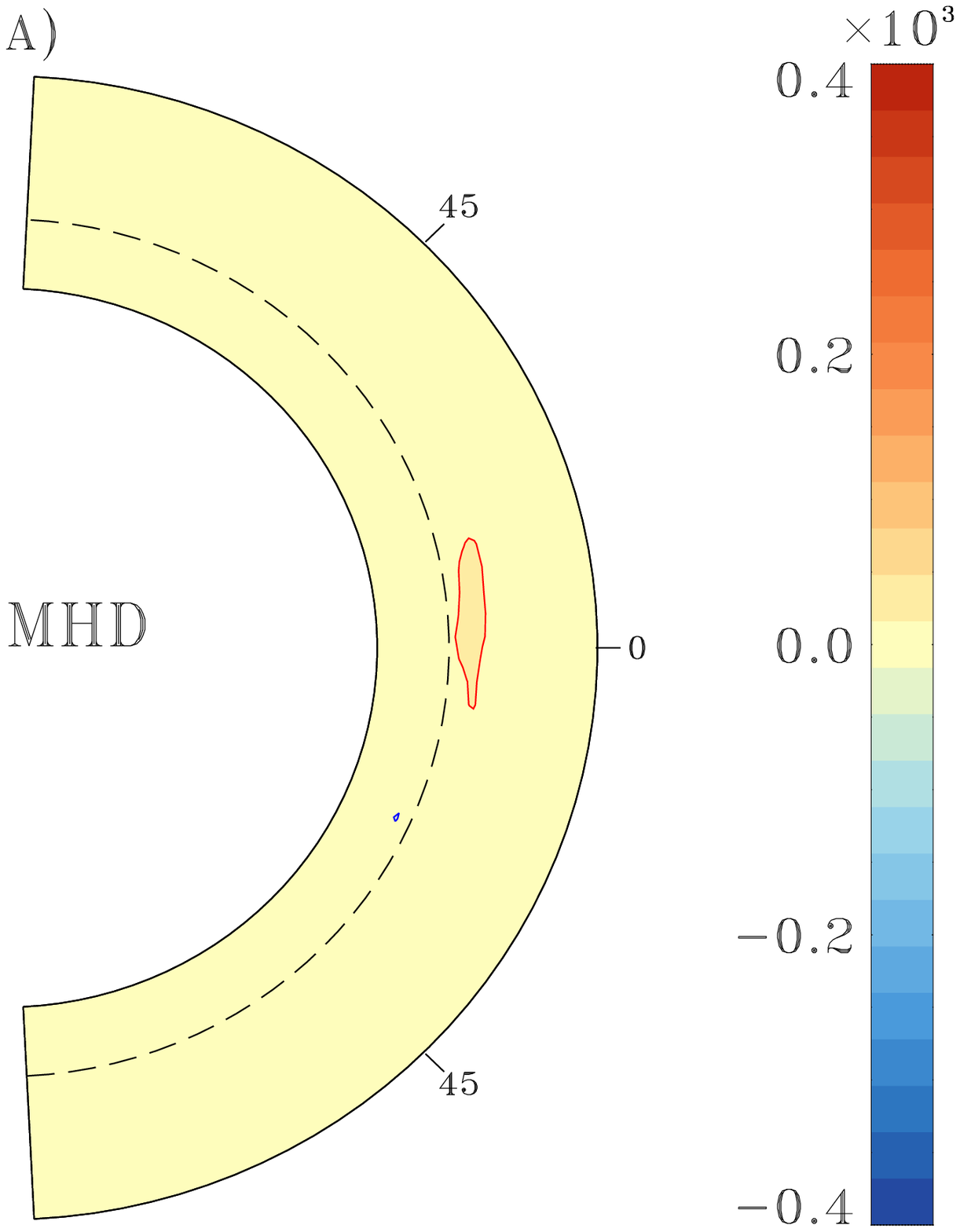}
     \includegraphics[height=6.5 cm]{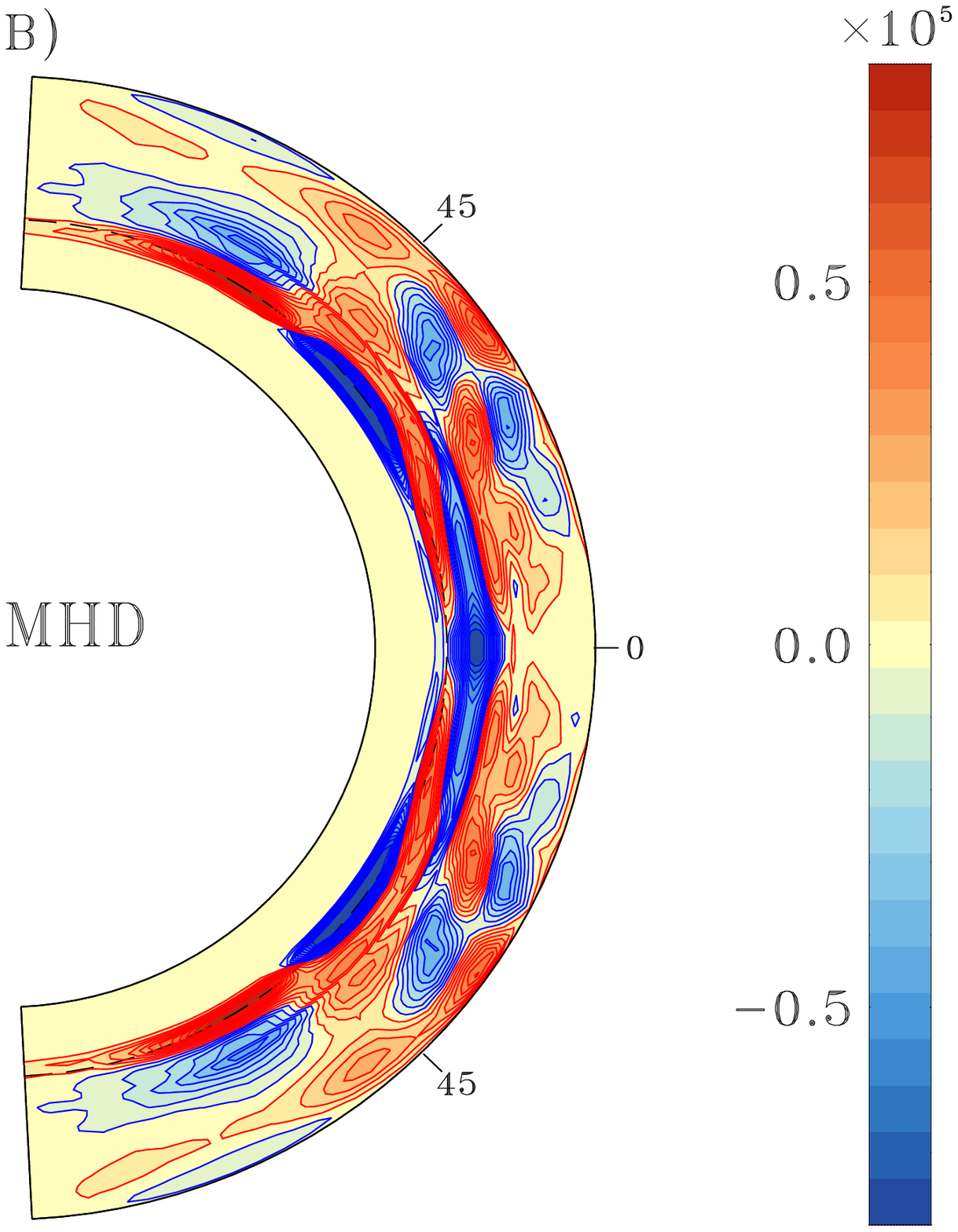}
     \includegraphics[height=6.5 cm]{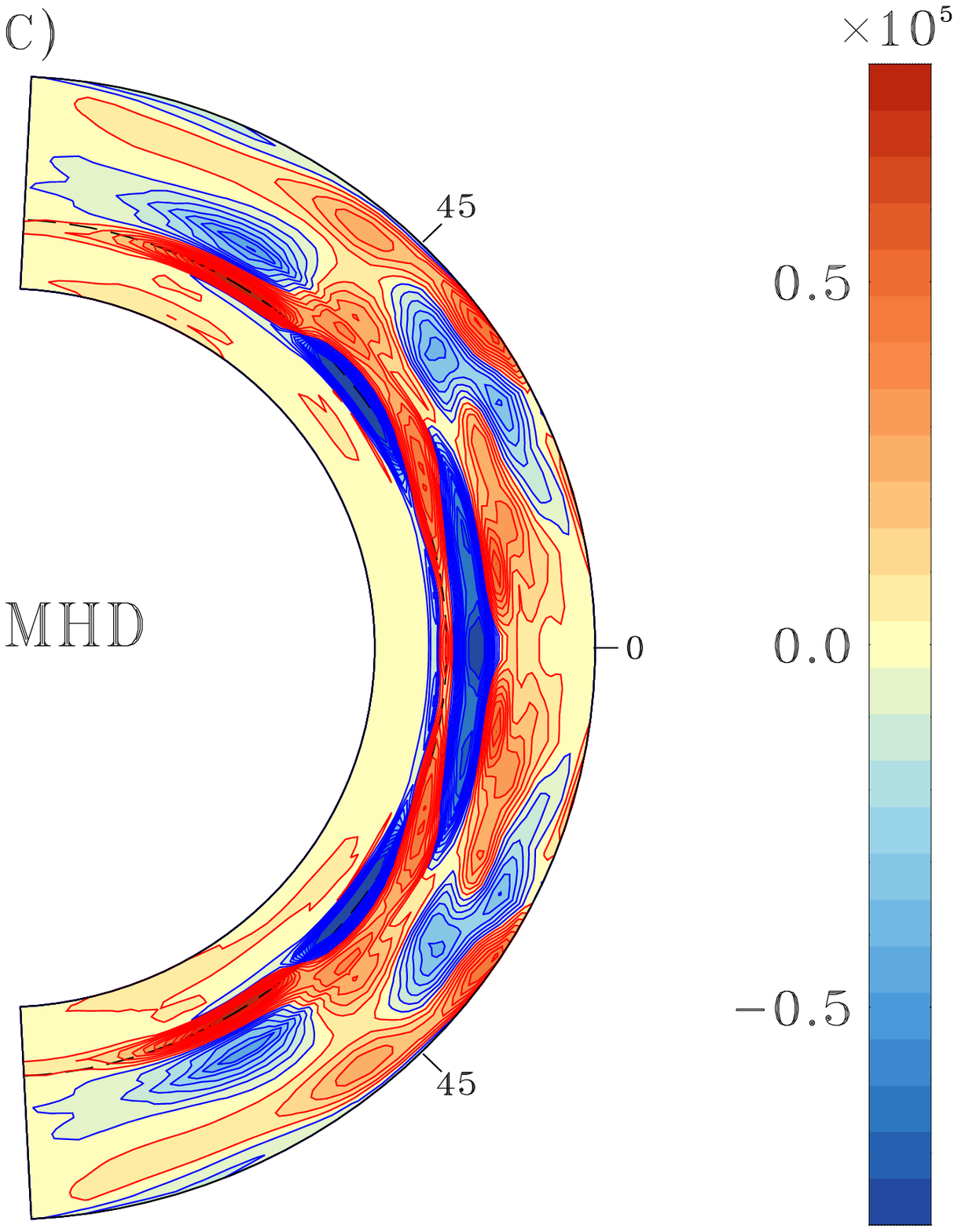}
     \caption{Panels A) and B) show the same terms as in Fig.\
     \ref{fig:allHD} for the MHD simulation also in kg m$^{-1}$ s$^{-2}$.
       Panel C shows the entire
     r.h.s.\ of eq.\ (\ref{eq:dLdt}) that include the contributions from
     $\mathbf{F}^\mathrm{RS}$, $\mathbf{F}^\mathrm{MS}$, and
     $\mathbf{F}^\mathrm{MT}$.  All quantities are averaged over 10 cycles
     (406 yrs).}
     \label{fig:allMHD}
\end{figure*}

\begin{figure*}[htb]
  \centering
     \includegraphics[height=6.5 cm]{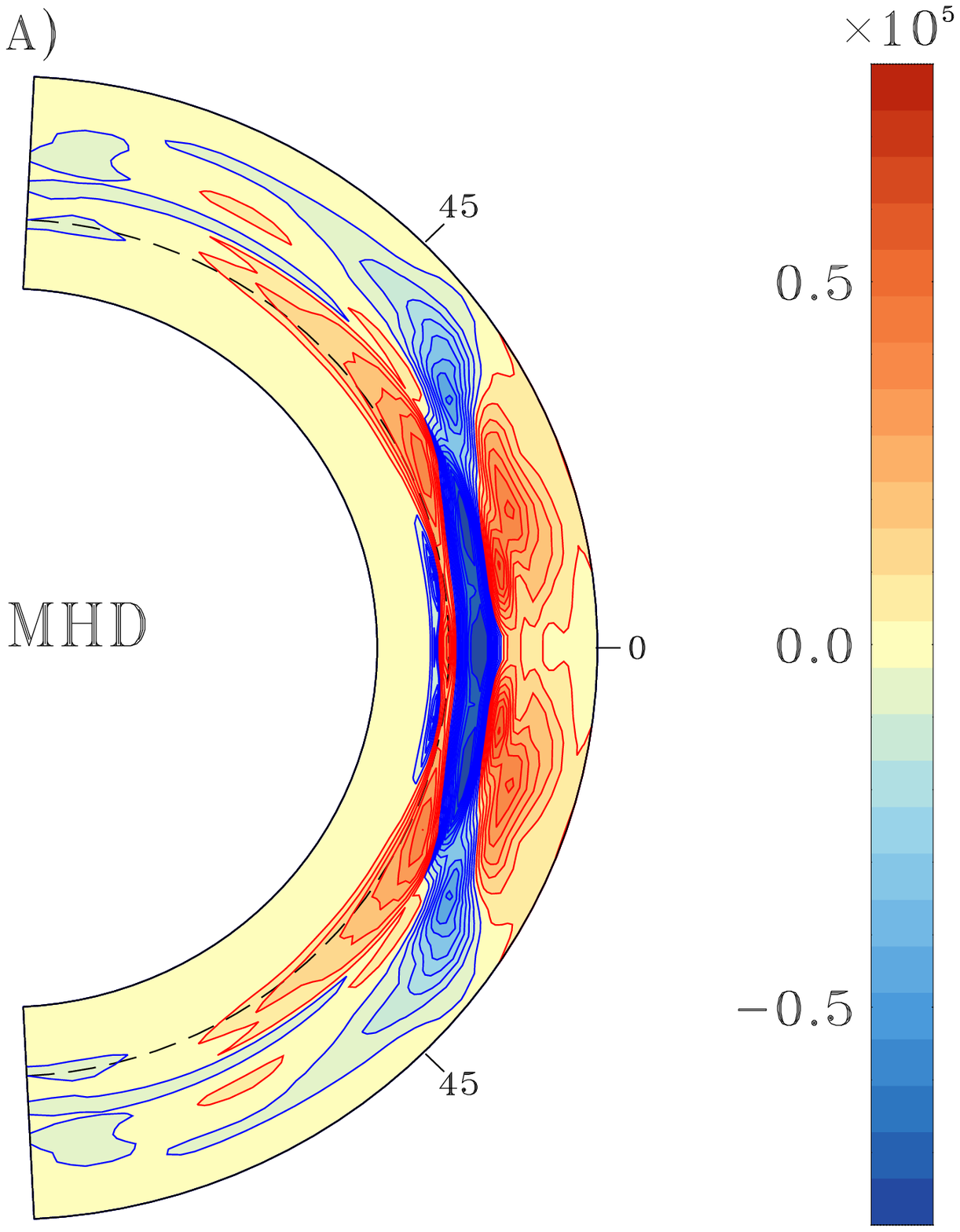}
     \includegraphics[height=6.5 cm]{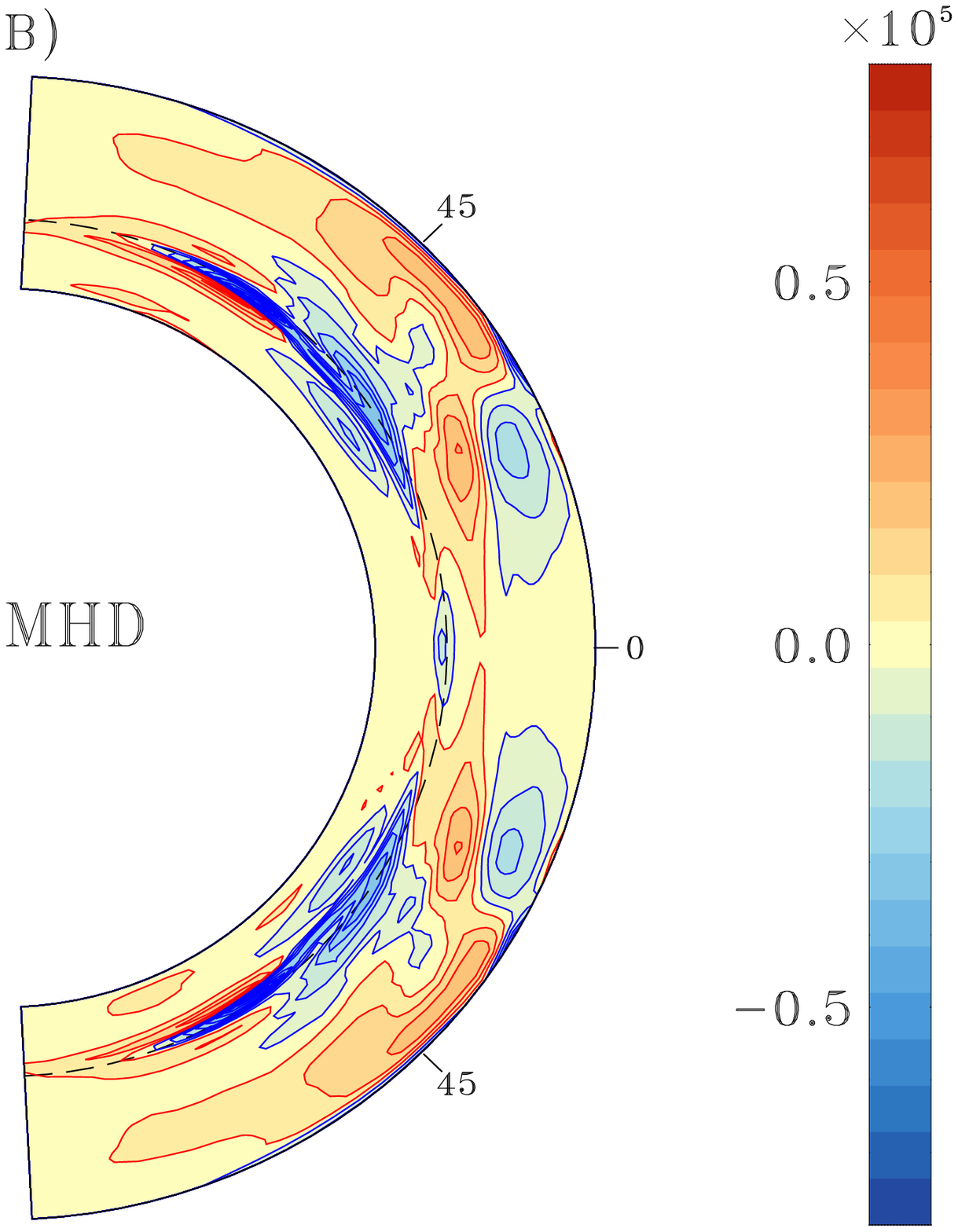}
     \includegraphics[height=6.5 cm]{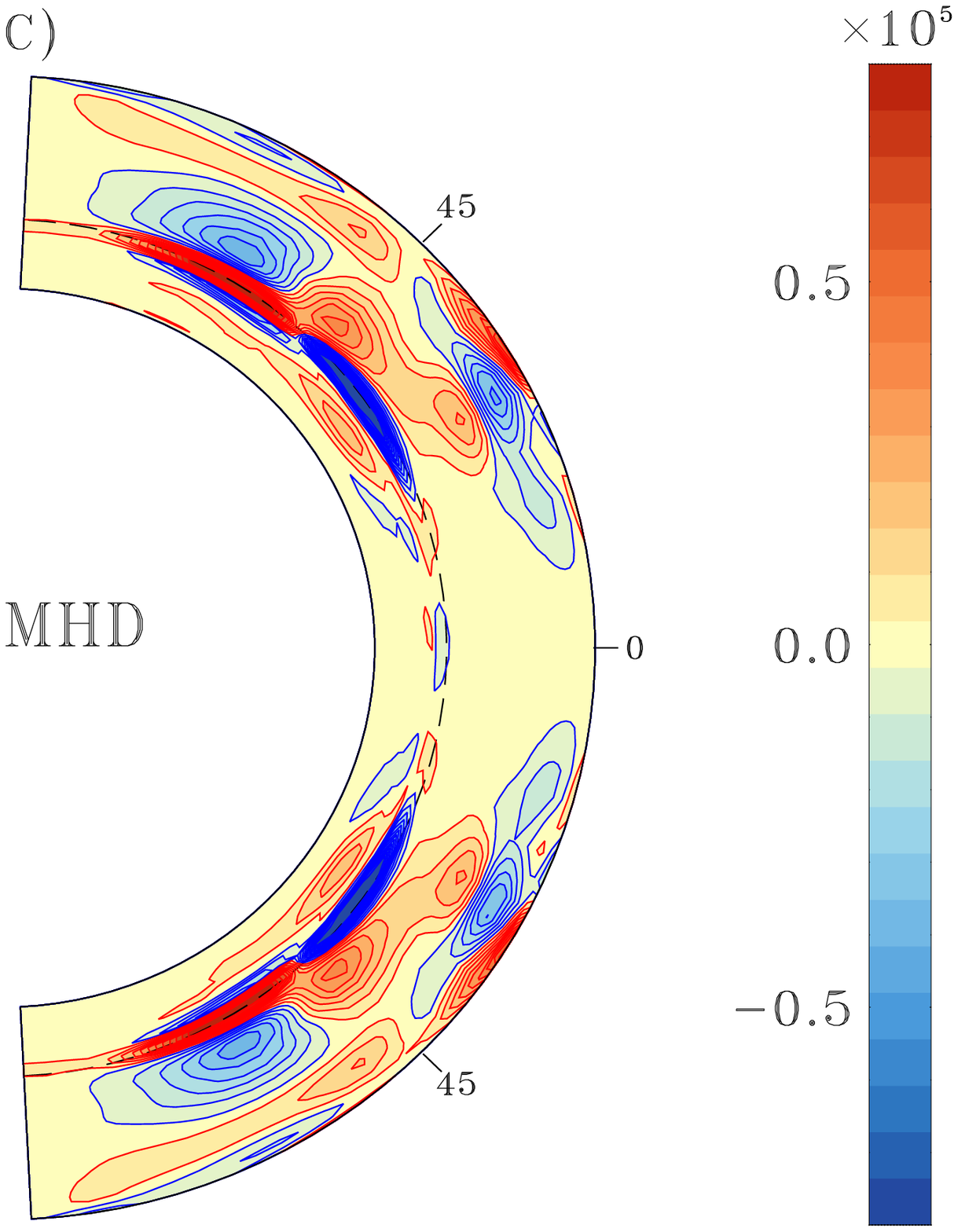}
     \caption{The components of the r.h.s. of equation (\ref{eq:Fmt})
     and Fig.\ \ref{fig:allMHD}C, plotted individually:
     A) $-\nabla \cdot \left( \mathbf{F}^{\mathrm{RS}}\right)$,
     B) $-\nabla \cdot \left( \mathbf{F}^{\mathrm{MS}}\right)$ and
     C) $-\nabla \cdot \left( \mathbf{F}^{\mathrm{MT}}\right)$}
     \label{fig:rhsMHD}
\end{figure*}

Next, we apply the same analysis procedure to the MHD simulation where
a dynamo generated large-scale magnetic cycle contributes to the transport
of angular momentum.
In Fig.\ \ref{fig:allMHD} we show a comparison between the l.h.s.\ and
r.h.s.\ of equation (\ref{eq:dLdt}) averaged over the 10 cycles interval. Here again, the first term of equation
(\ref{eq:dLdt}) shows only a small residual time dependence two orders
of magnitude smaller than the terms in the r.h.s..
As above, the numerical diffusion can account for most of the small
differences between panels \ref{fig:allMHD}B and \ref{fig:allMHD}C.
The three components that make up the
r.h.s.\ in Fig.\ \ref{fig:allMHD}C are shown individually in
Fig.\ \ref{fig:rhsMHD}.

The influence of magnetism on the net axial torque can be seen by comparing
Figs.\ \ref{fig:allHD}C and \ref{fig:allMHD}C.
The general pattern outside the TC is similar in both figures,
with a region of divergence (blue) in the lower CZ at low latitudes,
straddled above and below by regions of convergence (red).  This is
somewhat expected since the strong columnar flow behavior in this region
is mainly maintained by Reynolds stresses and the large-scale
magnetic field is
weak. The Reynolds stresses induce a series of three stacked circulation
cells at low latitudes. The small differences with the HD case in this
 region can be attributed mainly to the contribution of Maxwell stresses;
see Figs.\ \ref{fig:vmoy}C and \ref{fig:vmoy}F.  As in the HD case,
the angular momentum transport by these circulation cells largely balances
the net torque; compare Figs.\ \ref{fig:allMHD}B and \ref{fig:allMHD}C.
Still outside the TC, but in the upper part of the
CZ, around 35$^\circ$ there are some small differences that can be
attributed to local magnetic fields. We will address this further below.

As noted in section 3, the main difference between the HD and MHD cases is
the paired set of MC cells, inside the TC at mid latitudes,
seen for example in Fig.\ \ref{fig:vmoy}F in conjunction with an upwelling
in the lower CZ at a latitude of about 48$^\circ$.  We attribute
these circulation cells to GP induced by the magnetic part
of the axial torques seen in Fig.\ \ref{fig:allMHD}C.
At the base of the CZ, these torques are accelerating the rotation rate at
latitudes higher than 48$^\circ$ (red), and decelerating lower latitudes
(blue), inducing a convergent flow which produces the mid-latitude upwelling.
At slightly larger radii in the lower CZ, the pattern of torques reverses,
with deceleration (blue) and acceleration (red) at latitudes poleward and
equatorward of 48$^\circ$.  This establishes the horizontally diverging flow
in the mid CZ that closes off the pair of mid-latitude circulation cells.
Though such a closure is not required by GP (closed
circulation cells are always ensured by mass conservation), the quadrupolar
pattern of red-blue-red-blue serves to enhance the mid-latitude circulation
cells and to keep them localized in the lower CZ (see Fig. \ref{fig:bmoy_25}D).  This interpretation is
confirmed by studying the angular momentum transport by the MC in Fig.\
\ref{fig:allMHD}B, which shows a similar quadrupolar pattern at
mid-latitudes.

It is clear from Fig.\ \ref{fig:rhsMHD} that this quadrupolar pattern in
the net torque arises from the large-scale Lorentz force (panel C). However, in
order to interpret this, we must begin with the Reynolds stress component in
panel A). As noted above, $\mathbf{F}^\mathrm{RS}$ is dominated by banana
cells, which transport angular momentum cylindrically outward (away
from the rotation axis) and equatorward.  The equatorward component
of the transport leads to a divergence of $\mathbf{F}^\mathrm{RS}$ at
mid-latitudes in the upper CZ (blue), i.e., the Reynolds stress is extracting angular momentum from top half of the CZ at latitudes 30$^\circ$--55$^\circ$ in order to establish the solar-like
differential rotation. At the same time, in the bottom half of the CZ, angular momentum is also being transported downward.  At mid-latitudes, downflow plumes carry
angular momentum to the stable zone and their deceleration in the overshoot
region gives rise to a convergence of the angular momentum flux.  This is
visible in Fig.\ \ref{fig:rhsMHD}A as a red stripe near the base of the CZ at
latitudes between $\pm 60^\circ$ that becomes particularly prominent
between latitudes of $\pm 25^\circ$--$50^\circ$.  This acts to accelerate the
rotation rate near the base of the CZ and decelerate the rotation rate in the
upper CZ.
The magnetic torques respond to this Reynolds stress.  In particular,
the quadrupolar pattern of $\mathbf{F}^\mathrm{MT}$ at mid-latitudes arises
when the torques exerted by the Reynolds stress are extended poleward by
magnetic tension.  For example, in the upper CZ at a latitude of
40$^\circ$, the Reynolds stress is acting to decelerate $\Omega$ (blue).
Magnetic tension opposes this local deceleration (red) and spreads it
to higher latitudes (blue).  Similarly, when the Reynolds stress acts to
accelerate the fluid near the base of the convection zone, the rigidity
imparted by magnetic tension serves to ``drag'' higher latitudes along.  This
acts to decelerate mid latitudes (blue) and accelerate higher latitudes
(red).The Maxwell stress also opposes the Reynolds stress,
both in the overshoot region and in the upper CZ (Fig.\ \ref{fig:rhsMHD}B).
However, in contrast to the large-scale magnetic tension,
$\mathbf{F}^\mathrm{MS}$ is more diffusive in nature, and thus more localized.

The quadrupolar mid-latitude pattern of angular momentum divergence and
convergence bears an interesting resemblance to the angular momentum cycle
described by \cite{Gilman89b}; see their Fig.\ 1$a$.  They postulated a
poleward angular momentum transport by some unspecified process in the
solar tachocline that offset the equatorward transport by convective Reynolds
stresses in the CZ.  They even identified magnetic stresses as a possible
candidate process.  Poleward angular momentum transport by magnetic stresses
in the tachocline is an important component of several
tachocline confinement models \citep[reviewed by][]{Miesch2005}.
Our global MHD convection simulation clearly demonstrates this.
Though the magnetic stresses in our model vary over the course of a magnetic
cycle (see below), they do induce a net angular momentum transport toward
the poles near the base of the CZ, as seen in Fig.\ \ref{fig:rhsMHD}C.

Another minor difference in the axial torques distribution between
the HD and MHD cases is that the latter has a more prominent prograde
(positive) torque in the upper CZ at the equator (compare Figs.\
\ref{fig:allHD}C and \ref{fig:allMHD}C).
This is reflected in the MC profile of the MHD model where
an upflow seems to be induced (see in Fig.\ \ref{fig:vmoy}F the
paired set of circulation cells in the upper CZ near the equator,
red in the NH, blue in the SH).
This appears to be due not to the Lorentz force directly, but rather to
a modification of the Reynolds stress by magnetism (Fig.\ \ref{fig:rhsMHD}A).
This may be attributed to the diffusive nature of the Maxwell stress, which
tends to make the flow more laminar, enhancing the cylindrically outward
angular momentum transport by banana cells \citep{Brun2004,Nelson13,Fan14,Karak15,Hotta2016}.

Furthermore, in the upper CZ near a latitude of about 28$^\circ$, the MHD case
exhibits a torque pattern that is not present in the HD case (compare Figs.\
\ref{fig:allHD}C and \ref{fig:allMHD}C). This is due to the
presence of a secondary dynamo that exists in that region. A complete
characterization of this secondary (weaker and short period) dynamo mode
is presented in \cite{Beaudoin2016}. The authors study its interaction with
the primary mode and show how it is maintained by a latitudinal shear.
Although the magnetic field created in this region its
weaker than that produced by the main dynamo mode that operates near the
bottom of the convection zone around 50$^\circ$, it still contributes to
MC variability.

\begin{figure*}[htb]
  \centering
     \includegraphics[width=4.5 cm]{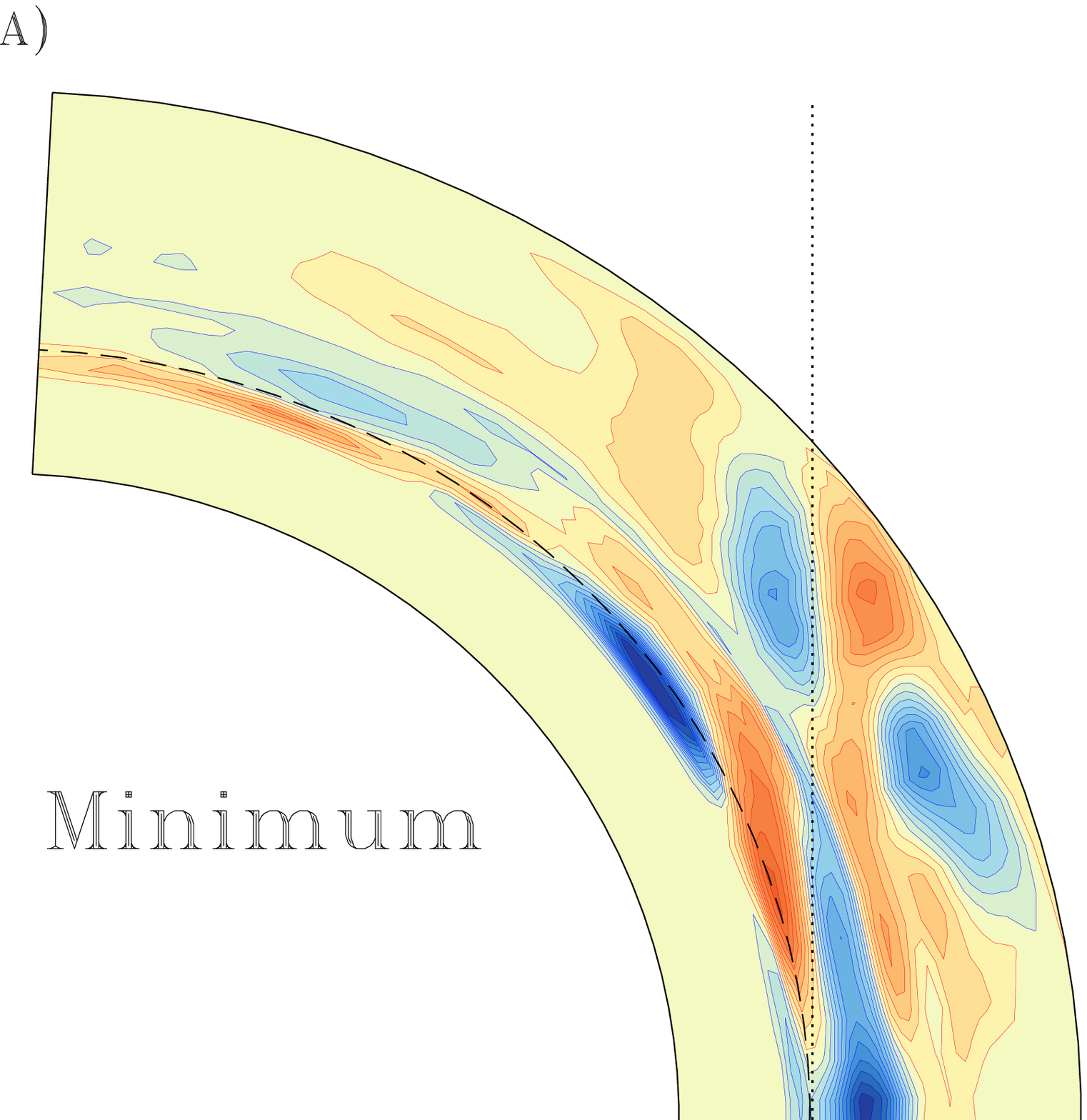}
     \includegraphics[width=4.5 cm]{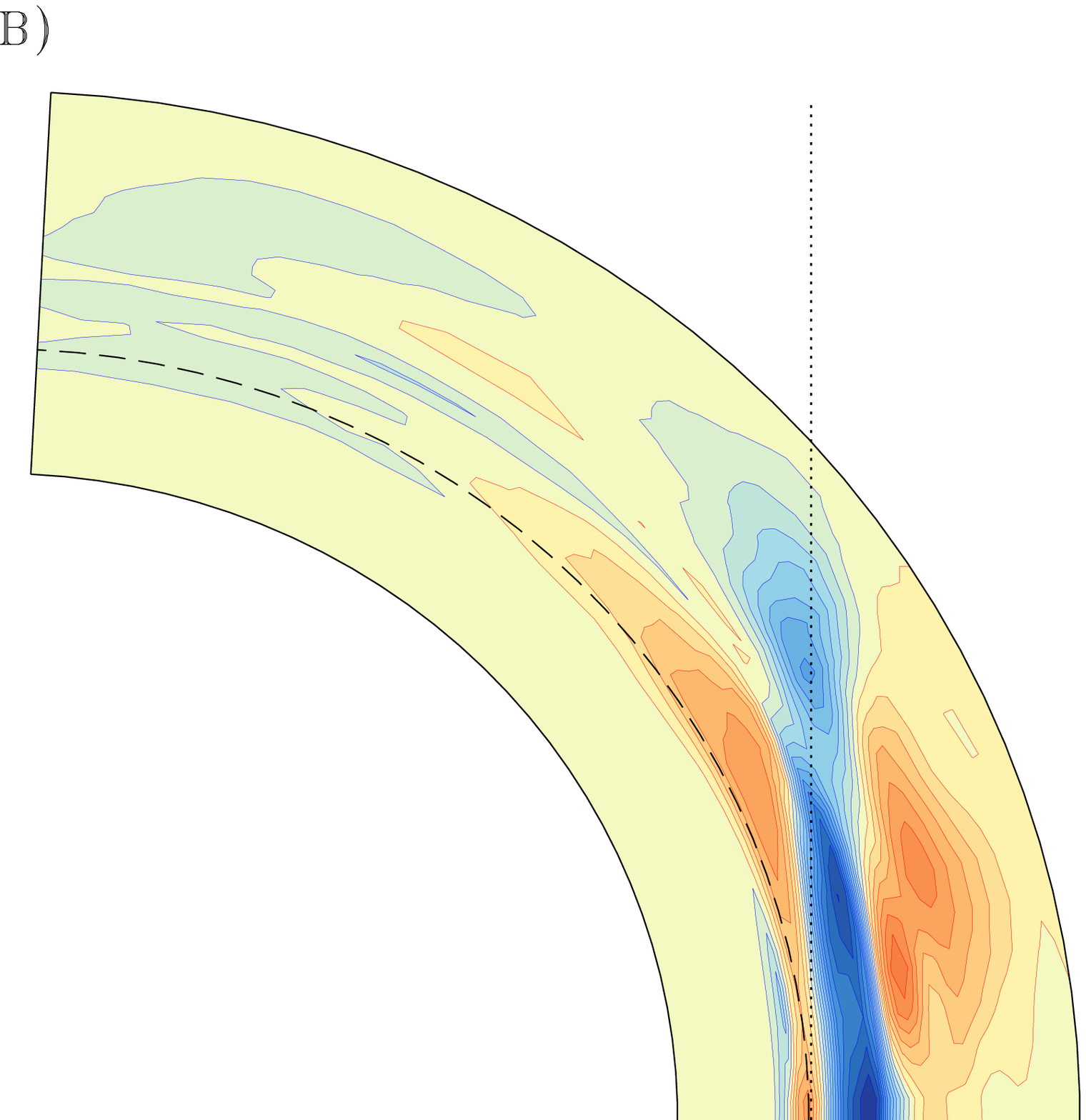}
     \includegraphics[width=4.5 cm]{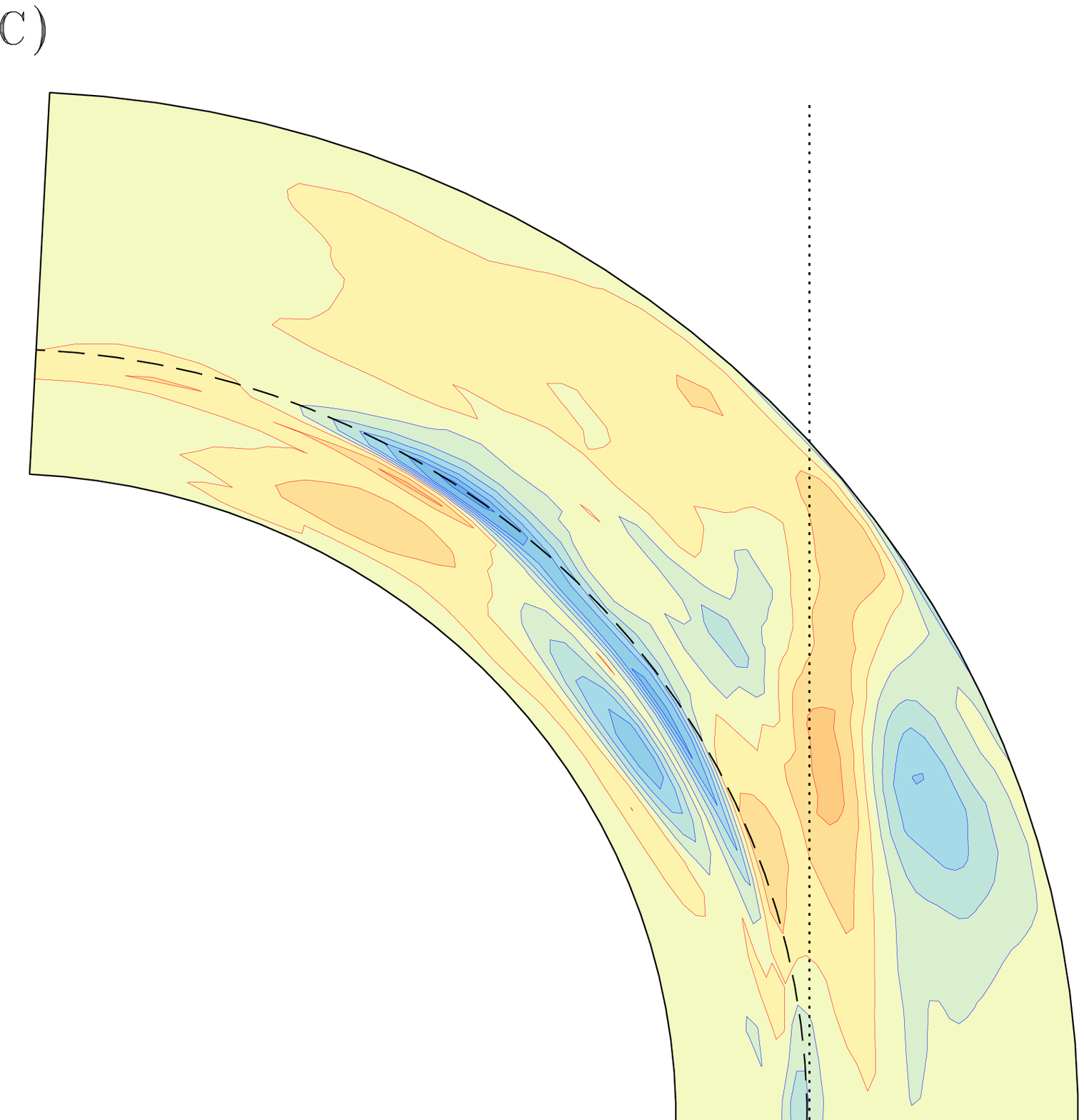}
     \includegraphics[width=4.5 cm]{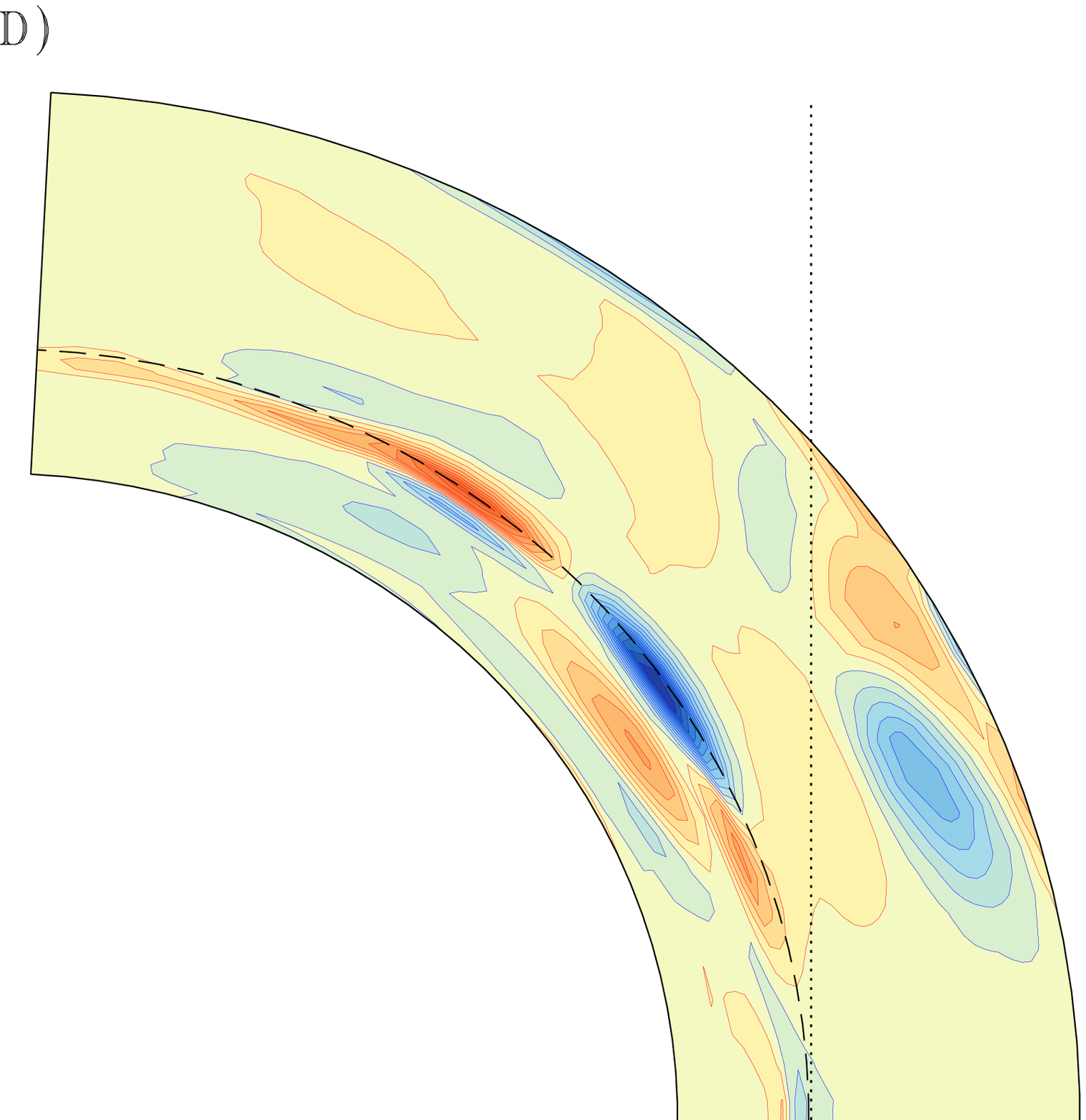}\\
     \includegraphics[width=4.5 cm]{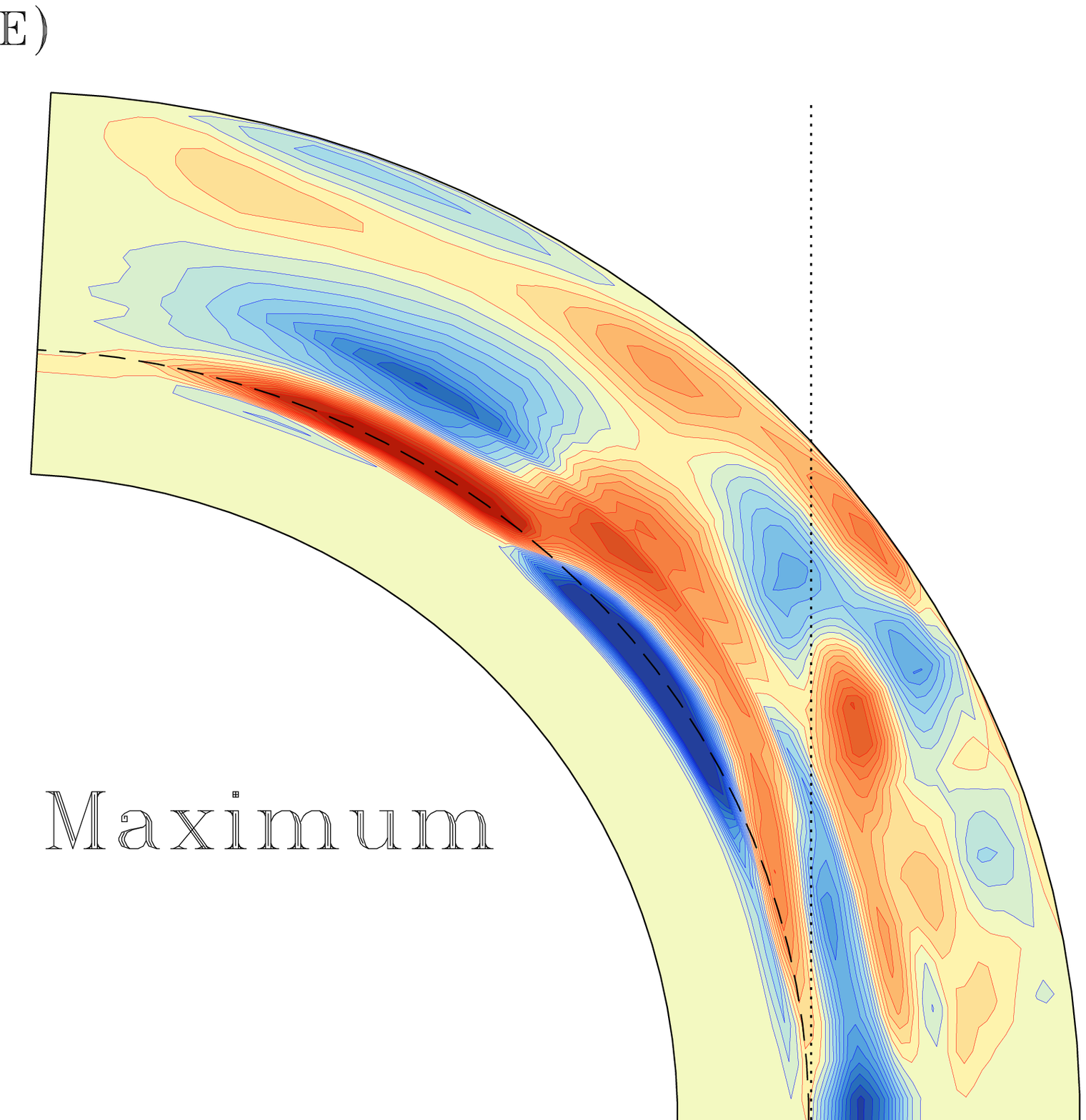}
     \includegraphics[width=4.5 cm]{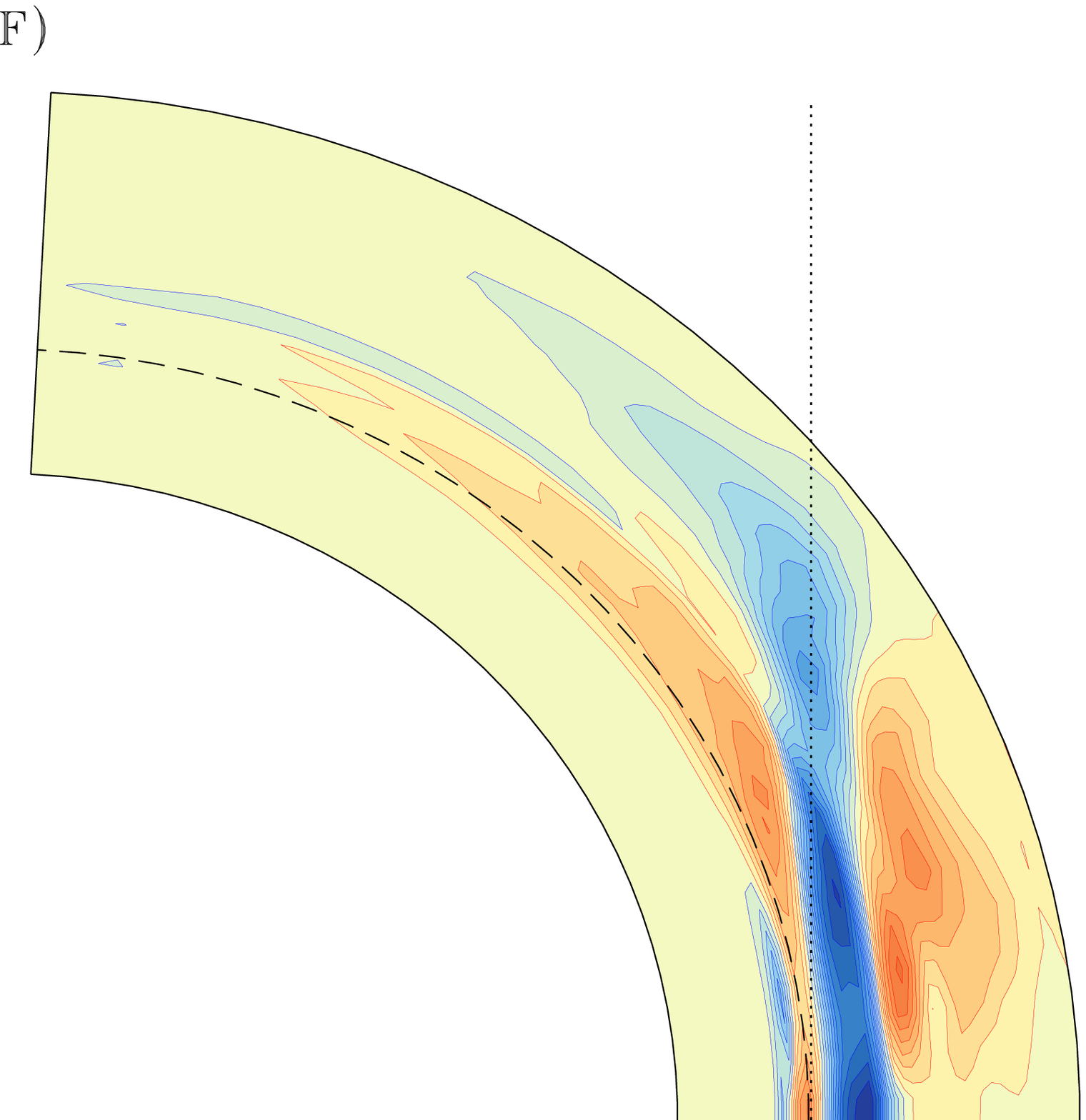}
     \includegraphics[width=4.5 cm]{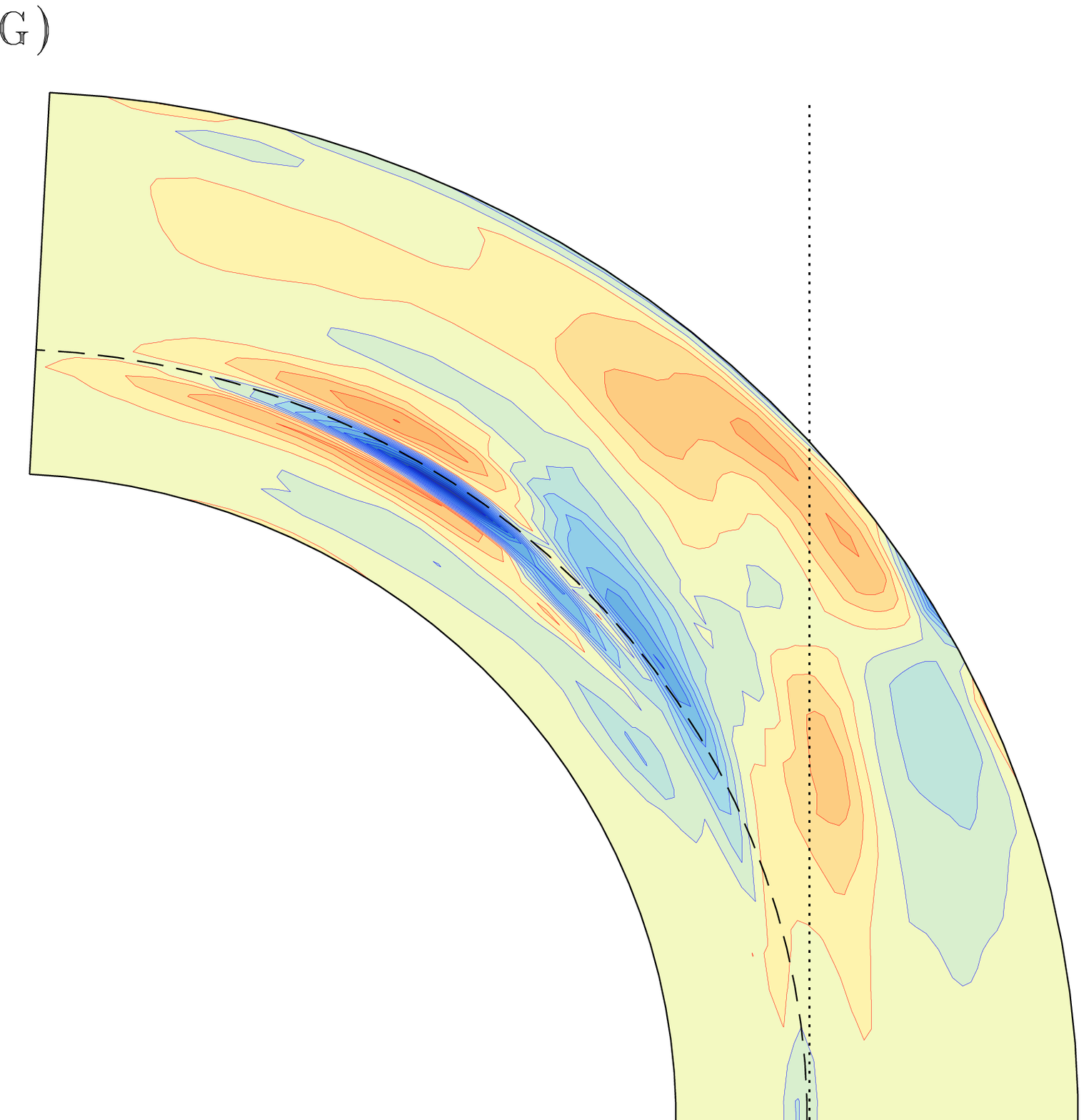}
     \includegraphics[width=4.5 cm]{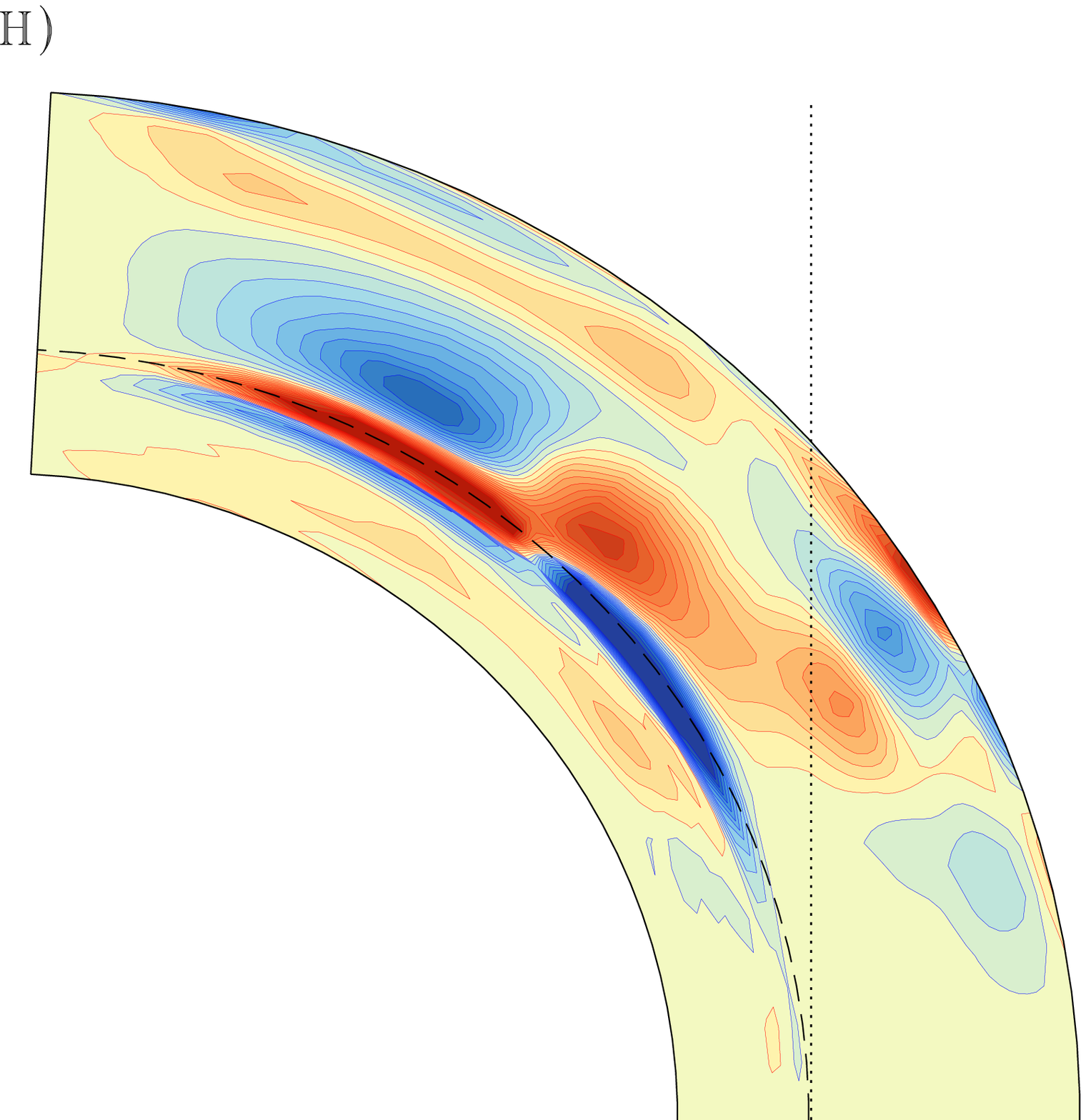}
     \caption{The several terms of equation (\ref{eq:dLdt}) averaged
     over 10 cycle minima (A, B, C, D) and 10 cycle maxima (E, F, G, H).
     We focus only on the NH for simplification.
     A) and D) show the l.h.s. of equation (\ref{eq:dLdt});
     B) and F),  $-\nabla \cdot \left(\mathbf{F}^{\mathrm{RS}}\right)$;
     C) and G), $-\nabla \cdot \left( \mathbf{F}^{\mathrm{MS}}\right)$; and
     D) and H), depict  $-\nabla \cdot \left(\mathbf{F}^{\mathrm{MT}}\right)$.
     All the contours have the same color scale as in Fig.\ \ref{fig:rhsMHD}. The vertical dotted lines indicate the TC for reference.}
     \label{fig:rhsMHD2}
\end{figure*}

The dominant role of the large scale magnetic torque in regulating
the MC inside the TC suggests that the cyclic variation of
the large scale field should
account for the cycle dependence of the MC, as discussed in Sec.\ 3.  We
find that this is indeed the case, as demonstrated in Fig.\
\ref{fig:rhsMHD2}. The most apparent difference between cycle minimum and
cycle maximum is in the mean magnetic torque, represented in panels D) and H).
Although the transport of angular momentum by the large scale
Lorentz force relies on the zonal poloidal field, the mid-latitude poloidal
fields in the lower CZ are generally strongest when the toroidal bands
are strongest.
Thus, the establishment of meridional flows by magnetic torques via GP is
most efficient at cycle maximum.  This is when the mid-latitude upwelling at
$\pm 48^\circ$ in the lower CZ is strongest. Signatures from the secondary
dynamo mode operating in the upper CZ near $\pm 28^\circ$ latitude are
also apparent in panels C), D, G) and F) of Fig.\ \ref{fig:rhsMHD2}.

As expected, $\mathbf{F}^{\mathrm{RS}}$ shows little difference between
cycle minima and maxima (Figs.\ \ref{fig:rhsMHD2}B and \ref{fig:rhsMHD2}F).
The Maxwell stress (Figs.\ \ref{fig:rhsMHD2}C and \ref{fig:rhsMHD2}G)
acts as a diffusive component (opposing the Reynolds stress below 48$^\circ$
in the CZ) with a time-varying component (opposing the large
scale magnetic torque in the stable layers and above 48$^\circ$ in the
lower CZ). This likely reflects non-axisymmetric structure in the bands,
as opposed to turbulent diffusion by smaller-scale, more chaotic motions.
The transport of angular momentum by the MC depicted by panels A) and
E) of Fig.\ \ref{fig:rhsMHD2} shows the spatial combination of the action
of the r.h.s. torques.

\begin{figure*}[htb]
  \centering
     \includegraphics[height=7 cm]{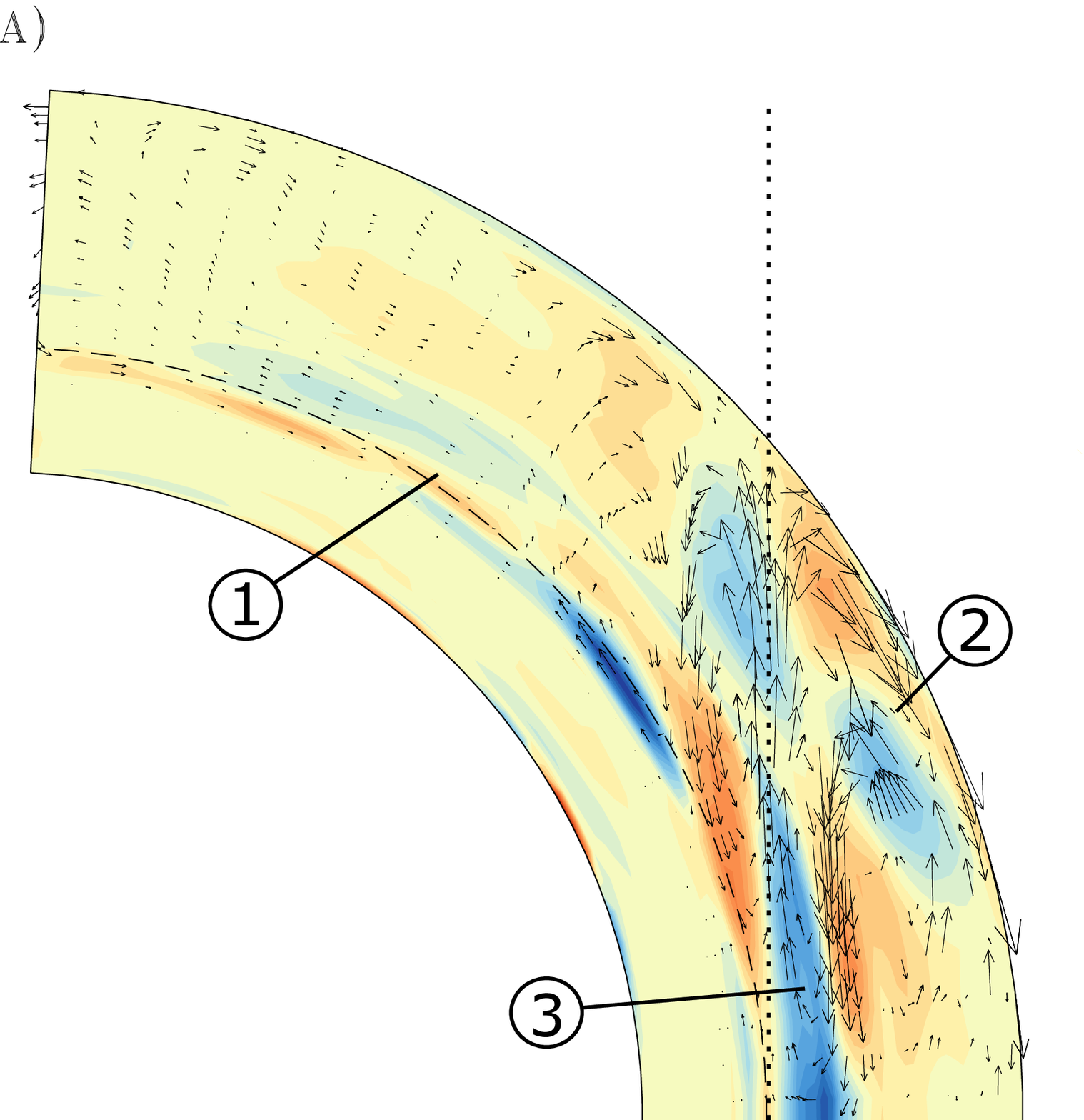}
     \includegraphics[height=7 cm]{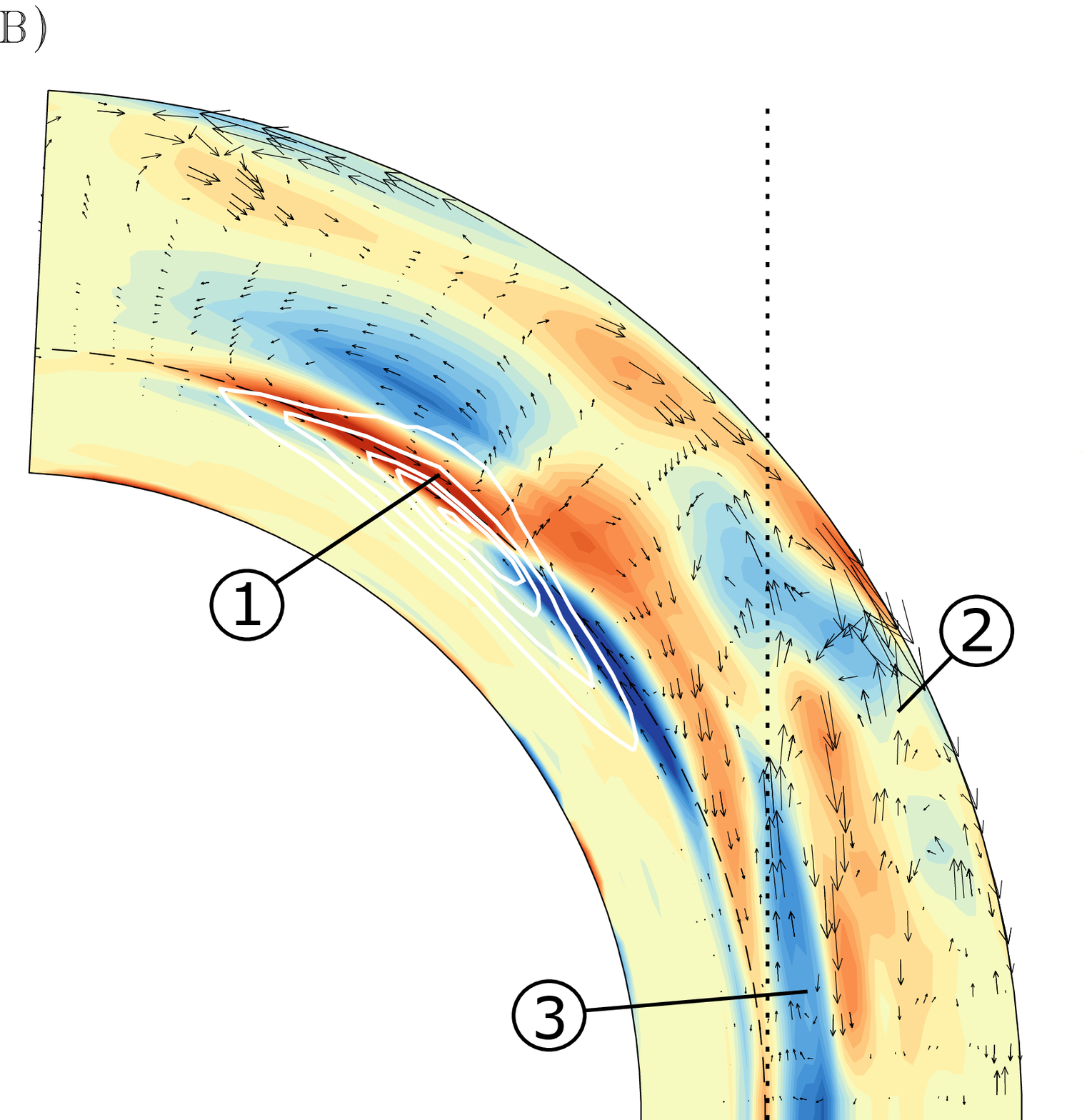}\\
     \vspace{0.3cm}
     \includegraphics[width=16 cm]{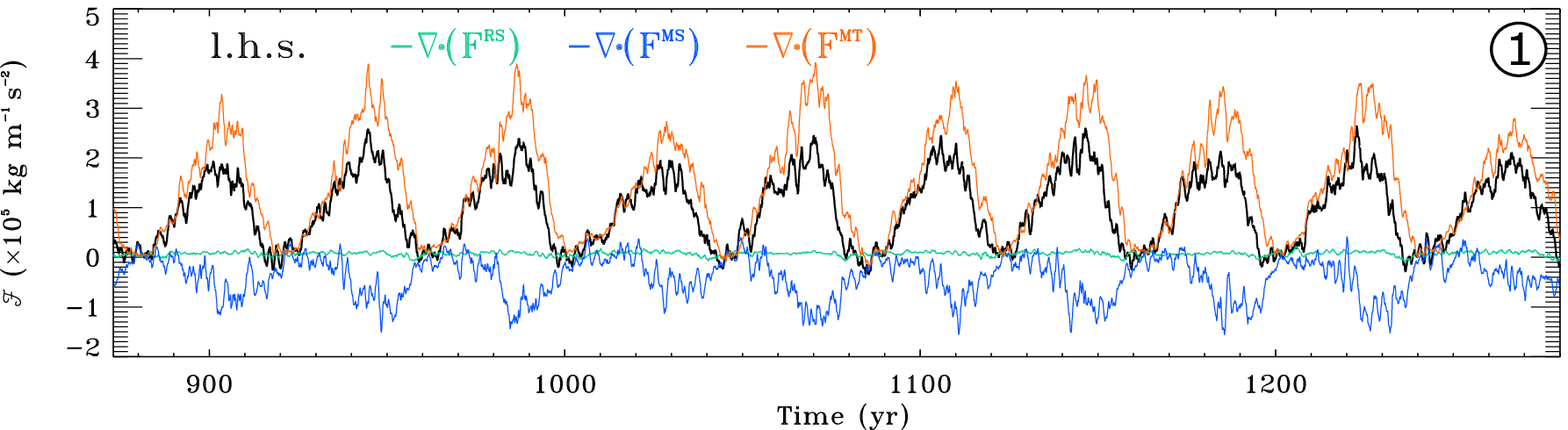}
     \includegraphics[width=16 cm]{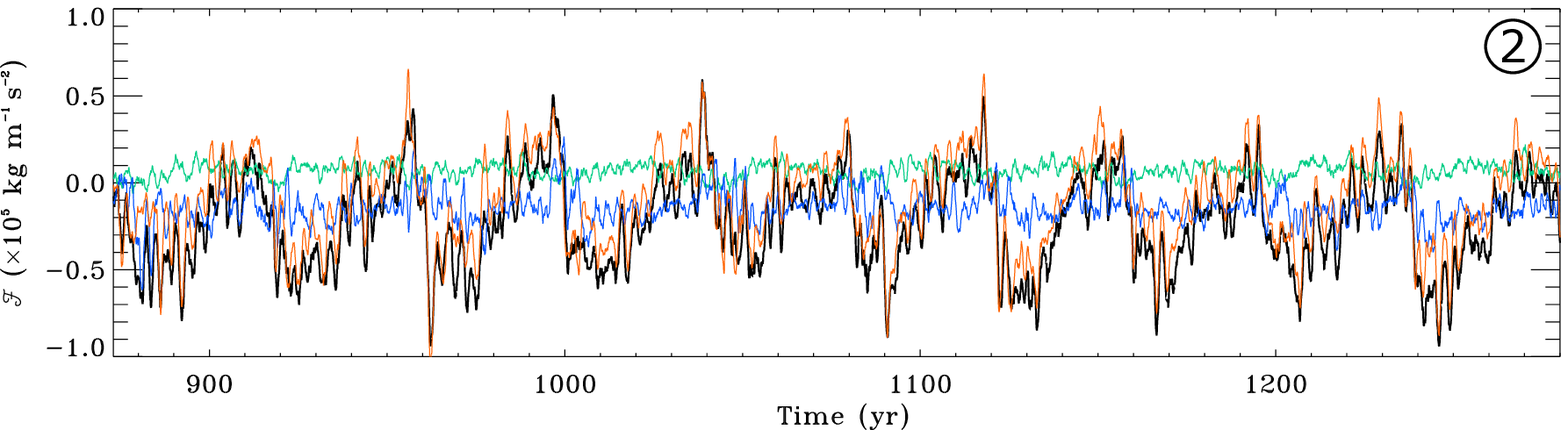}
     \includegraphics[width=16 cm]{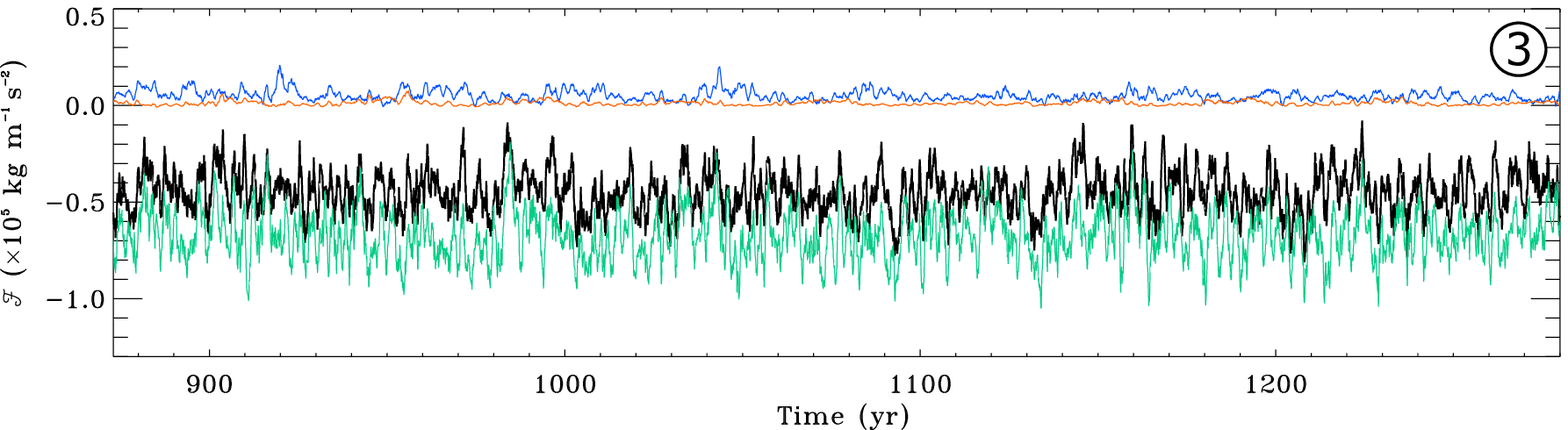}
     \caption{Panels A) and B) show the r.h.s.\ of equation (\ref{eq:dLdt})
     in the NH, averaged over cycle minima and cycle maxima. The
     vector field represents the direction of the meridional flow and the
     vertical dotted line represents the TC. The
     white contour lines in B) show the area where the toroidal field
     accumulates at cycle maximum. The bottom panels show the time
     evolution of the l.h.s.\ of equation (\ref{eq:dLdt}) and the individual
     terms on the r.h.s.\ (as labelled), sampled at the numbered
     locations. The data in the lower three panels were smoothed with a 1 year average filter.}
     \label{fig:rhsevolution}
\end{figure*}

In Figure \ref{fig:rhsevolution} we take a closer look at the time evolution
of the various torque components and the response of the MC.
Panels \ref{fig:rhsevolution}A and \ref{fig:rhsevolution}B show the net
axial torque overlaid with the direction and
amplitude of the meridional flow, shown as arrows.  This confirms our
interpretation in terms of GP; regions of negative torque
(flux divergence: blue) generally exhibit a flow component toward the rotation
axis (in addition to  an axial flow component sustained by mass conservation)
and regions of positive torque (flux convergence: red) generally exhibit a
flow component away from the rotation axis.
In panel \ref{fig:rhsevolution}B, it is clear that the mid-latitude upwelling in the lower CZ
during cycle max, and the associated torques, are induced by the presence
of the toroidal bands (white contours).

The bottom panels of Fig.\ \ref{fig:rhsevolution} show the evolution of the
l.h.s. of Eq. (\ref{eq:dLdt}) together with the individual components of
the net axial torque sampled at three different locations:
\textcircled{\tiny 1} the area of higher correlation between the amplitudes
of the toroidal field and the horizontal flow, \textcircled{\tiny 2} where
the secondary (weaker) dynamo mode is operating and \textcircled{\tiny 3}
an area outside the TC and away of the main influence of magnetic torques.

\begin{figure*}[htb]
  \centering
     \includegraphics[width=16 cm]{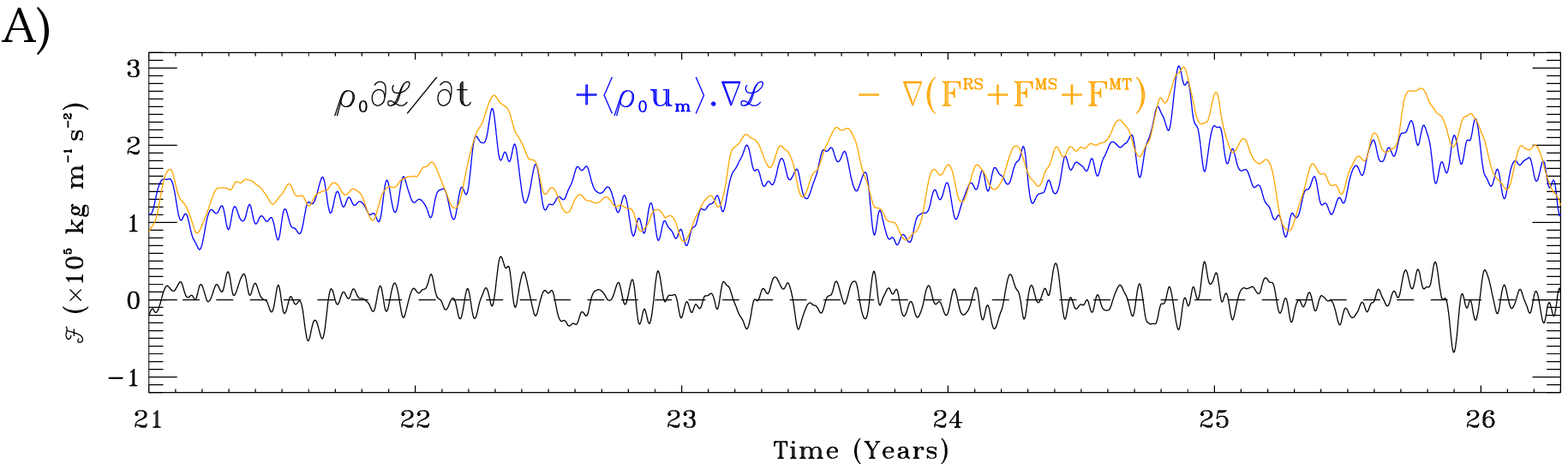}
     \includegraphics[width=16 cm]{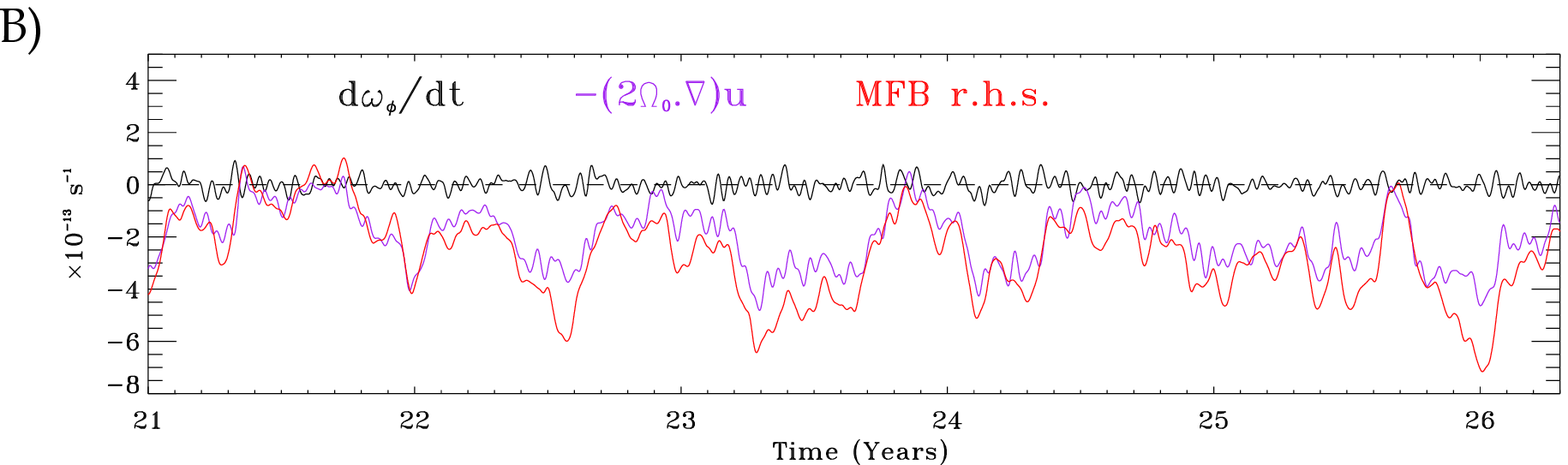}
     \caption{Panel A) shows the individual terms of the l.h.s.\ of
     equation (\ref{eq:dLdt}) (in black and blue) and the total contribution
     of the r.h.s. (orange) sampled at region \textcircled{\tiny 1} for a 5
     years interval, taken at cycle maximum.
     Panel B) shows the individual terms of the l.h.s. (black and purple)
     and r.h.s. (red) of the meridional force balance (equation
     (\ref{eq:thermalwind}) in section \ref{sec:MFB}) for the same location
     and for the same time interval.}
     \label{fig:rhsevolution2}
\end{figure*}

The variability in region \textcircled{\tiny 1} is clearly driven by the
large scale Lorentz force, $\textbf{F}^\textrm{MT}$ (orange line),
as discussed above.
In the heart of the toroidal bands, magnetic tension is accelerating the
rotation rate by extracting angular momentum from lower latitudes.  Maxwell
stresses resist this acceleration (blue line).
The sum of these two components (not show here) closely matches the black line
that represents the l.h.s.\ of equation (\ref{eq:dLdt}). The Reynolds stress
is almost negligible here, showing signs of cycle modulation but at 1 to 2
orders of magnitude below the amplitude of the other signals.  The small
amplitude of the Reynolds stress reflects the location where the toroidal
bands accumulate; inside the TC and close to the base of the
CZ where the convective amplitude is weak.
In Fig. \ref{fig:rhsevolution2}A we present the individual
contributions of the two terms of the l.h.s. of equation (\ref{eq:dLdt}) and
the joint contribution of the torques of the r.h.s. sampled at
in region \textcircled{\tiny 1}, for an interval of 5 years taken around a
cycle maximum\footnote{These graphics where produced using a 40 years
extension (half magnetic cycle) to the \textit{millennium simulation} with
a higher temporal data cadence of 24h.}. This figure clearly shows that the angular
momentum response due to variations in the zonal flows happens at a time
scale of the order of one month (black line) and that the long term
variations in the MC (blue line) follow the GP forcing exerted by the r.h.s.
torques.

\begin{figure*}[!htb]
  \centering
     \includegraphics[width=16 cm]{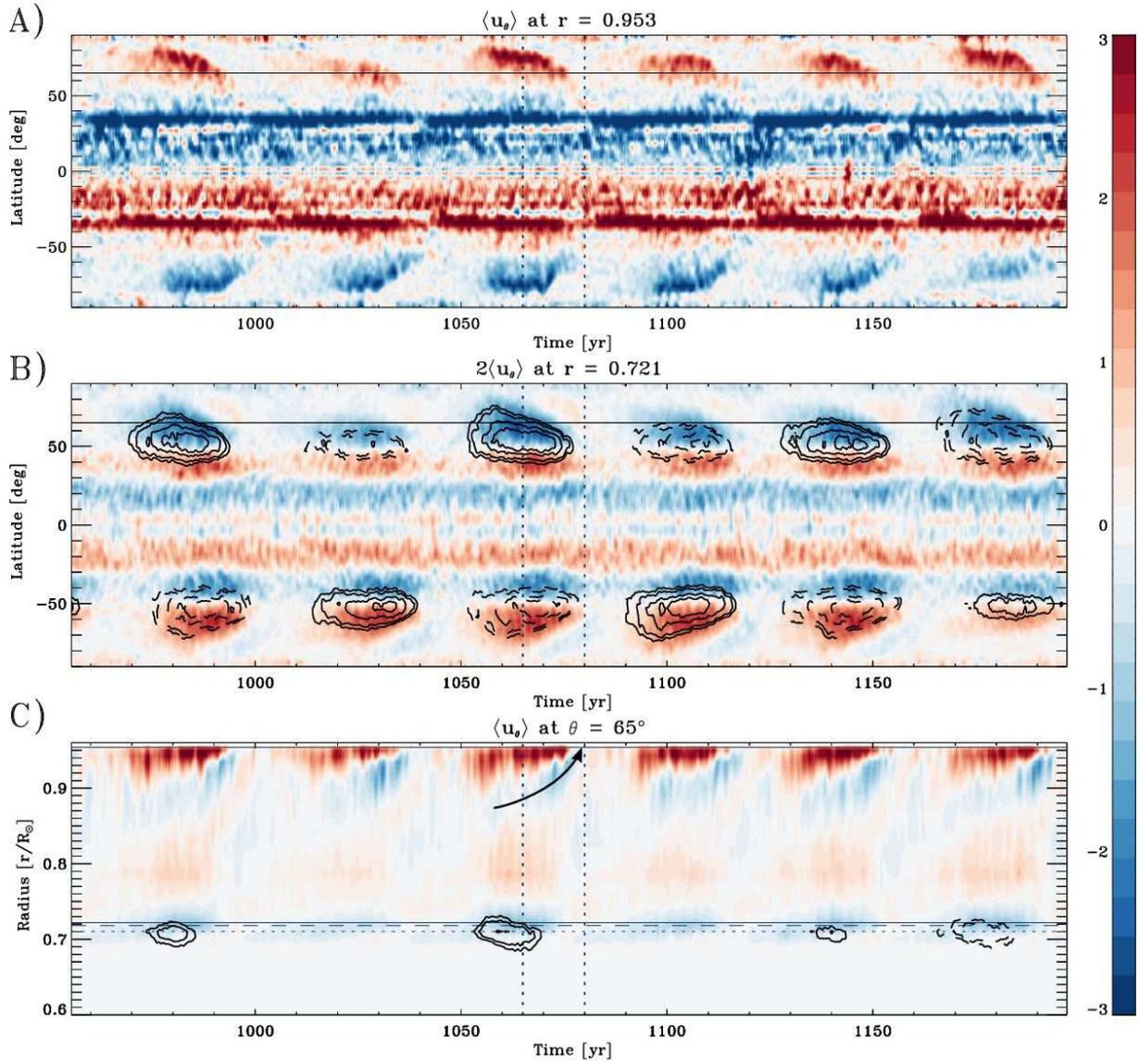}
     \caption{Mean latitudinal flow $\langle u_\theta \rangle$ as
        a function of latitude and time at A) the top layers ($r=0.95
        R_\odot$) and B) near the base of the CZ ($r=0.72 R_\odot$). In order to have a better contrast, in panel B) the displayed quantity is actually 2$\langle u_\theta \rangle$. The black contour lines represents $\langle B_\phi\rangle$ (solid and dashed for positive and negative polarities respectively). The horizontal solid line represents the latitude at which panel C) is sampled.
        The time interval covers cycles 23 to 28. The color scale
        saturates at $\pm 3$ m s$^{-1}$ (red northward, blue southward).
        Panel C) is a radius vs. latitude plot of $\langle u_\theta \rangle$
        taken at 65$^\circ$ north.
        Both  $\langle u_\theta \rangle$ and $\langle B_\phi\rangle$  are smoothed over 6 months. The vertical dotted lines represent the maximum and minimum of cycle 25. The dashed line marks the tachocline depth, the dotted line below it is the depth where $\langle B_\phi\rangle$ contours are computed and the solid line is the depth where panel B) is sampled.}\label{fig:mcbfly}
\end{figure*}

In terms of dynamics, near the bottom of the CZ, the acceleration of
the fluid within the heart of the toroidal bands induces an equatorward
flow while the extraction of angular momentum from lower latitudes induces
a poleward flow. This establishes a meridional flow that
converges horizontally into the magnetic toroidal bands
and then turns upward (see Figs. \ref{fig:rhsevolution}B and \ref{fig:mcbfly}).
The location of this deep convergence and upwelling
is shifted slightly toward the equatorward edge of the bands, as
shown in Fig.\ \ref{fig:mcbfly}B.
It is also interesting to notice from panel \ref{fig:mcbfly}C
that for latitudes higher then 50$^\circ$ (above region
\textcircled{\tiny 1}) we can observe an apparent migration of
equatorward flows (blue in the north) from the middle of the CZ to the upper
layers as the cycle unfolds (marked with an arrow).
This is not actually a migration. It is a decrease in the radius of the
near surface CCW MC cell that can be found at those latitudes (see Fig. 3)
caused by the decrease of the torque near the surface. As the magnetic torque
becomes weaker from the cycle maximum to minimum, the near
surface CCW cell becomes thinner and its equatorward flow (blue)
approaches the top layers.
At the same time there is a CW MC cell near the pole (see
Fig. \ref{fig:bmoy_25}) that gradually expands to lower latitudes.
The surface equatorward section of this higher latitude cell associated
with the previous behaviour is what finally establishes the
observed superficial dynamical pattern above 50$^\circ$.

\pagebreak

It is clear from Figs.\ \ref{fig:rhsevolution} to \ref{fig:mcbfly} that
the accumulation of strong toroidal fields induces cyclic variations in
the meridional flow. Though MC
variability induced by Lorentz force feedbacks has been studied within the
context of non-kinematic mean field models \citep[e.g.][]{Rempel06},
it is worth noting that this line of causality is in stark contrast to
the kinematic assumption that is often adopted in mean-field solar dynamo
modeling.

The time evolution of the axial torques and meridional flow in region
\textcircled{\tiny 2} is more erratic (see Fig.\ \ref{fig:rhsevolution}).
In this small border-like region the large scale magnetic torque
and the Maxwell stress have the same sign. The meridional flow compensates this changes
in the axial torque and a small CCW circulation cell appears during
the maximum.
In this region, the magnetic torque peaks (at negative values) around
the time when the deep toroidal field in region \textcircled{\tiny 1} starts
to rise.  Meanwhile, there is also a pronounced shorter-term variability
that is associated with the periodicity of the secondary dynamo mode.

Finally, we also highlight region \textcircled{\tiny 3} in Fig.\
\ref{fig:rhsevolution}. Here, where the large scale magnetic field has a
small influence, we observe almost no variation with the magnetic cycle.
Like in the HD case, the angular momentum transport is dominated by
banana cells ($\mathbf{F}^\mathrm{RS}$).

\section{Meridional force balance in the presence of large scale magnetic fields}
\label{sec:MFB}

Besides the GP mechanism presented in the previous section,
the MC might also be influenced by the presence of
entropy gradients throughout the CZ. These thermal gradients imply
that surfaces of constant mean pressure and density do not overlap completely,
which gives rise to a baroclinicity contribution in the vorticity equation.
This contribution influences the whole system causing the MC to readjust
in order to achieve the so called thermal wind balance (TWB).
In the context of purely HD simulations, e.g.\
\cite{Brun2002,Miesch2006,Brun2010,Brun2011} have shown that the action
of baroclinicity in TWB, together with the Reynolds stresses are
the key factors for the system to achieve such equilibrium.
In addition, by studying fully MHD models of solar-like stars,
\cite{Varela2016} unveiled the relevant role of a large-scale magnetic
field in the TWB influence in differential rotation.
In this section, we assess the influence of the magnetic field in the
on the MC by comparing the balance conditions in HD and MHD models.

We find useful to start by defining an equation for the evolution of
the vorticity, $\mathbf{\omega}=\nabla \times \mathbf{u}$. Applying
$\nabla \times$ to the momentum equation (\ref{eq:momentum}) yields
\bea
      \frac{\partial\mathbf{\omega}}{\partial t}&=&
      (\mathbf{\omega_a} \cdot \nabla) \mathbf{u}
      -(\mathbf{u} \cdot \nabla) \mathbf{\omega_a}
      - \mathbf{\omega_a}(\nabla \cdot \mathbf{u})
      - \nabla \times \left(\mathbf{g}\frac{\Theta^\prime}{\Theta_0}\right)
      + \frac{1}{\mu_0} \left(\nabla \frac{1}{\rho_0}\right)
        \times (\mathbf{B} \cdot \nabla) \mathbf{B}
      + \frac{1}{\mu_0 \rho_0} (\nabla \times
      (\mathbf{B}\cdot \nabla)\mathbf{B}) \;,
\label{eq:dvorticitydt}
\eea
where $\mathbf{\omega_a}=(\nabla \times \mathbf{u}) + 2 \mathbf{\Omega_0}$
is the absolute vorticity.

As mentioned before, TWB studies have been carried out mainly for HD
simulations and generalized under the argument that the magnetic field
influence can be neglected. However, as we have shown in the previous section
the magnetic field has an important role in the GP forcing mechanism.
Therefore, in order to gauge how important this magnetic
contribution is for TWB, we develop an equation for the meridional force balance (MFB) by computing the zonally averaged $\hat{\mathbf{e}}_\phi$
component of the vorticity evolution equation (\ref{eq:dvorticitydt}). A
similar equation is also presented \cite{Strugarek2011} and in
the recent work of
\cite{Varela2016}\footnote{Written for polar spherical coordinates, where
$\theta$ is the co-latitude.} but used in a different context.
More details are presented in the appendix.
The MFB equation assumes the form

\begin{small}
\begin{eqnarray}
    \frac{\partial \langle \omega_\phi\rangle}{\partial t}
    - \left\langle \, 2 \Omega_0 \left(\sin \theta \frac{\partial
    u_\phi}{\partial r} + \frac{\cos \theta}{r} \frac{\partial u_\phi}{\partial
    \theta}\right)
    \right\rangle &=&
     \underbrace{
    \left\langle \mathbf{\omega} \cdot \nabla u_\phi
    + \frac{\omega_\phi u_r}{r} -\frac{\omega_\phi u_\theta
    \tan\theta}{r}\right\rangle }_{Stretching} 
    +\, \underbrace{
    \left\langle  - \mathbf{u} \cdot \nabla \omega_\phi
    -\frac{u_\phi \omega_r}{r} -\frac{u_\phi \omega_\theta
    \tan\theta}{r}\right\rangle}_{Advection} 
    +\, \underbrace{
     \left \langle -\omega_\phi(\nabla \cdot \mathbf{u})   \right\rangle}_{Compressibility} \nonumber \\
    &&+\, \underbrace{
    \left\langle \frac{g(r)}{r}\frac{\partial}{\partial \theta} \left(
     \frac{\Theta'}{\Theta_0} \right)\right \rangle}_{Baroclinicity}
     + \underbrace{\left \langle \frac{1}{\mu_0}\frac{\partial }{\partial r}
    \left(\frac{1}{\rho_0}\right)\left[-\mathbf{B}\cdot\nabla B_\theta
    -\frac{B_\phi^2}{r} \tan\theta - \frac{B_\theta B_r}{r}\right]
    \right\rangle}_{Magnetic\,\, contribution\, 1}\nonumber \\
     &&+ \underbrace{\left\langle \frac{1}{\mu_0 \rho_0}\frac{1}{r} \left[
    \frac{\partial }{\partial r}
     \left(-r \mathbf{B} \cdot \nabla  B_\theta
     - B_\phi^2  \tan\theta
     - B_\theta   B_r  \right)
     +\,\frac{\partial}{\partial \theta} \left(
    \mathbf{B} \cdot \nabla B_r -
    \frac{ B_\theta^2}{r} -
    \frac{ B_\phi^2 }{r} \right)
    \right] \right\rangle}_{Magnetic\,\, contribution \,2}\, ,
    \label{eq:thermalwind}
\end{eqnarray}
\end{small}
\noindent where we can consider a stationary state (system in equilibrium) for
$\partial \langle\omega_\phi\rangle / \partial t =0$.
The second term on the l.h.s. is  the $\mathbf{\hat{e}}_\phi$ component
of $(2 \mathbf{\Omega_0} \cdot \nabla) \mathbf{u}$ or more commonly
represented by
$2 \Omega_0 \,\,\partial \langle u_\phi\rangle / \partial z$ where
$\partial / \partial z$ is the derivative in the direction parallel to
the rotation axis.
We follow \cite{Brun2011} for the naming for the several terms in
equation (\ref{eq:thermalwind}). The main differences between
theirs and ours MFB equation (11) is that their baroclinic term is
written in terms of the entropy while ours is written in terms of
potential temperature, the viscous term is absent in our case because
our simulation has no explicit viscosity and we take into account
the contribution of the magnetic field. We would like to note
that $\partial \omega_\phi / \partial t$ is not strictly zero but
it oscillates around zero on a time scale of commensurate with rotation
as shown in figure \ref{fig:rhsevolution2}B. As expected, this term
varies with the same frequency of the accelerations and decelerations
of the zonal flows depicted by $\partial \mathcal{L}/\partial t$.
Therefore, using previous arguments, we can consider that this
system is in a quasi-static equilibrium.

\subsection{MFB analysis for the HD case}

\begin{figure*}[htb]
  \centering
     \includegraphics[height=6 cm]{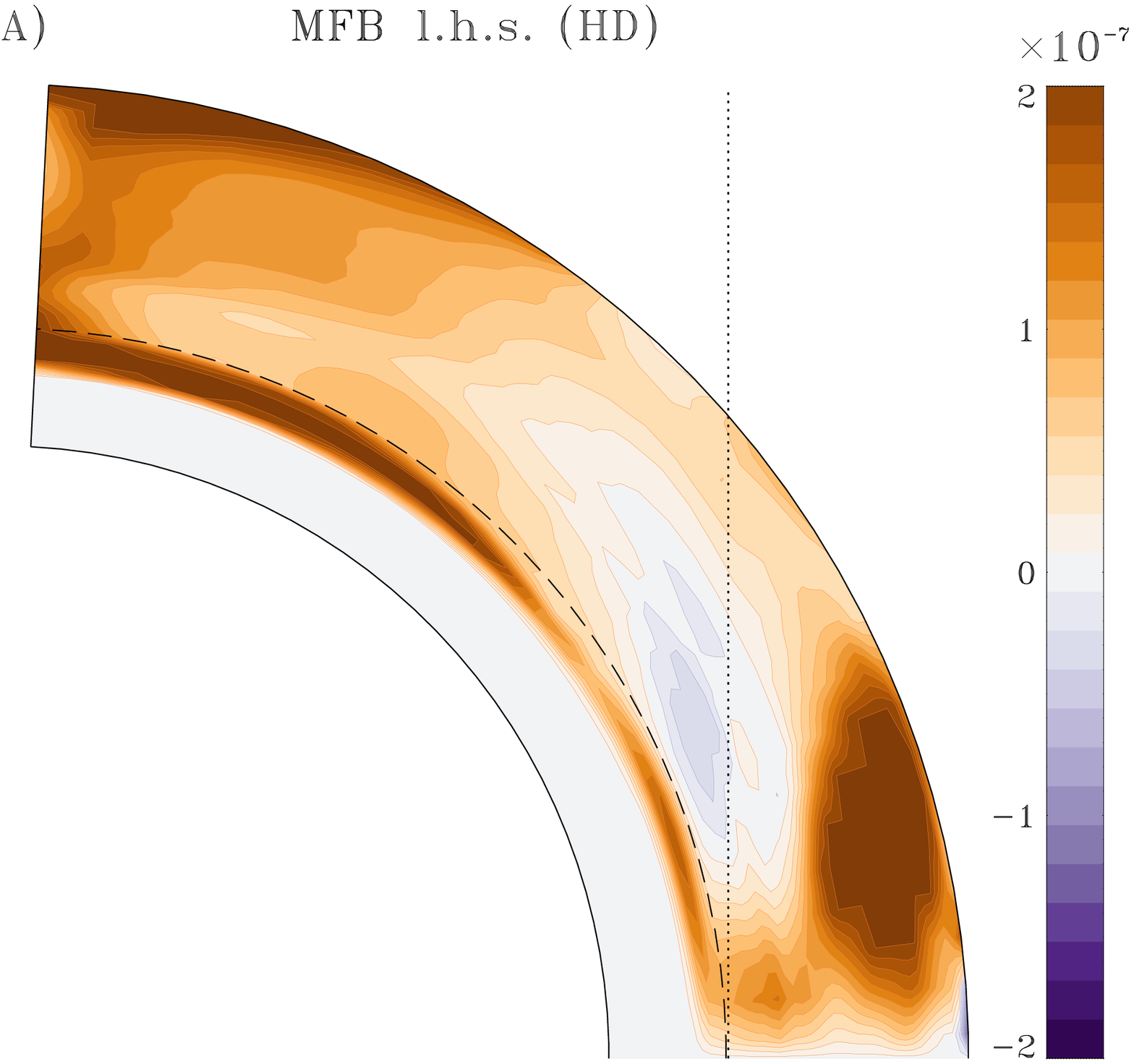}
     \includegraphics[height=6 cm]{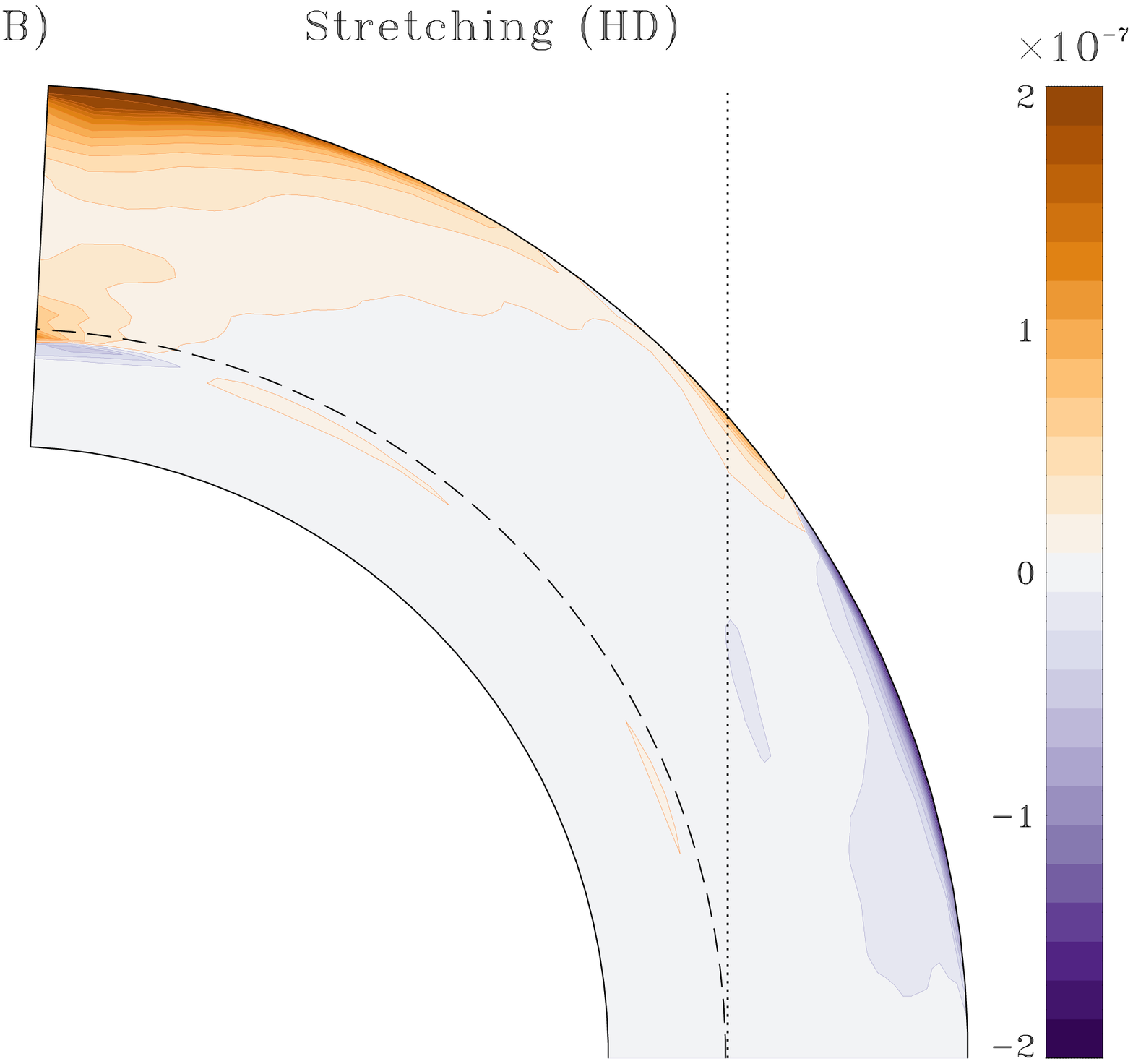}
     \includegraphics[height=6 cm]{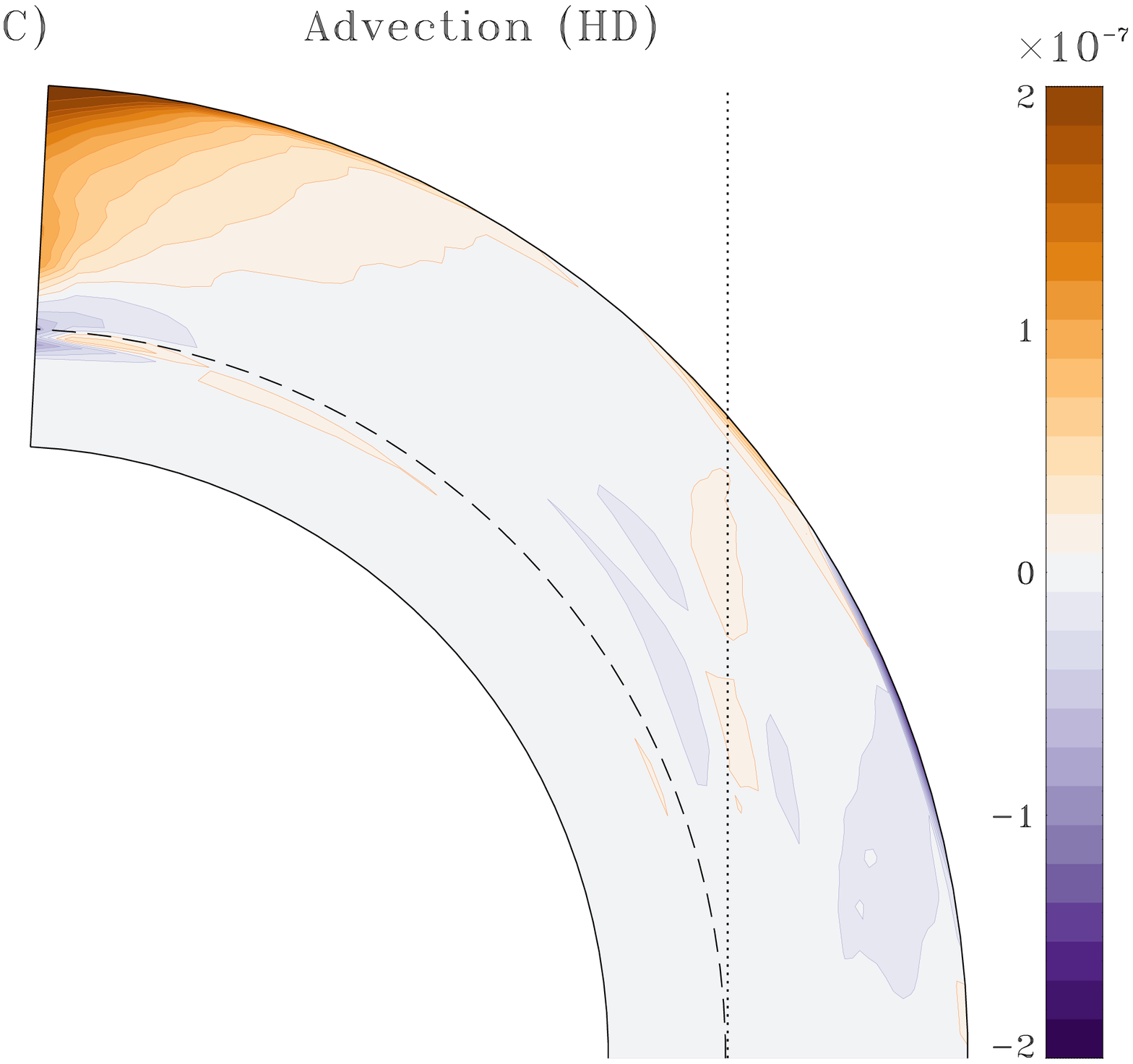}
     \includegraphics[height=6 cm]{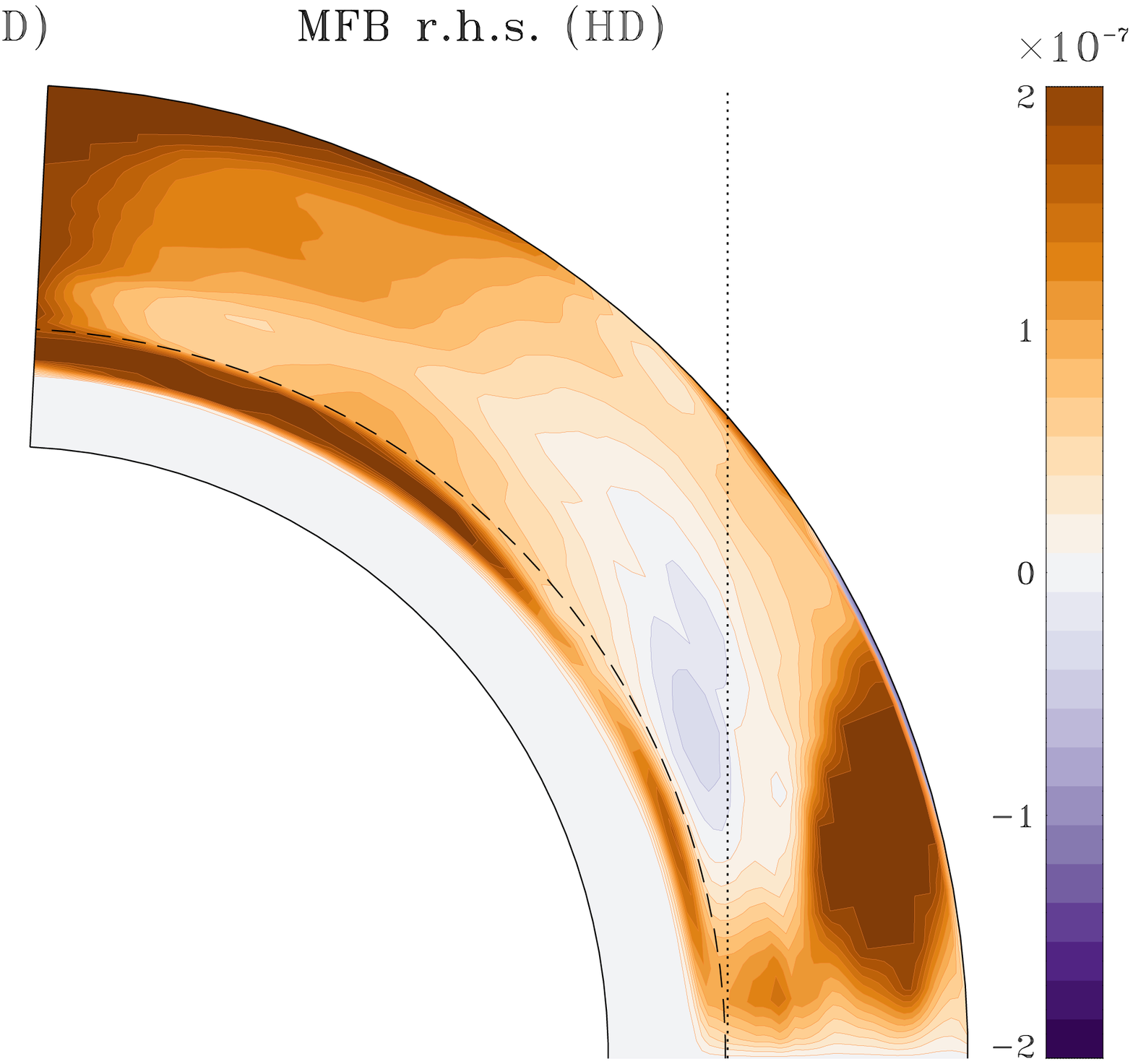}
     \includegraphics[height=6 cm]{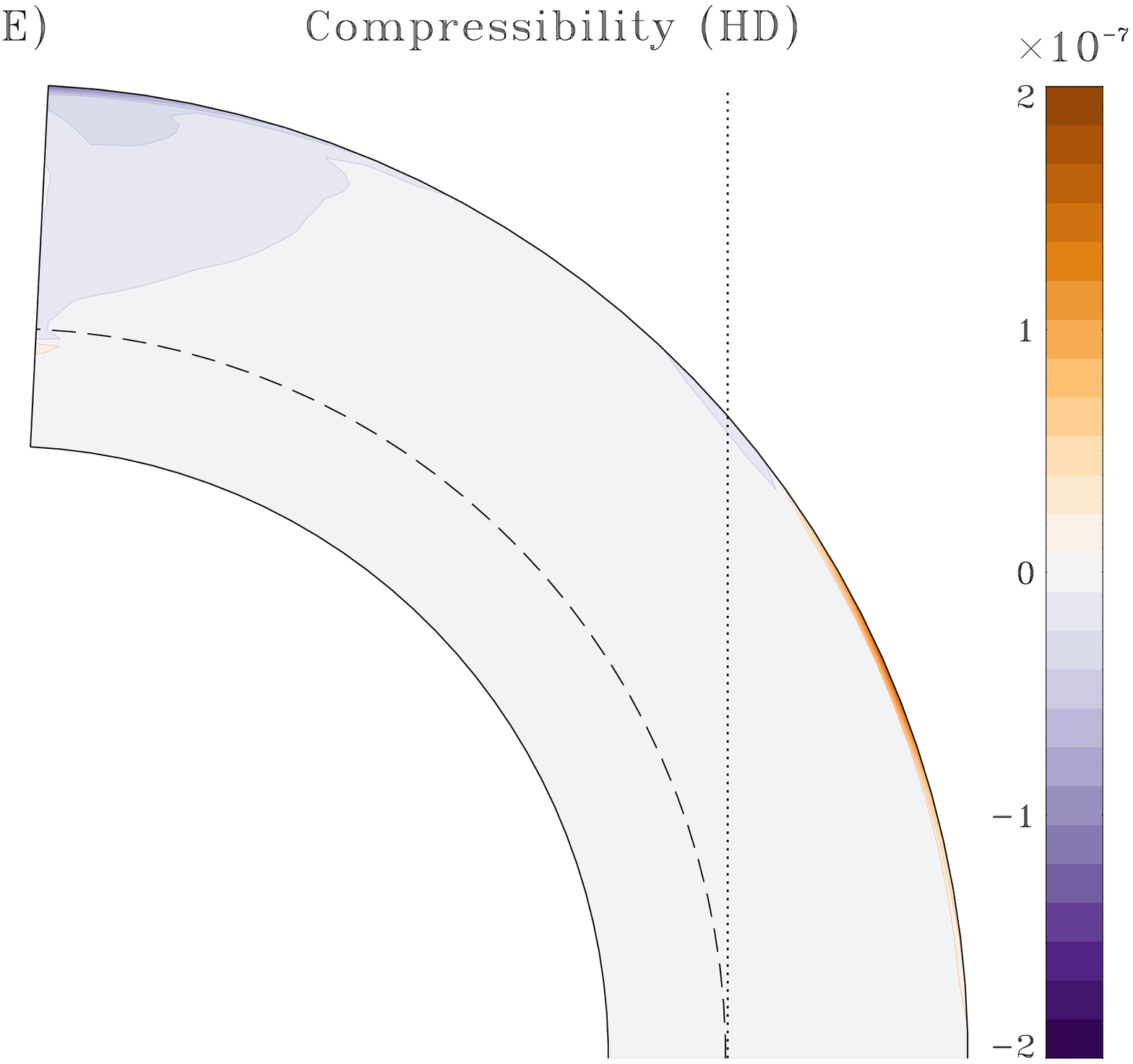}
     \includegraphics[height=6 cm]{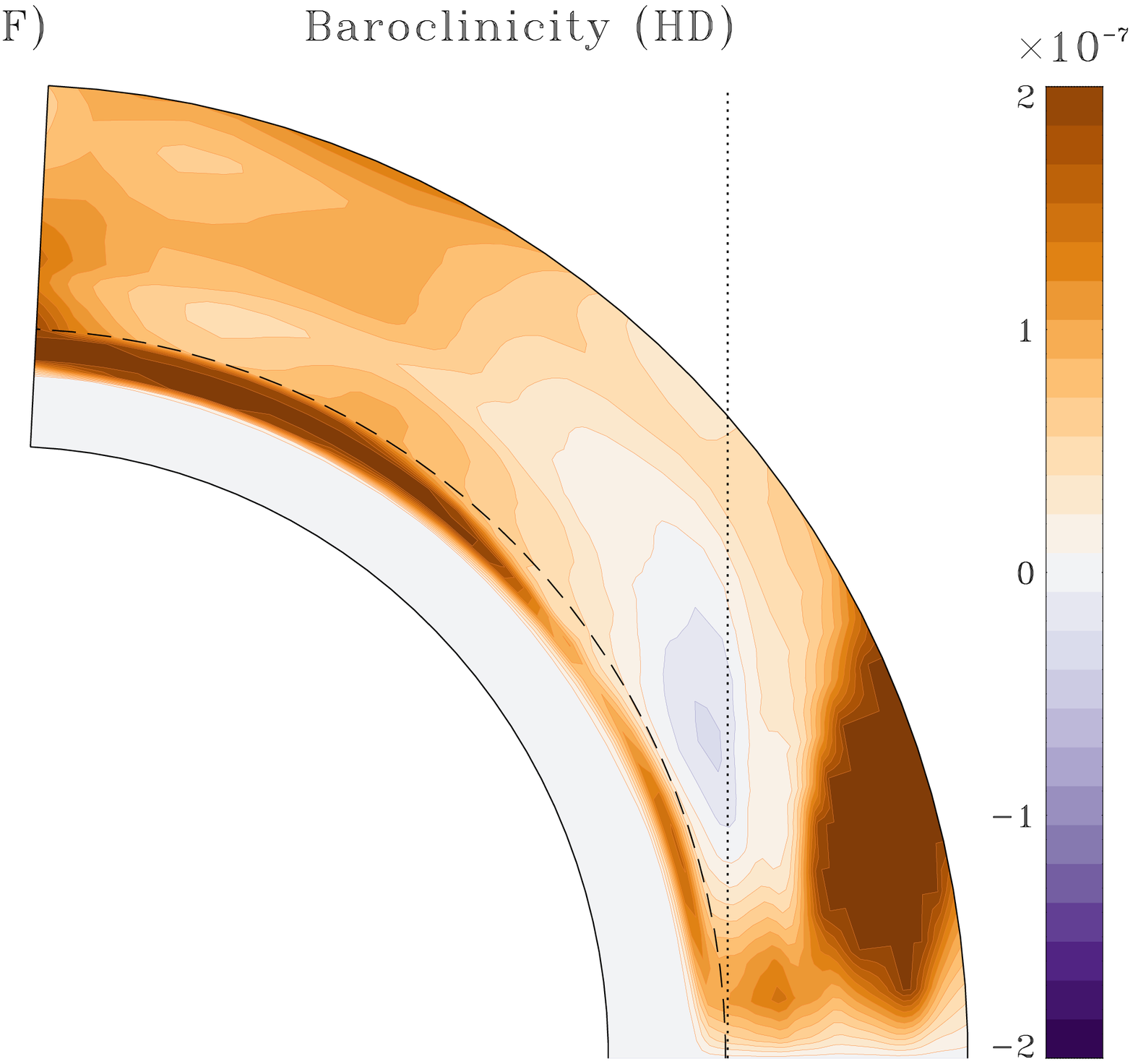}
     \caption{Panels A) and D) show the l.h.s. and r.h.s. of the
     MFB equation  for the HD simulation.
 	The terms were computed for the NH only, and with
	color saturation at $\pm\,3\times10^7$ s$^{-1}$.
	The vertical dotted line represents the TC. On the other
     4 panels we present the individual contribution on the r.h.s.,
     namely B) Stretching, C) Advection, E) Compressibility and F) Baroclinicity
	(with color saturation at $\pm\,2\times10^{-7}$ s$^{-1}$).}
     \label{fig:MFB_HD}
\end{figure*}

All the quantities needed to compute equation (\ref{eq:thermalwind}) can be
extracted directly from our numerical simulations. For comparison purposes
we start by showing the meridional force balance computed for the
HD case. In Fig.\
\ref{fig:MFB_HD} we show meridional plots for the l.h.s. and r.h.s.
(and its four first individual terms) of equation (\ref{eq:thermalwind})
averaged over the 246 yr interval and for the NH. For simplicity we
restrict our discussion to the NH and divided all terms by $2\Omega_0$.

The good agreement between the l.h.s and r.h.s. of (\ref{eq:thermalwind})
represented by panels A) and D) of Fig.\ \ref{fig:MFB_HD} is a clear
indication that in fact the system is very close to stationarity.
Although panel D) is the sum of the individual contributions of the r.h.s.
of equation (\ref{eq:thermalwind}), we can see that for most of the CZ
the baroclinic term (panel F)) is dominant.
It has a large positive contribution
in regions of strong rotational shear, such as the base of the CZ and
at low latitudes outside the TC. This is in agreement with
the results of \cite{Brun2011} who associate this behavior
with the presence of strong thermal gradients.
The baroclinic term is only weakly negative in the inner
border of the TC, a region where rotation has almost no
spatial variation. The contributions of stretching and advection are
only relevant near the pole, while vorticity compressibility is negligible.

\subsection{MFB analysis for the MHD case}

\begin{figure*}[htb]
  \centering
     \includegraphics[height=6 cm]{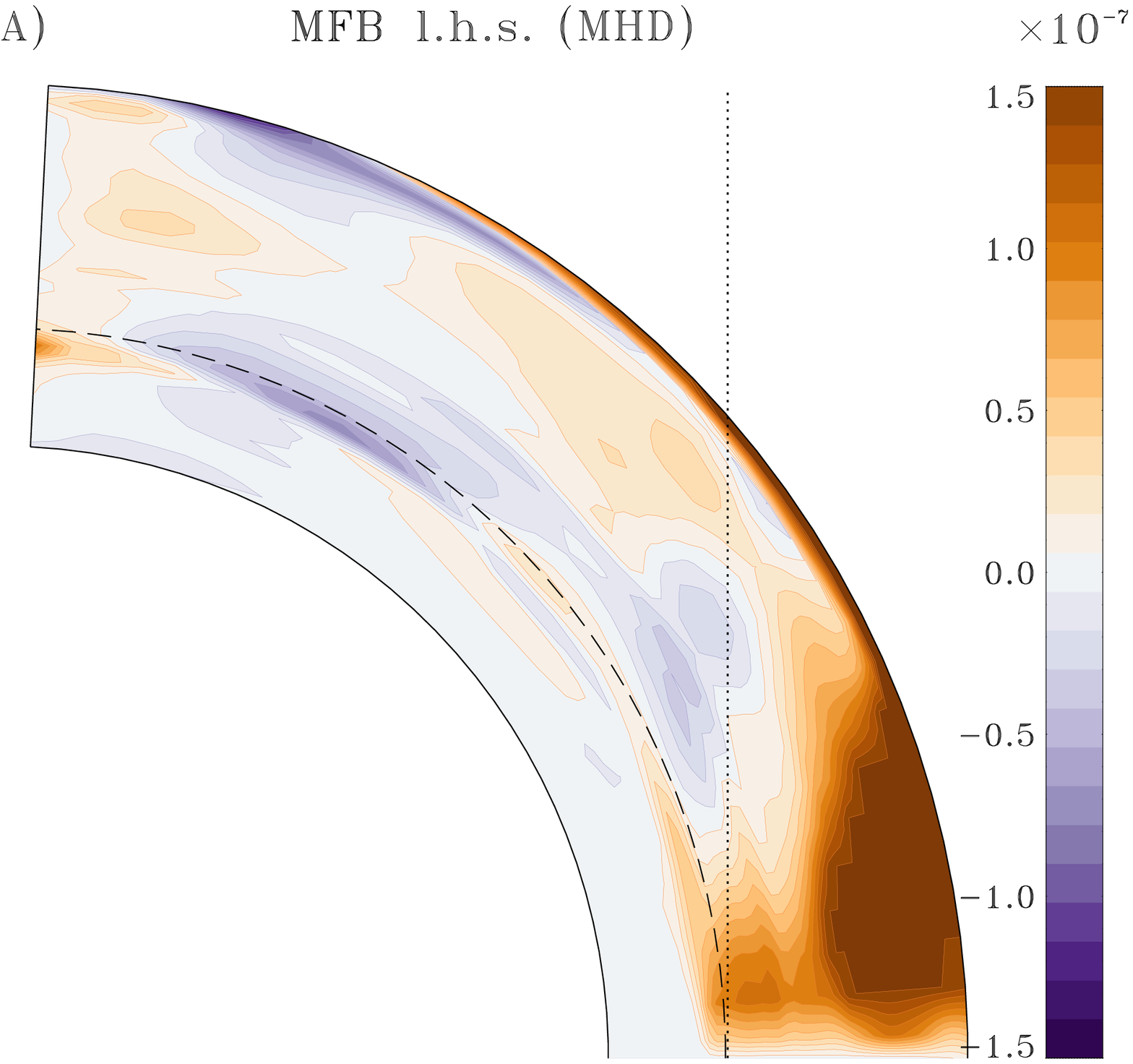}
     \includegraphics[height=6 cm]{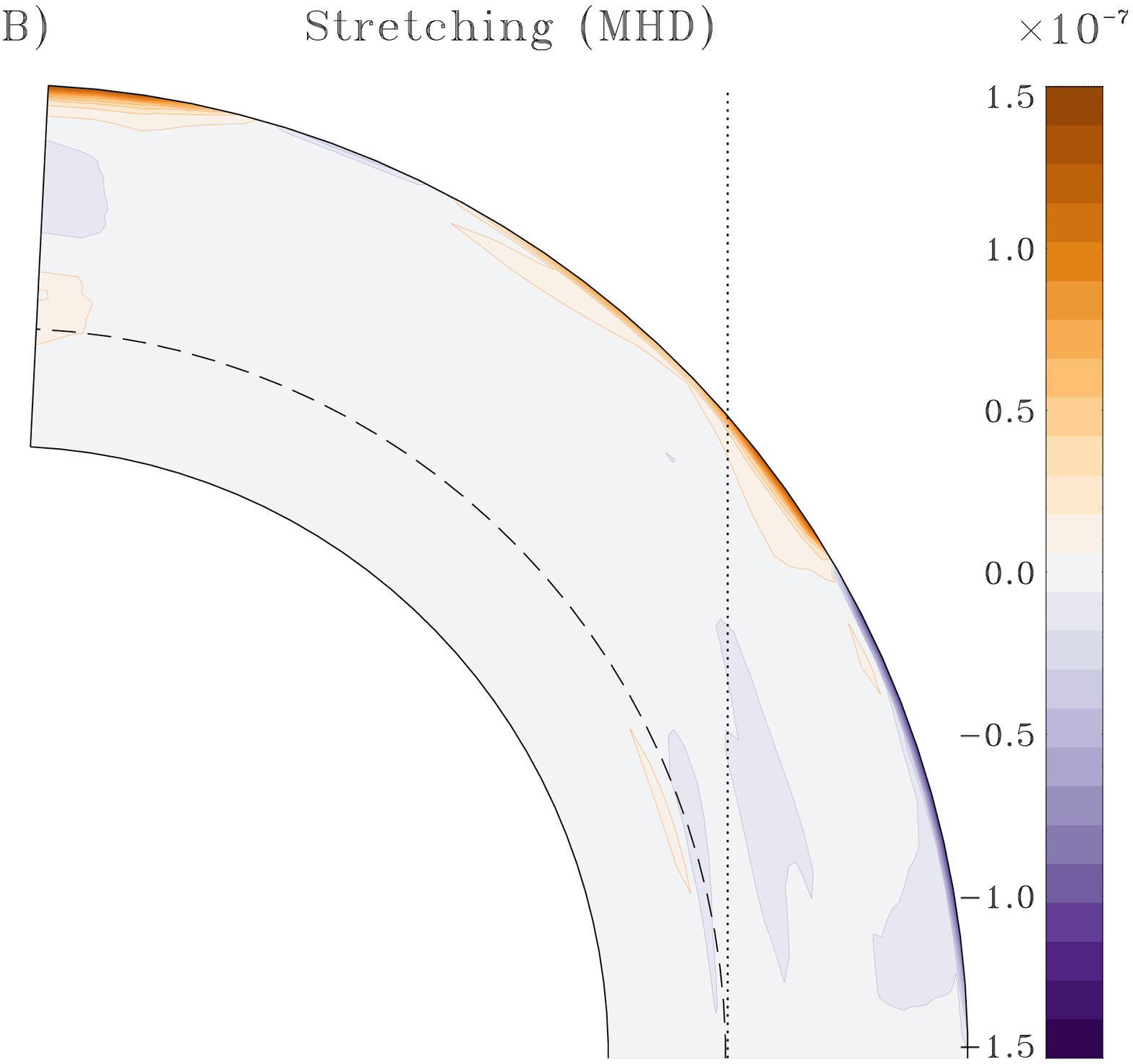}
     \includegraphics[height=6 cm]{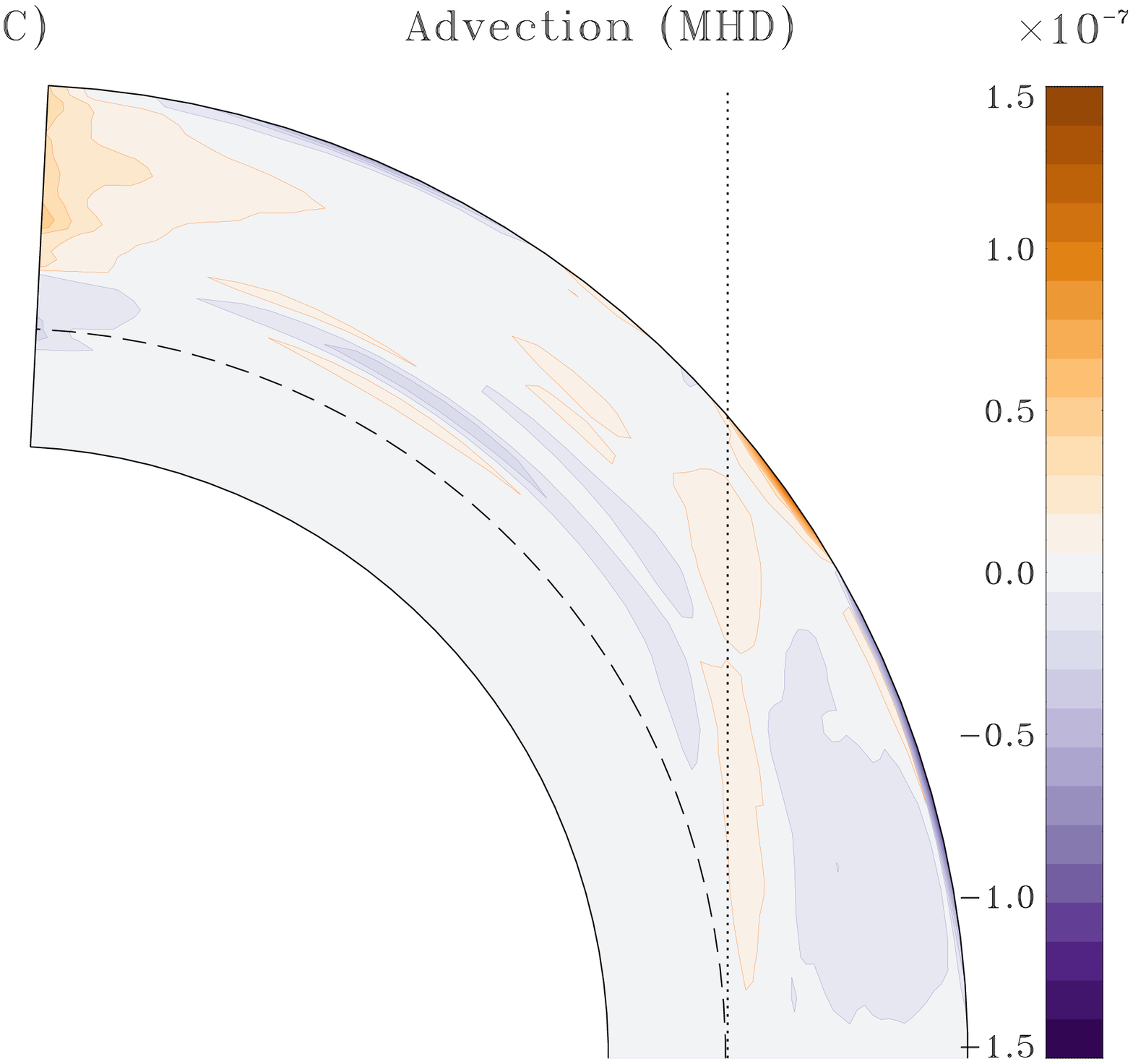}
     \includegraphics[height=6 cm]{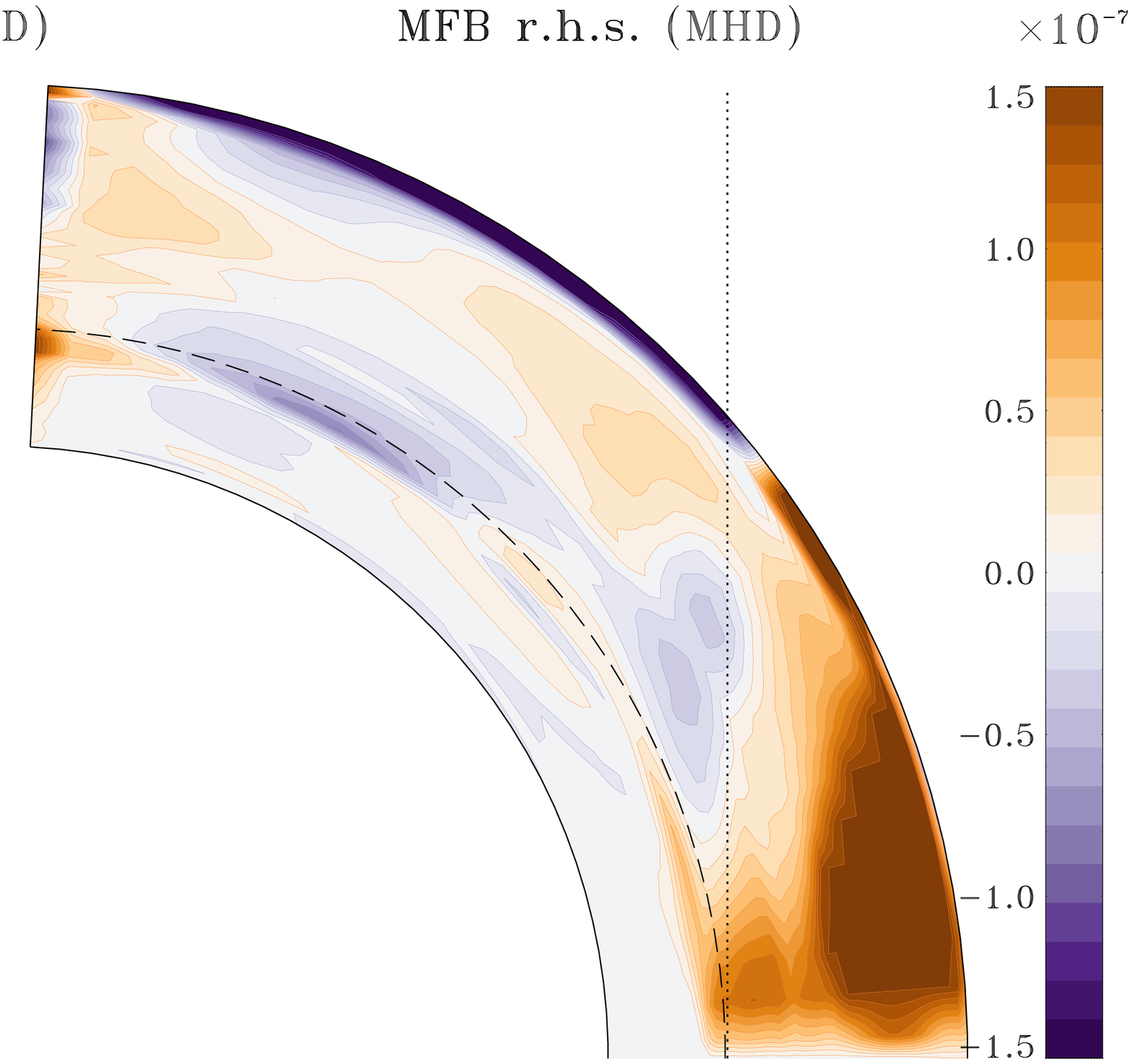}
     \includegraphics[height=6 cm]{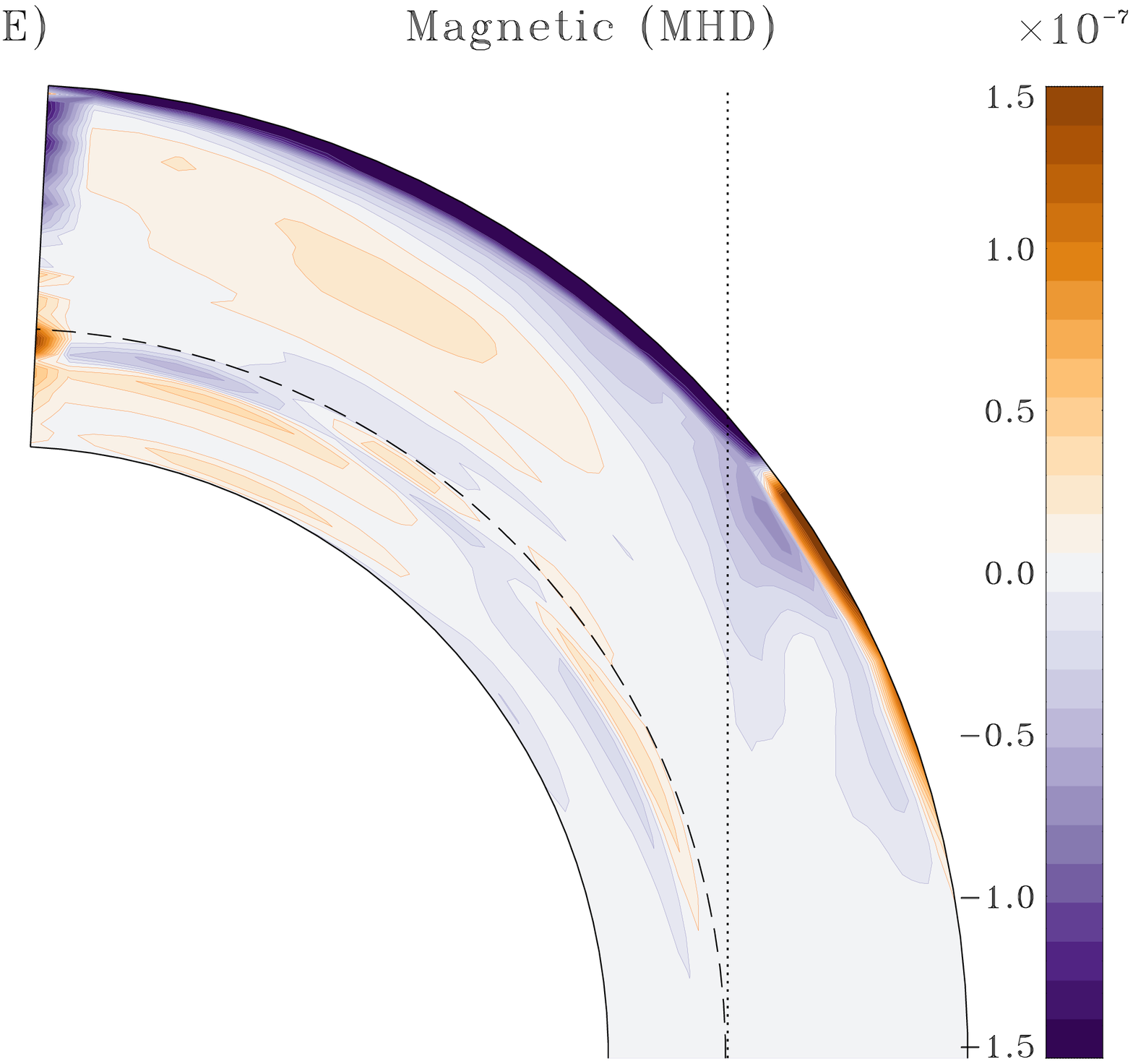}
     \includegraphics[height=6 cm]{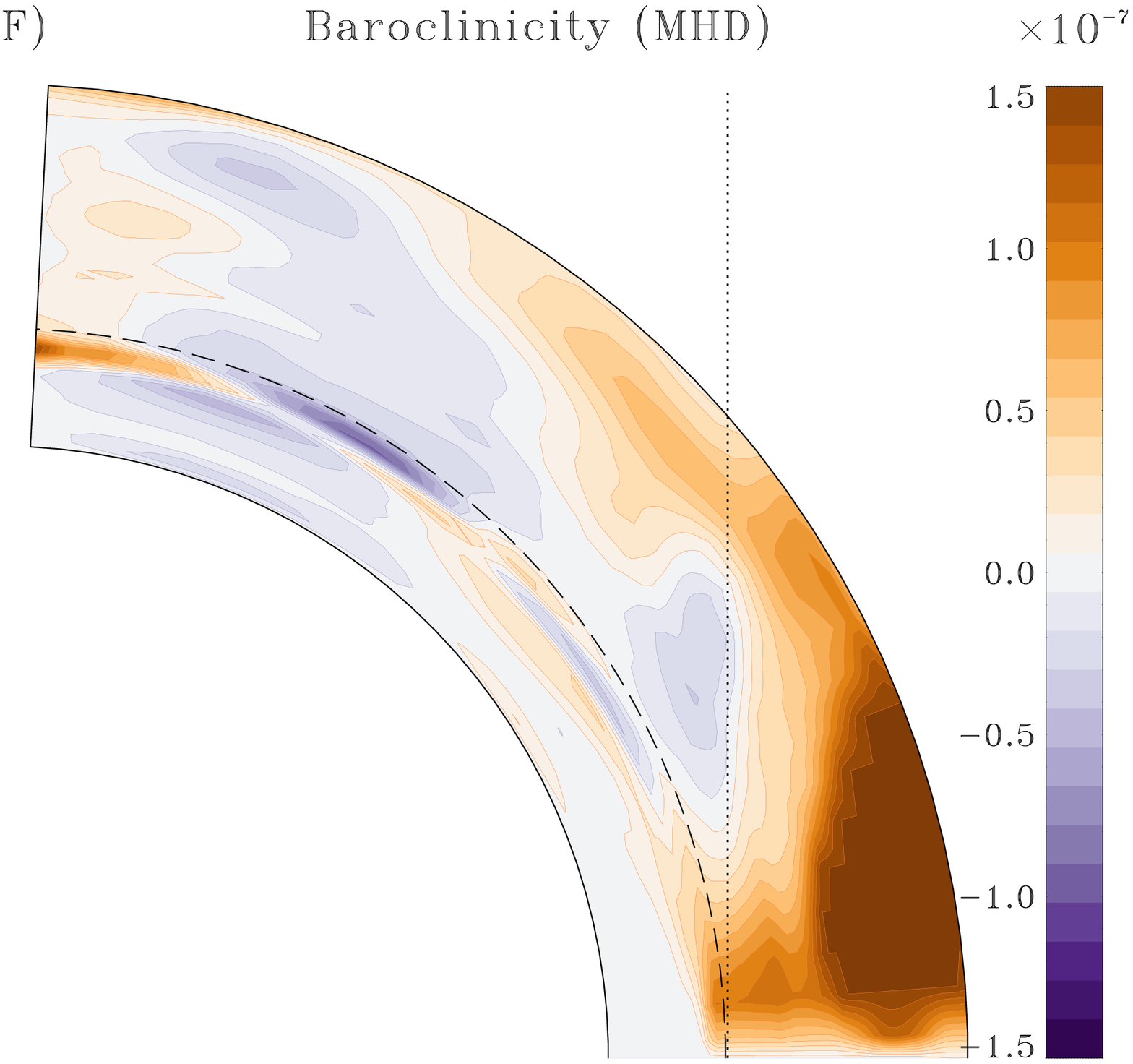}
     \caption{Panels A) and D) show the l.h.s. and r.h.s. the
     MFB equation in the NH for the MHD case. The individual
     terms of the r.h.s. are shown in panels B), C), E) and F). The
     magnetic contribution is the sum of the two individual magnetic
     terms.	The vertical dotted line represents the TC. The
     color scale saturates at $\pm\, 1.5 \times 10^{-7}$ s$^{-1}$.}
     \label{fig:MFB_MHD}
\end{figure*}

The same analysis is now applied to the MHD simulation, but this time taking
into consideration the complete form of equation (\ref{eq:thermalwind}).
We use the same time interval of 10 cycles as in the previous section
for the averaging.
In Fig.\ \ref{fig:MFB_MHD} we opted not to show the compressibility
contribution because, like in the HD case, it is negligible when
compared to the other terms.

Outside the TC, the MHD simulation behaves
similarly to its HD counterpart, with baroclinicity being the dominant
contribution.
In terms of balance
between l.h.s. and r.h.s. of (\ref{eq:thermalwind}), panels A) and D) show
a very good agreement in most of the CZ.
Differences are only evident inside the TC
in the top layers and slightly near the poles.
We attribute these differences to the upper and polar boundary conditions
for the magnetic field.
The difference observed between panels
\ref{fig:MFB_MHD}A) and \ref{fig:MFB_MHD}D) in the top boundary
might come from the fact that we inforce a radial magnetic field at
the surface. This means that when we have a strong poloidal field located
near the surface, it will be forced over a couple grid points by the
boundary condition into the radial direction.
This is exactly the case of the poloidal field configuration during
cycle maximum (see Fig.\ \ref{fig:bmoy_25}E). This
problem might be alleviated in future simulations by introducing more
realistic top boundary conditions like the one used in \cite{warne16}.
The differences between l.h.s. and r.h.s. during
cycle minimum (not shown here) are much smaller. It is the radial
derivative of the poloidal field present in the second magnetic term
of equation (\ref{eq:thermalwind}) that is responsible for this
"artificial" contribution.
There are two other possible sources of error that can explain
the minute differences we find in this balance calculation (and in the
previous section as well). The first
is numerical diffusivity which we cannot measure directly.
The other issue is
related with the different numerical methods used to compute derivatives
and other composite quantities in the main code during the simulation
and \textit{a posteriori}. During the simulation run, EULAG numerics
computes central cell values and fluxes across the cell borders, while
the type of analysis that we perform \textit{a posteriori} assumes values
computed in the cell corners using
centered finite differences.
Differentiation across the poles can also introduce some artifacts.
Nevertheless,
the very good match obtained in the HD case indicates that these two sources
of error are in fact very small, and that the main issue here seems related to
the magnetic field boundary conditions. Despite these possible
sources of uncertainty, we consider that there is a general good
agreement between l.h.s. and r.h.s. for most of the CZ.

By comparing Figs. \ref{fig:MFB_MHD}D--F, we can see that
inside the TC the baroclinic term also has an important role in establishing
MFB, especially in the area close to the inner TC border
(left side of the vertical dotted line in the panels).
In the remainder of the CZ (roughly above 40$^\circ$), the MFB is
maintained by a combination of baroclinic and magnetic contributions
(especially from the mean magnetic field).  \cite{Miesch2006} argued that
MFB (baroclinicity mainly) should have important effects in the tachocline
and lower CZ except in regions where this equilibrium can be disrupted
by strong magnetic fields. Our analysis also points in that
direction. The stretching and advection terms are almost completely
overshadowed by the contributions of the two previous terms.

\begin{figure*}[htb]
  \centering
     \includegraphics[height=6 cm]{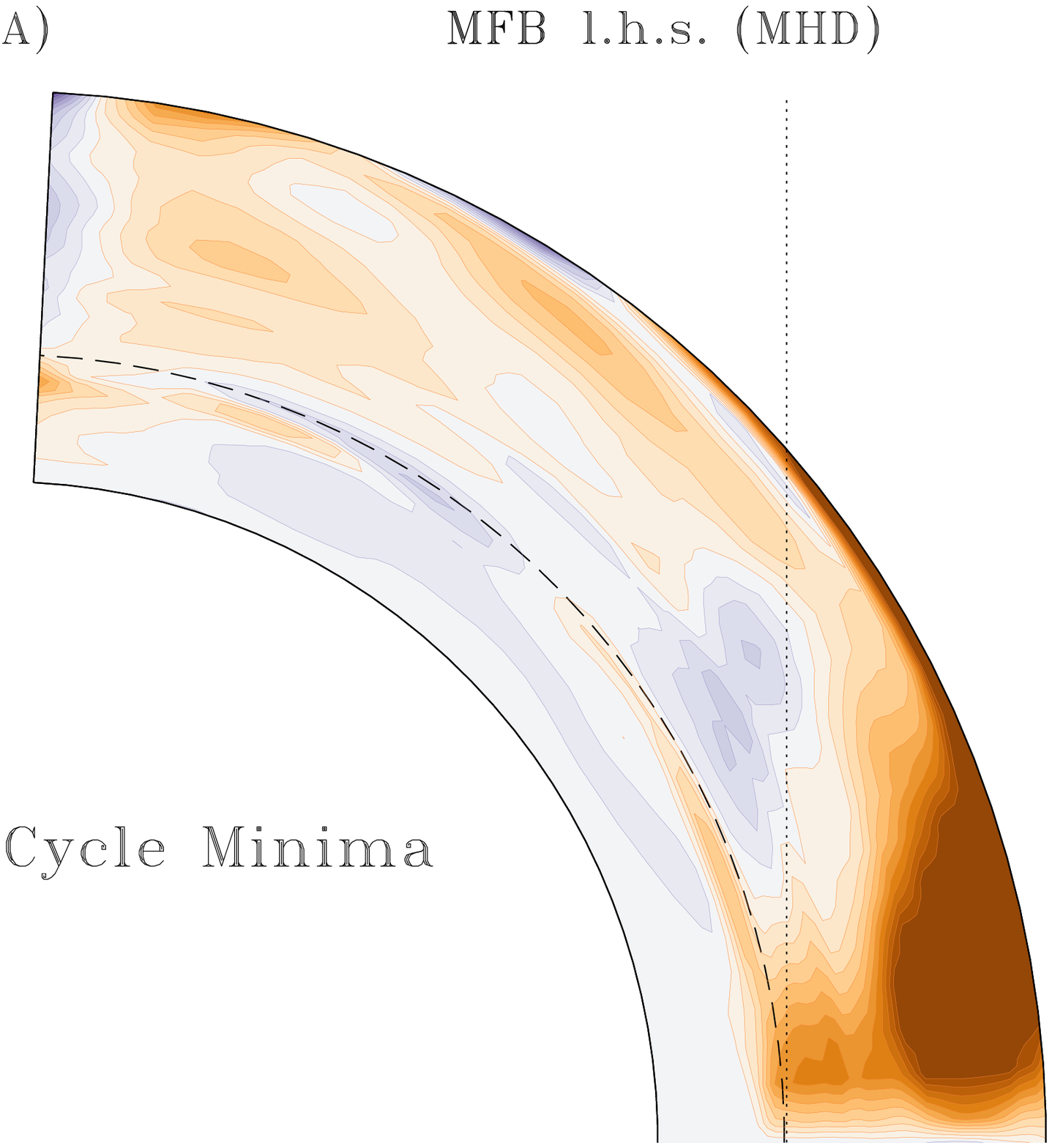}
     \includegraphics[height=6 cm]{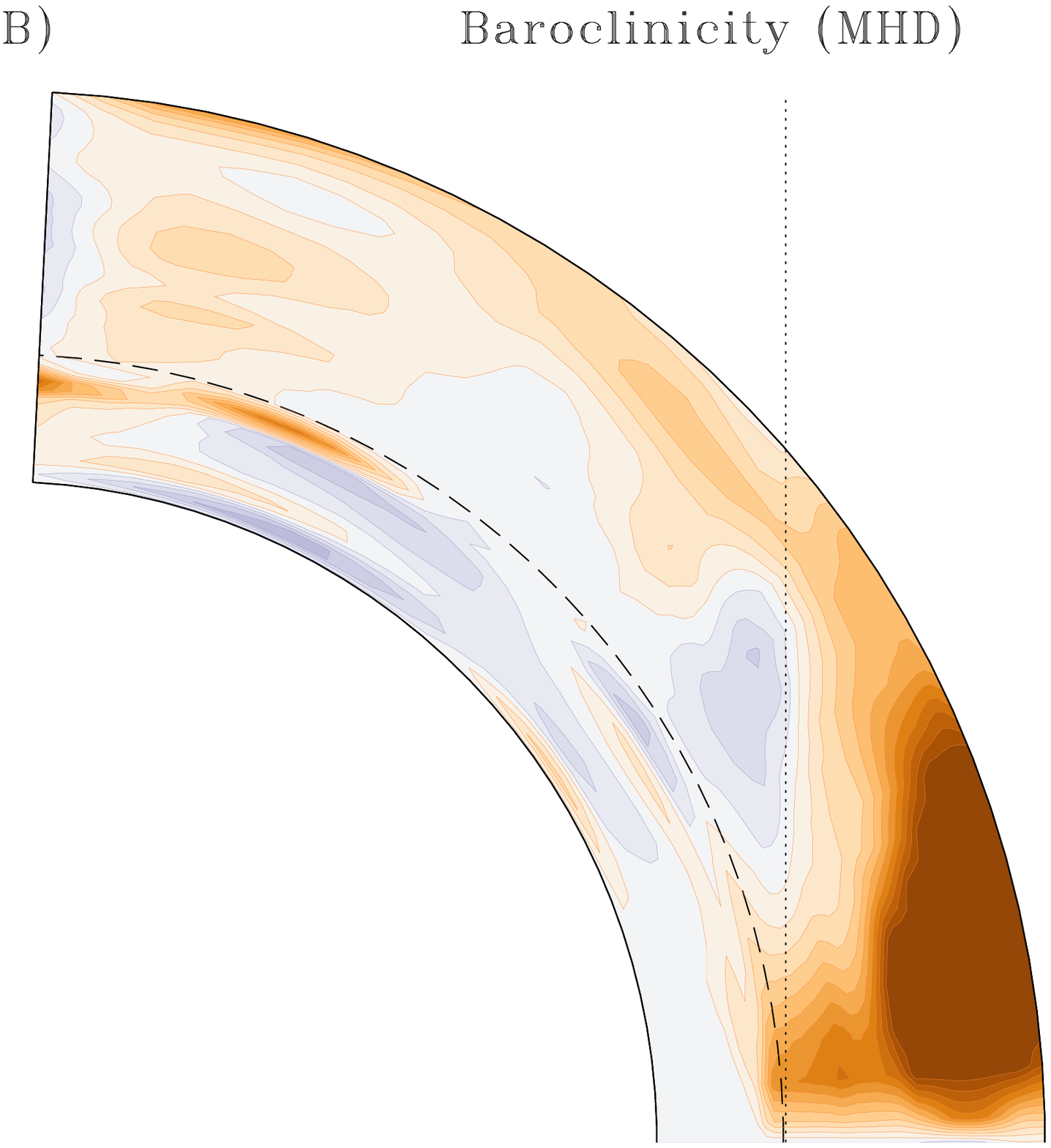}
     \includegraphics[height=6 cm]{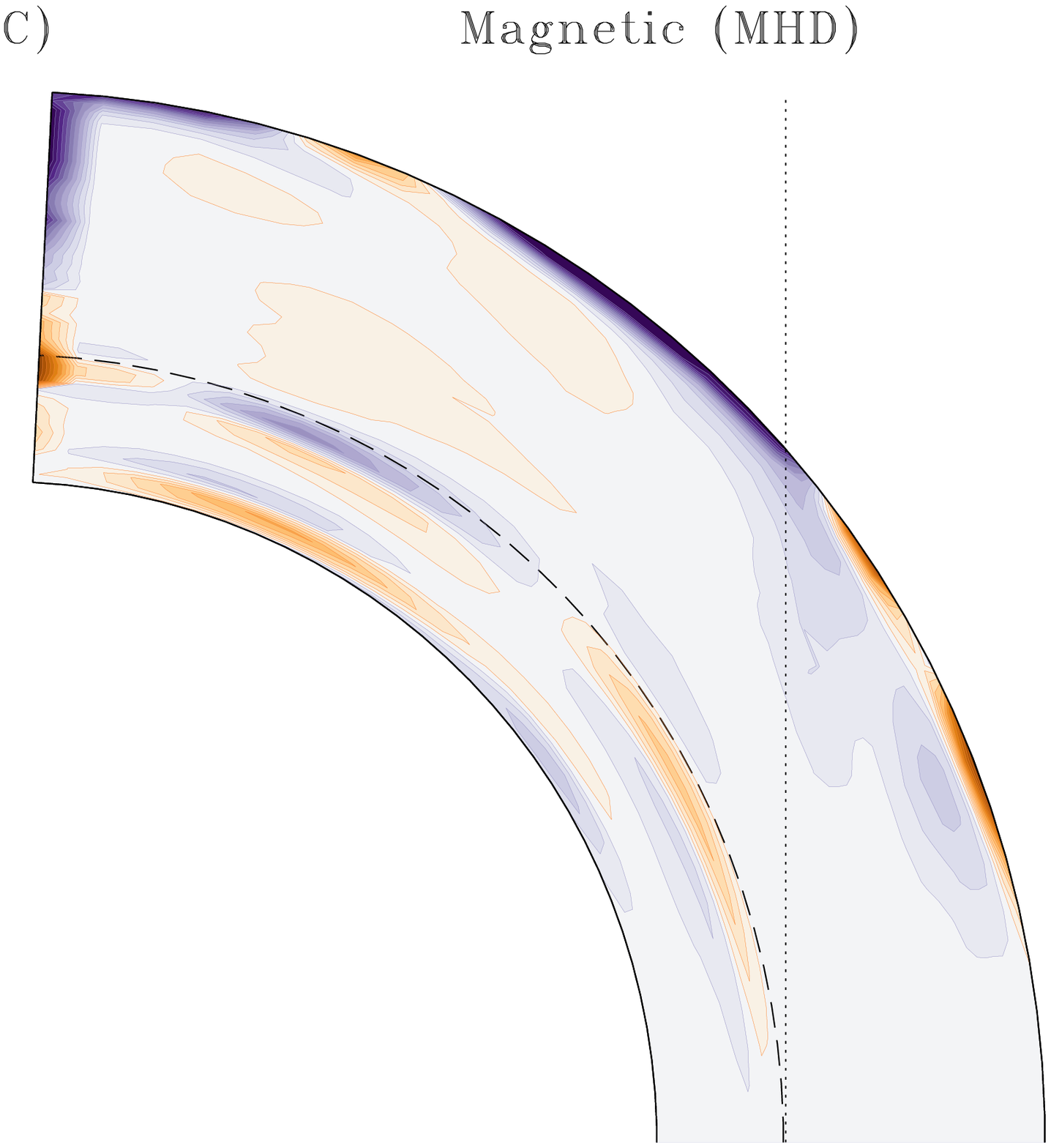}
     \includegraphics[height=6 cm]{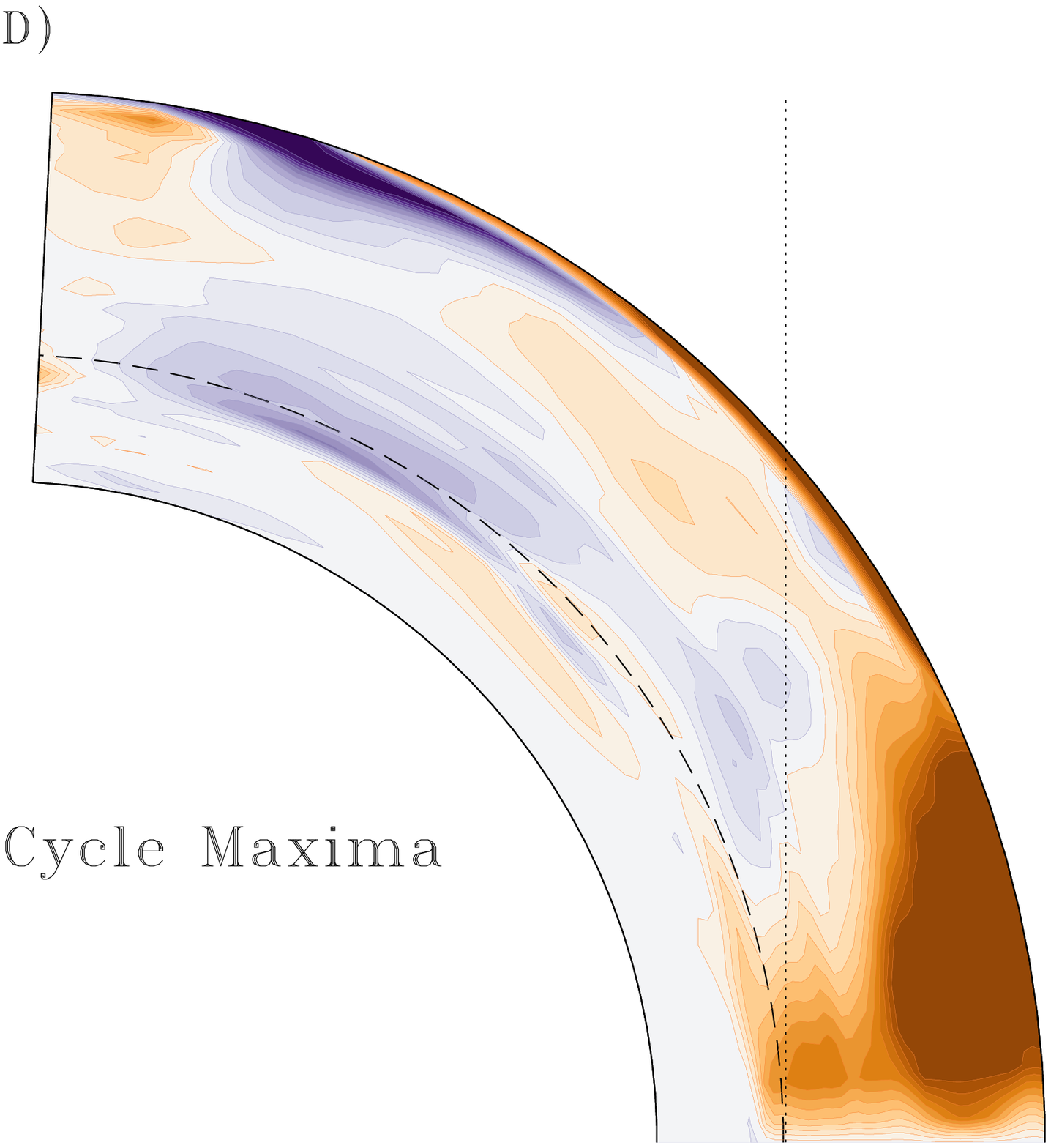}
     \includegraphics[height=6 cm]{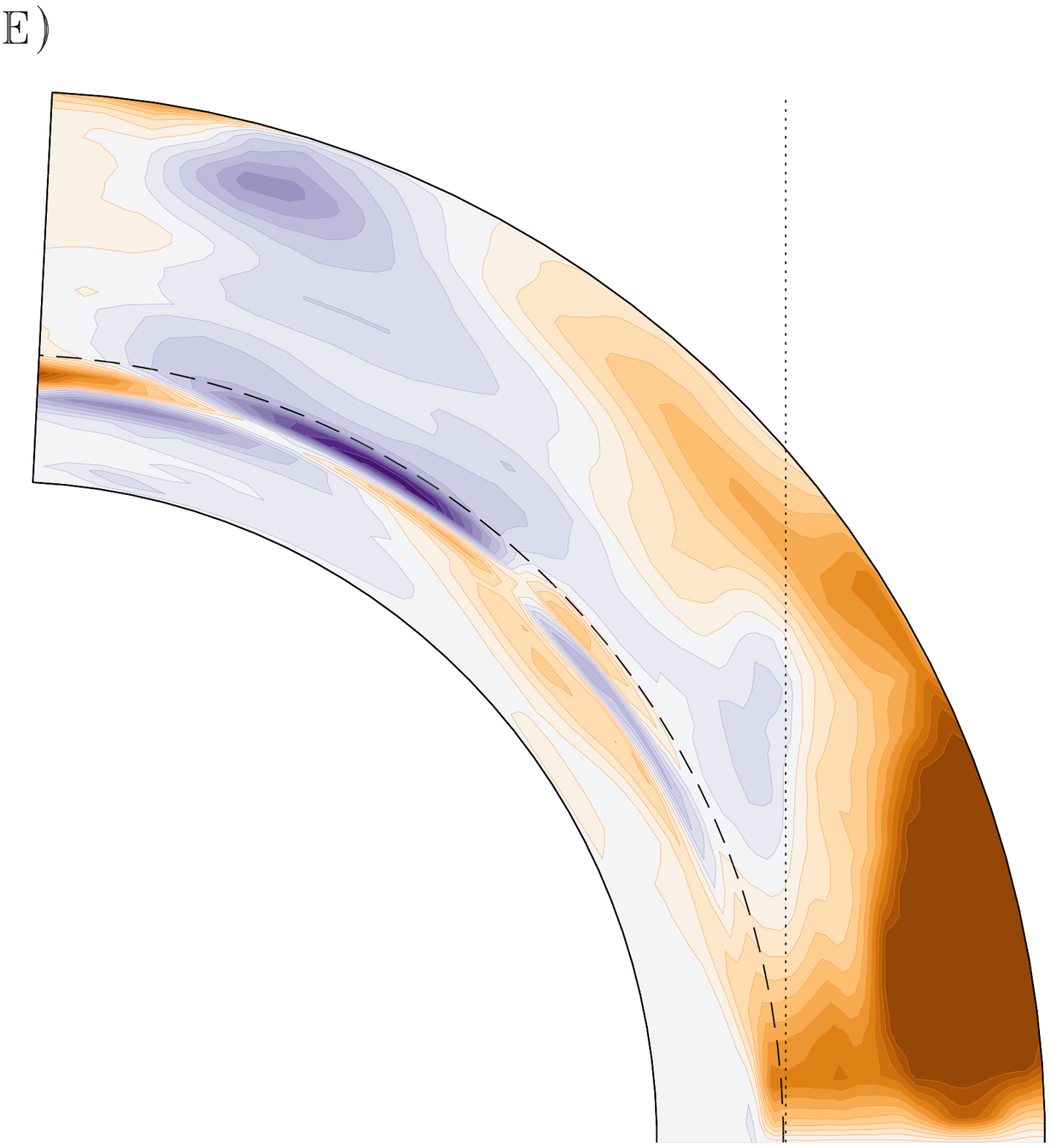}
     \includegraphics[height=6 cm]{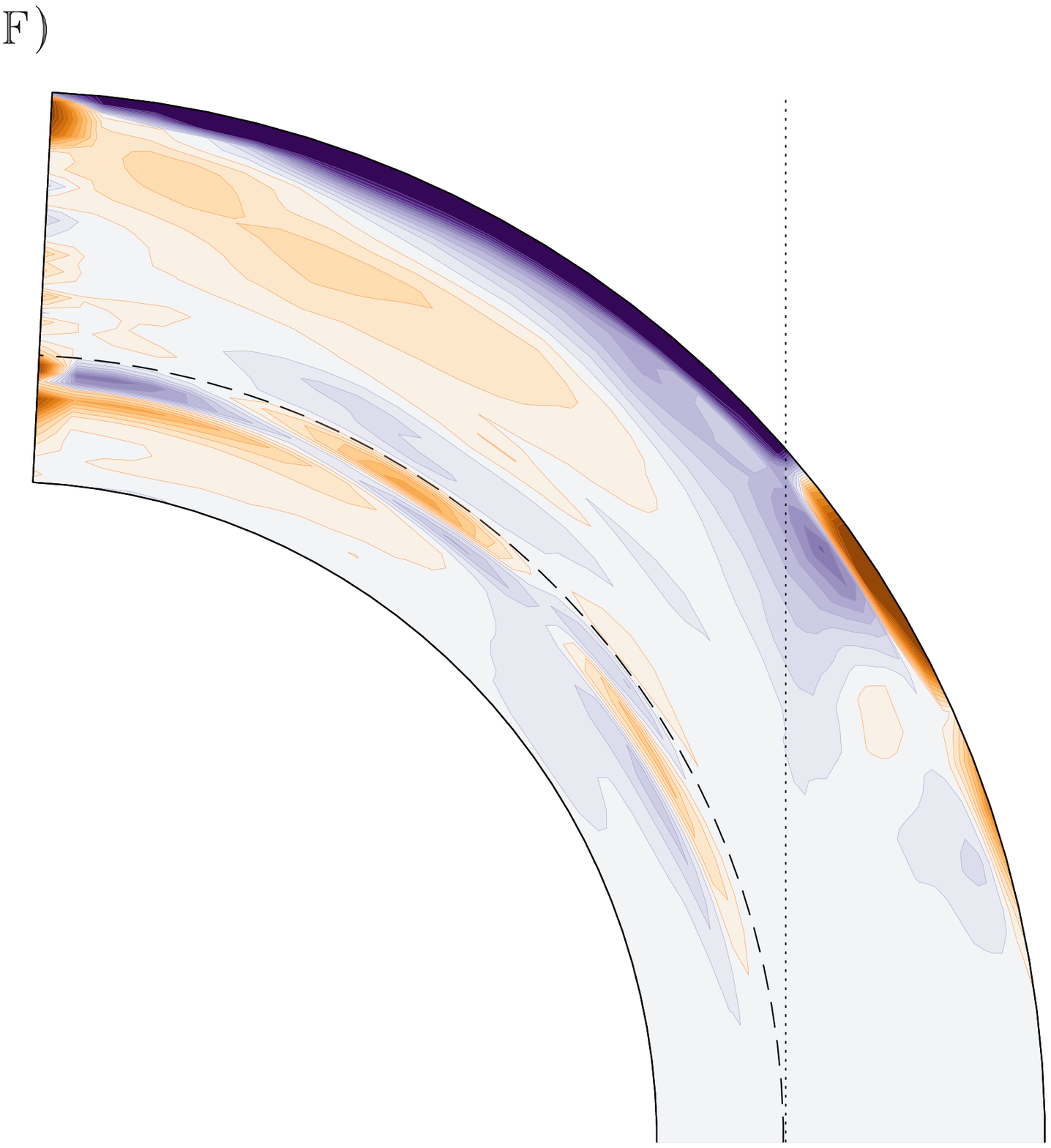}
     \caption{Panels A) to C) present the l.h.s., baroclinic and magnetic
      contributions of the MFB averaged over 10 cycle minima.
      Panels D) to F) show the same quantities of the top row averaged over
      10 cycle maxima. The color scale saturates at $\pm\,
      1.5 \times 10^{-7}$ s$^{-1}$.}
     \label{fig:MFB_MHD_MinMax}
\end{figure*}

We can highlight the influence of the magnetic
field by comparing the MFB at times of cycle minima and maxima.
In Fig.\ \ref{fig:MFB_MHD_MinMax} we present an average of the l.h.s.,
baroclinic, and magnetic contributions over 10 cycle minima (top row)
and maxima (bottom row).

Inside the TC, panels \ref{fig:MFB_MHD_MinMax}A) and
 \ref{fig:MFB_MHD_MinMax}D) show that for latitudes
between 0$^\circ$ and 41$^\circ$ (approximately where the TC
intersects the top boundary) the variation between minima and maxima for the
l.h.s. is small and almost restricted to the stable layers. This is also
applicable for the other two terms and is somewhat expected because
the magnetic field does not have a significant presence at low latitudes in
the bulk of the CZ.
Inside the TC, during cycle maxima there is an enhanced
negative (purple) baroclinic region that spreads from $\sim$48$^\circ$
to 85$^\circ$ in latitude (panel E)). During minimum the baroclinicity tends
to "relax" to a profile more close to the HD case, with a positive
enhancement in the shear region at the bottom of the CZ at high latitudes
and in most of the high latitude range of the bulk of the CZ (panel B)).
The "quadrupolar cell pattern" that we find in the stable layers
in the magnetic term is similar (with opposite sign)
to the one found in the baroclinic term during both maxima and minima,
indicating that these two quantities tend to balance each other.

The existence of this pattern in the stable layers in both terms as well as
the cyclic variation of the baroclinicity between maximum and minimum
are clear indications that the magnetic field is influencing the entropy
(temperature) distribution. This is in close agreement with the physical
mechanism of magnetic modulation of the thermal flux transport
recently proposed by \cite{Cossette2017}. We are currently investigating
the evolutionary patterns of this thermal modulation.

The "quadrupolar cell pattern" in the stable layers is also visible in
Fig.\ \ref{fig:rhsMHD2} and is associated with the large scale magnetic
torque. Its presence in Fig.\ \ref{fig:MFB_MHD_MinMax}
reflects the adjustments in the MC in response to GP.

\section{Conclusions and final remarks}

In this paper we have examined in detail the dynamical driving of meridional
circulation in numerical simulations of solar convection, with and without
a large-scale magnetic cycle. This work is motivated by the fact that
the internal meridional flow in the sun is a weak flow
(a few m s$^{-1}$) compared to convection and differential rotation,
and therefore
quite difficult to measure helioseismically. Yet, this large-scale flow
is believed to play a key role in the class of solar cycle models known
as flux transport dynamos: it sets the cycle period in the advection-dominated
regime, and its temporal variations on these decadal timescales
are believed to influence markedly the amplitude of activity cycles.
Moreover, the surface component of this flow contributes to the poleward
transport of photospheric magnetic flux released by the decay of active
regions, and thus has a direct ---and observed--- impact on the reversal
and amplitude of the surface dipole moment.

In a thick, rotating, stratified turbulent fluid layer,
the meridional flow dynamics are closely coupled
to the balance of angular momentum. Consequently, we analysed
the evolution of the angular momentum
as well as the meridional force balance in two analog simulations
of global solar convection. The first
is purely hydrodynamic (i.e., unmagnetized), while the second includes
magnetic fields and self-consistently generates
a large-scale magnetic cycle undergoing
regular polarity reversals on a multi-decadal timescale.
The comparison between both simulations highlights the role of magnetism as
a driver of meridional flow, and of its spatiotemporal variations.

The results obtained in the HD regime are in good agreement with
previous analyses, and indicate that the convective angular momentum
transport is responsible for the establishment of the large scale
meridional flow through the mechanism of gyroscopic pumping (GP).
This mechanism reflects mainly the action of Reynolds stresses in
areas where the rotation profile presents strong gradients, here
primarily outside the TC.

The MHD simulation used in this study is the \textit{millennium simulation}
presented in PC14. It exhibits a number of solar-like features,
including cyclic large-scale magnetic activity as well as cyclic
variability patterns in the large-scale flows (torsional oscillations and
MC variations) and convective heat flux. However,
in this simulation, these characteristics appear
in a range of latitudes that spans from 45$^\circ$ to 85$^\circ$, i.e.
inside the TC, away from the low latitude strong cylindrical
differential rotation,
while in the Sun they occur at lower latitudes (where cylindrical
rotation is not observed). However, if we compare the
active latitudinal range in the simulation to the Sun's active latitudes,
then the spatiotemporal patterns of rotational torsional oscillations
become quite solar like \citep[see, e.g.][]{Beaudoin2013}, and
the temporally-averaged meridional flow pattern becomes remarkably
similar to the helioseismic measurements of
\cite{Zhao2013} \citep[see][]{PCM15}.

Using the same methodology used to analyse the HD simulation we find that, in
the MHD case, gyroscopic pumping is strongly influenced by the large scale
magnetic field. In section 4 we showed that this influence materializes via
a magnetic torque located at mid latitudes at the bottom of the CZ.
This magnetic torque accelerates and decelerates bands of rotation
situated poleward and equatorward of the toroidal field bands.
Essentially, the magnetic field briefly changes the differential
rotation, and the system re-establishes equilibrium by continuously altering
the MC cell morphology in all of the CZ. The time scale associated with the
recovery of this balance, i.e. the MC response, is of the order of 1 or 2
rotations (months). This is to be contrasted with the time scale associated
with the variation of the large scale magnetic field (several years).
Since the system readjusts quickly,
we can interpret the variation along the magnetic cycle as a quasi-static
process where the system is always very close to equilibrium.
We also note that the area where the magnetic torque is concentrated,
and where the core of the GP mechanism takes place, is situated at the
bottom of the CZ, away from the boundaries of the simulation domain,
and therefore unlikely to be influenced by boundary conditions.

As the MC readjusts its structure to maintain angular momentum balance, it
also has to satisfy the meridional force balance condition.
In section 5 we showed how this MFB is influenced not only by baroclinicity
but also by a magnetic contribution.
Moreover, we find evidence
that the baroclinic term itself is being modulated by magnetism as well.
One of the ways this may happen is by the modulation of heat flux
transport as discussed in \cite{Cossette2017}. The presence of magnetic field
in certain areas alters the way heat is transported by convection and
establishes thermal gradients. In the millennium simulation
we observe a cyclic temperature variation pattern where the poles get cooler
during cycle maxima and hotter than average at cycle minima. We are
currently investigating whether the small CW rotation cell that appears near
the poles at cycle maximum is a consequence of this
MFB constraint in the adjustment of the MC.

One of the most prominent features resulting from these MC variations
is a horizontal convergence of fluid into the equatorward
edge of the toroidal magnetic field bands building up
at the base of the CZ, and the associated mid-latitude upwelling it generates.
This upwelling waxes and wanes in phase with the cyclic evolution
of the magnetic field, becoming most prominent at cycle maximum.
It may interfere with equatorward flux transport, promote flux emergence,
and more generally affect the dynamical coupling between the convection
zone and underlying radiative core over long timescales.

Another noteworthy specific spatiotemporal
MC variation pattern (from the several) that can be extracted from this
numerical experiment merits attention.
In the top layers of our simulation domain, between 50 degrees and the
poles, we observe a poleward flow (see Fig.\ \ref{fig:mcbfly}A).
This surface flow exhibits a characteristic temporal modulation pattern,
due to the appearance of an equatorward flow at high latitudes,
peaking at cycle minima. This pattern is associated with a
decrease of the magnetic torque at these these latitudes
near the surface and the appearance of a CW rotating MC cell (in the NH)
near the poles (see Figs. 3B and 3C).
A somewhat similar pattern is observed at the solar surface. The
observational evidence
\citep{Haber2002, Ulrich2005, Ulrich2010,
Hathaway2014, Bogart2015} indicates that this so-called counter-cell
tends to appear in the descending and minimum phase of the cycle.
This cannot be explained by localized surface
inflows because at those latitudes there are no active
regions.

Generally speaking, the magnetic field alters the characteristics of
convection and mean flows at both small and large scales. What we observe in this simulation
is a general magnetic modulation of convection and its associated
dynamics during cycle rise and maximum, and a subsequent relaxation
towards an HD-like profile when the cycle drops to a minimum.
This raises concerns
regarding the kinematic approach generally
used in mean field and mean field-like
axisymmetric dynamo models of the solar cycle. If the MHD effects
that we see in
this simulation scale up to solar conditions, then the kinematic
approximation might be missing important physical effects.

Our analyses have led to a dynamically consistent scenario for the
spatiotemporal evolution of the large-scale meridional flow in a simulated
solar convection zone.
However, this scenario is established on the basis of
numerical simulation results
carried out in a physical parameter regime far removed from solar internal
conditions, so that one may legitimately question
their relevance to the real Sun and stars.
As a magnetized fluid system, our simulation does generate
behaviors resembling solar observations, most notably decadal large-scale
magnetic cycles, a reasonably solar-like internal differential rotation
and pattern of torsional oscillations, and high-latitude pattern of
surface meridional flow variations. This suggests
---certainly without proving--- that the overall
dynamical interactions between cyclic magnetism, angular momentum balance,
and thermal wind balance taking place in the simulation do capture
similar effects taking place in the solar interior. One potentially
testable prediction emerging from our analysis is the buildup of
a large-scale upwelling starting deep with the convection zone
at active latitudes and peaking at cycle maximum.
Because it is sustained over a time period commensurate with that of the
magnetic cycle, such a spatiotemporally coherent
upflow may actually be detectable helioseismically.
We leave open the search for an associated helioseismic signature
in extant data as an interesting observational challenge.

\begin{acknowledgements} D. Passos is thankful to Sandra Braz for support, and
 acknowledges the financial support from the Funda\c{c}\~{a}o para a
 Ci\^{e}ncia e Tecnologia grant SFRH/BPD/68409/2010 (POPH/FSE), CENTRA-IST,
 the GRPS-UdeM and the University of the Algarve for providing office space.
 P. Charbonneau is supported primarily by a Discovery Grant from the Natural
 Sciences and Engineering Research Council of Canada. All EULAG-MHD
 simulations reported upon in this paper were performed on the computing
 infrastructures of Calcul Qu\'{e}bec, a member of the Compute Canada
 consortium. The National Center for Atmospheric Research is sponsored by
 the National Science Foundation.
\end{acknowledgements}

\bibliographystyle{aa} 
\bibliography{references_v0.9}  

\clearpage

\begin{appendix}
\section{Meridional force balance equation}

The equation for the meridional force balance (\ref{eq:thermalwind})
is derived by computing the zonally averaged $\hat{\mathbf{e}}_\phi$ component
of (\ref{eq:dvorticitydt}). In this appendix we detail the derivation of
each individual term of (\ref{eq:thermalwind}) for our spherical
latitudinal coordinate system.

\subsection*{i) Vorticity stretching}
\bea
    (\mathbf{\omega_a} \cdot \nabla) \mathbf{u}
    &=&  (\mathbf{\omega} \cdot \nabla) \mathbf{u} + (2\mathbf{\Omega_0} \cdot \nabla) \mathbf{u} \nonumber \\
    \langle \hat{\mathbf{e}}_\phi \cdot [(\mathbf{\omega_a} \cdot \nabla) \mathbf{u}]\rangle &=&
     \left\langle \mathbf{\omega}_r \frac{\partial u_\phi}{\partial r}
    - \frac{\mathbf{\omega}_\theta}{r}\frac{\partial u_\phi}{\partial \theta}
    +\frac{\mathbf{\omega}_\phi}{r\, \cos\theta} \frac{\partial u_\phi}    {\partial \phi}
    + \frac{\omega_\phi u_r}{r} -\frac{\omega_\phi u_\theta \tan\theta}{r}\right.
    \nonumber \\
    &&\left. +\, 2 \Omega_0 \left(\sin \theta \frac{\partial u_\phi}{\partial r} + \frac{\cos \theta}{r} \frac{\partial u_\phi}{\partial \theta}\right)
    \right\rangle
    \label{eq:vort_stretch}
    \eea

\subsection*{ii) Vorticity advection}
\bea
     -(\mathbf{u}\cdot \nabla) \mathbf{\omega_a} &=&  -(\mathbf{u} \cdot \nabla)\mathbf{\omega} - 2(\mathbf{u} \cdot \nabla)\mathbf{\Omega_0}\nonumber \\
     \langle -\hat{\mathbf{e}}_\phi \cdot[(\mathbf{u}\cdot \nabla) \mathbf{\omega_a}]\rangle &=& \left\langle - u_r \frac{\partial \omega_\phi}{\partial r}
     -\frac{u_\theta}{r} \frac{\partial \omega_\phi}{\partial \theta}
     -\frac{u_\phi}{r \,\cos\theta} \frac{\partial \omega_\phi}{\partial \phi}
      -\frac{u_\phi \omega_r}{r} -\frac{u_\phi \omega_\theta \tan\theta}{r}
     \right\rangle
     \label{eq:vort_advec}
\eea

\subsection*{iii) Vorticity "compressibility"}
\bea
    -\mathbf{\omega_a}(\nabla \cdot \mathbf{u})
    &=& -\omega (\nabla \cdot \mathbf{u}) - 2\mathbf{\Omega_0} (\nabla \cdot \mathbf{u})\nonumber \\
    \left \langle - \hat{\mathbf{e}}_\phi \cdot [\mathbf{\omega_a}(\nabla \cdot \mathbf{u})]\right \rangle &=&  \left \langle -\omega_\phi \left(\frac{1}{r^2}\frac{\partial (r^2 u_r)}{\partial r}
    + \frac{1}{r \cos \theta}
    \frac{\partial (u_\theta \cos\theta)}{\partial \theta}
    +\frac{1}{r \cos \theta} \frac{\partial u_\phi}{\partial \phi}\right)\right     \rangle
    \label{eq:vort_compress1}
\eea

\subsection*{iv) Baroclinicity }
\bea
    - \nabla \times \mathbf{g}\frac{\Theta'}{\Theta_0} &=&
     - \nabla \left(\frac{\Theta'}{\Theta_0} \right) \times \mathbf{g}
     \nonumber \\
    \left\langle -\hat{\mathbf{e}}_\phi \cdot \left[
     \nabla \left(\frac{\Theta'}{\Theta_0} \right) \times \mathbf{g}
     \right]\right\rangle
     &=&\left\langle \frac{g(r)}{r}\frac{\partial}{\partial \theta} \left(
     \frac{\Theta'}{\Theta_0} \right)\right \rangle
     \label{eq:vort_grav}
\eea

\subsection*{v) Magnetic contribution 1}
\bea
    \left\langle  \hat{\mathbf{e}}_\phi \cdot \frac{1}{\mu_0}
    \left(  \nabla \frac{1}{\rho_0}\right)
    \times (\mathbf{B} \cdot \nabla ) \mathbf{B} \right\rangle
    &=& \left \langle \frac{1}{\mu_0} \frac{\partial}{\partial r}
    \left(\frac{1}{\rho_0}\right) \left[ -B_r \frac{\partial B_\theta}
    {\partial r} - \frac{B_\theta}{r}\frac{\partial B_\theta}{\partial \theta}
    \right. \right. \nonumber \\
    && \left. \left. -\frac{B_\phi}{r\, \cos\theta} \frac{\partial B_\theta}{\partial \phi}
    -\frac{B_\phi^2}{r} \tan\theta - \frac{B_\theta B_r}{r}\right]
    \right\rangle
    \label{eq:vort_mag1}
\eea

\subsection*{vi) Magnetic contribution 2}
\bea
     \left \langle \hat{\mathbf{e}}_\phi \cdot \frac{1}{\rho_0 \mu_0}
     (\nabla \times (\mathbf{B}\cdot \nabla)\mathbf{B}) \right \rangle
     &=&\left \langle \frac{1}{r\, \mu_0\, \rho_0} \left[
    \frac{\partial }{\partial r}
    \left( -r\, B_r \frac{\partial  B_\theta }{\partial r}
    -  B_\theta  \frac{\partial  B_\theta }{\partial \theta}
    - \frac{B_\phi}{\cos\theta} \frac{\partial B_\theta}{\partial \phi}
    - B_\phi^2 \tan\theta
    - B_\theta  B_r  \right) \right. \right. \nonumber \\
    && + \left. \left. \frac{\partial }{\partial \theta}
    \left(
     B_r  \frac{\partial  B_r }{\partial r}
    + \frac{ B_\theta }{r} \frac{\partial B_r }{\partial \theta}
    +\frac{B_\phi}{r\,\cos\theta}\frac{\partial B_r}{\partial \phi}
    - \frac{B_\theta^2}{r}
    - \frac{B_\phi^2 }{r} \right)
    \right] \right \rangle \,\,\, ,
    \label{eq:vort_mag2}
\eea

\end{appendix}

\end{document}